\documentclass{emulateapj}
\usepackage{amssymb,amsmath}
\newcounter{address}
\newcommand{\ie}{i.e.}
\newcommand{\etal}{et al.}
\newcommand{\dd}{\mathrm{d}}
\newcommand{\eg}{e.g.}
\newcommand{\eqnname}{equation}
\newcommand{\equationname}{\eqnname}
\renewcommand{\tablename}{Table}
\renewcommand{\figurename}{Figure}
\newcommand{\figurenames}{\figurename~s}
\newcommand{\sectionname}{$\mathsection$}
\newcommand{\segue}{SEGUE}

\newcommand{\apoor}{\ensuremath{\alpha}-young}
\newcommand{\aenhanced}{\ensuremath{\alpha}-old}

\newcommand{\feh}{\ensuremath{[\mathrm{Fe/H}]}}
\newcommand{\afe}{\ensuremath{[\alpha\mathrm{/Fe}]}}

\newcommand{\dens}{\ensuremath{\nu_*}}

\renewcommand{\vec}[1]{\ensuremath{\mathbf{#1}}}
\newcommand{\vecx}{\ensuremath{\vec{x}}}
\newcommand{\vecv}{\ensuremath{\vec{v}}}
\newcommand{\vecj}{\ensuremath{\vec{J}}}
\newcommand{\df}{\ensuremath{f}}
\newcommand{\qdf}{\ensuremath{\mathrm{qDF}}}
\newcommand{\paramsdf}{\ensuremath{\vec{p}_{\mathrm{DF}}}}
\newcommand{\paramspot}{\ensuremath{\vec{p}_\Phi}}

\newcommand{\map}{MAP}
\newcommand{\dex}{\ensuremath{\,\mathrm{dex}}}
\newcommand{\Gyr}{\ensuremath{\,\mathrm{Gyr}}}
\newcommand{\kpc}{\ensuremath{\,\mathrm{kpc}}}
\newcommand{\pc}{\ensuremath{\,\mathrm{pc}}}
\newcommand{\kms}{\ensuremath{\,\mathrm{km\ s}^{-1}}}

\newcommand{\inv}{\ensuremath{^{-1}}}

\newcommand{\sigoneone}{\ensuremath{\Sigma_{1.1}}}
\newcommand{\kzoneone}{\ensuremath{K_{Z,1.1}}}
\newcommand{\jr}{\ensuremath{J_R}}
\newcommand{\jphi}{\ensuremath{J_\phi}}
\newcommand{\jz}{\ensuremath{J_Z}}
\newcommand{\lz}{\ensuremath{L_Z}}

\newcommand{\bo}{BO12}
\newcommand{\iv}{I08}
\newcommand{\an}{An09}

\submitted{}

\begin{document}

\title{A direct dynamical measurement of the Milky Way's disk surface
  density profile, disk scale length, and dark matter profile at
  $4\kpc \lesssim R \lesssim 9\kpc$}
\author{Jo~Bovy\altaffilmark{1,2,3} \&
  Hans-Walter~Rix\altaffilmark{4}}
\altaffiltext{\theaddress}{\label{IAS}\stepcounter{address} Institute
  for Advanced Study, Einstein Drive, Princeton, NJ 08540, USA}
\altaffiltext{\theaddress}{\label{Hubble}\stepcounter{address} Hubble
  Fellow}
\altaffiltext{\theaddress}{\label{email}\stepcounter{address}
  Correspondence should be addressed to bovy@ias.edu~.}
\altaffiltext{\theaddress}{\label{MPIA}\stepcounter{address}
  Max-Planck-Institut f\"ur Astronomie, K\"onigstuhl 17, D-69117
  Heidelberg, Germany}

\begin{abstract} 
 We present and apply rigorous dynamical modeling with which we infer
 unprecedented constraints on the stellar and dark matter mass
 distribution within our Milky Way (MW), based on large sets of
 phase-space data on individual stars.  Specifically, we model the
 dynamics of 16,269 G-type dwarfs from SEGUE, which sample $5\kpc <
 R_{GC} < 12\kpc$ and $0.3\kpc$ $\lesssim |Z| \lesssim 3\kpc$. We
 independently fit a parameterized MW potential and a three-integral,
 action-based distribution function (DF) to the phase-space data of 43
 separate abundance-selected sub-populations (\map s), accounting for
 the complex selection effects affecting the data.  We robustly
 measure the total surface density within $1.1\kpc$ of the mid-plane
 to $5\%$ over $4.5 < R_{GC} < 9\kpc$. Using metal-poor \map s with
 small radial scale lengths as dynamical tracers probes $4.5 \lesssim
 R_{GC} \lesssim 7\kpc$, while \map s with longer radial scale lengths
 sample $7 \lesssim R_{GC} \lesssim 9\kpc$.  We measure the
 mass-weighted Galactic disk scale length to be $R_d =
 2.15\pm0.14\kpc$, in agreement with the photometrically inferred
 spatial distribution of stellar mass.  We thereby measure dynamically
 the mass of the Galactic stellar disk to unprecedented accuracy: $M_*
 = 4.6\pm0.3+3.0\,(R_0/\kpc-8)\times10^{10}\,M_\odot$ and a total
 local surface density of $\Sigma_{R_0}(Z=1.1\kpc) =
 68\pm4\,M_\odot\pc^{-2}$ of which $38\pm4\,M_\odot\pc^{-2}$ is
 contributed by stars and stellar remnants. By combining our surface
 density measurements with the terminal velocity curve, we find that
 the MW's disk is maximal in the sense that $V_{c,\mathrm{disk}} /
 V_{c,\mathrm{total}} = 0.83$$\pm$$0.04$ at $R=2.2\,R_d$.  We also
 constrain for the first time the radial profile of the dark halo at
 such small Galactocentric radii, finding that $\rho_{\mathrm{DM}}
 (r;\approx R_0) \propto 1 / r^{\alpha}$ with $\alpha < 1.53$ at
 95\,\% confidence. Our results show that action-based
 distribution-function modeling of complex stellar data sets is now a
 feasible approach that will be fruitful for interpreting Gaia data.
\end{abstract}

\keywords{
	Galaxy: abundances
	---
	Galaxy: disk 
	---
	Galaxy: fundamental parameters
	---
        Galaxy: kinematics and dynamics
        ---
	Galaxy: structure
        ---
        solar neighborhood
}

\section{Introduction}

The mass distribution of the Milky Way inside $R \sim 10\kpc$ is
uncertain in many respects. While good measurements exist of the
rotation curve, these do not allow the separation of the major
Galactic components: bulge, halo, and stellar and gas disks. Hence the
relative contributions as well as the density profiles of these
components are still very uncertain. Determining these is fundamental
to our understanding of the Milky Way's formation and evolution and to
evaluating the importance of dark matter within the visible extent of
galaxies. Because the Milky Way is in many ways an anchor point for
empirical relations used to determine the physical properties of
external galaxies, this ultimately has a large impact on the study of
the low-redshift Universe \citep{Rix13a}.

The structure, size, and mass of the Milky Way's disk in particular
are poorly known. Measurements of the radial scale length of the disk
over the last few decades have produced values anywhere in between
$2\kpc$ and $5\kpc$ (see, \eg, \citealt{Sackett97a} for an
overview). While photometric measurements have improved with the
advent of wide-area surveys with accurate photometry, systematic
uncertainties in the distance scale significantly limit the precision
and accuracy of photometrically measured scale lengths
\citep[\eg,][]{Juric08a}, and such measurements only trace the radial
luminosity profile of the tracer stars rather than the mass profile of
the disk. This is especially problematic as it has recently become
clear that the stellar disk contains abundance-distinct subcomponents
with scale lengths ranging from $2\kpc$ to $>4.5\kpc$
\citep{BovyMAPstructure}, and in particular that the thicker disk
components in the Milky Way have a much shorter scale length than the
thin disk \citep{Bensby11a,BovyMAPstructure,Cheng12a}. Thus, more than
ever it is important to measure the mass-weighted radial profile,
which can be achieved using the dynamics of tracer populations. There
are currently no plausible dynamical measurements of the scale length
of the disk. While there are global potential fits to a variety of
dynamical data \citep[\eg,][]{McMillan11a}, these cannot measure the
scale length without strong prior assumptions as there have been no
dynamical data and models that constrain the disk scale length to
date; because of this, the disk's scale length is the most important
unknown in disentangling the contributions from the disk and the dark
halo to the mass distribution near the disk \citep{Dehnen98a}.

In external galaxies, rotation curves are often decomposed into disk
and halo contributions using the maximum-disk hypothesis
\citep{vanAlbada86a}: the disk component is assumed to contribute the
maximum amount possible without exceeding the rotation curve anywhere;
in practice this means that such maximal disks contribute $85\pm10\%$
of the rotation velocity at the peak of the disk rotation curve,
allowing for the contribution of a bulge and a cored halo
\citep{Sackett97a}. Observational evidence for or against the
maximum-disk hypothesis is scant: \eg, \citet{Courteau99a} argues
against spiral galaxy disks being generically maximal based on the
residuals from the Tully-Fisher relation, while detailed gas
kinematics shows that some disks, particularly fast rotators, are
maximal \citep[\eg,][]{Weiner01a,Kranz03a} (see
\sectionname~\ref{sec:cfexternal} for more discussion of
this). Ultimately, this is due to the fact that directly measuring and
decomposing the mass profile of external disks is close to
impossible. However, because mass-to-light ratios for external spiral
galaxies are calibrated using the maximum-disk hypothesis
\citep[\eg,][]{Bell01a}, the fate of the maximum-disk hypothesis is
intimately connected with many conclusions regarding the stellar-mass
content of the low-redshift Universe
\citep[\eg,][]{Bell03a,Li09a}. Whether the Milky Way has a maximum
disk is unknown \citep{Sackett97a} and determining whether it does is
important since it can provide a robust data point for the discussion
of the maximum-disk hypothesis and because it could be used to
calibrate dynamical measurements of the radial profile of external
galaxies. Because the local midplane density and the local surface
density are reasonably well measured
\citep[\eg,][]{Kuijken89b,Holmberg00a}, whether the Milky Way's disk
is maximal hinges on whether it has a short stellar disk mass scale
length.

Because the rotation curve reflects the sum of the radial in-plane
accelerations contributed by the various Galactic components, precise
measurements of the radial profile and mass of only the disk also
allows the radial profile of the dark-matter halo to be
constrained. Measurements of the local density of dark matter have
significantly improved recently \citep{BovyTremaine,Zhang13a}, but
there are no dynamical constraints on the radial profile of dark
matter at $R \lesssim10\kpc$. The large microlensing optical depth
measured toward the bulge argues for a relatively shallow halo profile
\citep{Binney01a}, although this constraint has been relaxed compared
to \citet{Binney01a} due to the downward revision of the optical depth
\citep[\eg,][]{Popowski05a} and of the local dark matter density since
2001 \citep{BovyTremaine,Zhang13a}. Measuring the radial profile of
dark matter in the inner Milky Way would be invaluable for analyses of
dark-matter indirect detection toward the Galactic center and for
comparing the Milky Way to the predictions from cosmological
simulations: A Milky-Way-like galaxy is presumed to have an inner
dark-matter profile close to the Navarro-Frenk-White (NFW;
\citealt{Navarro97a}) profile seen in simulations or steeper due to
the adiabatic contraction that is expected to occur when the Milky
Way's massive disk forms within the dark matter halo.

In this paper, we constrain the mass distribution in the inner Milky
Way by measuring the vertical force at $|Z|\approx1\kpc$ as a function
of radius between $R\approx 4\kpc$ and $\approx9\kpc$. Because the
vertical force and the surface density are approximately proportional
near the disk \citep{Kuijken89a,BovyTremaine}, this quite directly
measures the radial surface-density profile near the disk, allowing
the mass-weighted scale length of the disk to be obtained without any
prior assumptions derived from stellar number counts; we measure the
radial profile of the mass rather than that of the luminosity.

We measure the vertical force as a function of radius using modeling
based on simple, parameterized, six-dimensional (phase-space),
three-action distribution-functions (DFs), developed by
\citet{Binney10a,Binney12b} and \citet{McMillan13a}. In
\citet{Ting13a} we showed that this approach should work well when
applied to subsets of Galactic disk stars with very similar elemental
abundances, so called mono-abundance populations (\map s). In a recent
series of papers, we \citep{BovyMAPkinematics,BovyMAPstructure}
demonstrated that these constitute structurally-simple disk
populations. Such populations can hence be described well by
few-parameter families of distribution functions, which allows a full
likelihood-based exploration of the joint posterior distribution
function (PDF) of dynamical (gravitational potential) and DF
parameters. In this paper, we apply this methodology for the first
time to a real data set, G-type dwarfs from the SDSS/SEGUE survey
\citep{Yanny09a}. Each \map, \ie, each distinct, abundance-selected
stellar subset, can act as a separate tracer population of the
gravitational potential. We focus our efforts in this paper on
measuring the dynamical quantity of interest that is most robustly
determined for each \map. This turns out to be the vertical force at
$1.1\kpc$ at a \emph{single} Galactocentric radius for each \map,
where the radius is determined directly from the dynamical PDF for
each \map; because the \map s span a wide range in orbital
distributions, this leads to measurements between $R\approx4\kpc$ and
$R\approx9\kpc$ for 43 \map s.

We use these new measurements in combination with constraints on the
rotation curve, primarily from terminal-velocity measurements, and
with measurements of the local vertical profile of the vertical force
to constrain the mass distribution at $R \lesssim10\kpc$ and its
decomposition into disk and dark halo contributions. Because our
measurements are at $1.1\kpc$, we strongly constrain the properties of
the Milky Way's disk, finding a relatively short mass scale length of
$2.15\pm0.14\kpc$ and deriving precise measurements of the local
column density of stars and of the total mass in the disk. We find
that the Milky Way's disk is maximal, providing about $70\%$ of the
rotational support at 2.2 radial scale lengths ($V_{c,\mathrm{disk}} /
V_{c,\mathrm{total}} = 0.83$$\pm$$0.04$; the rotational support is
$(V_{c,\mathrm{disk}} / V_{c,\mathrm{total}})^2 = 0.69 \pm 0.06$). As
the halo contributes very little to the rotation curve and the
vertical force at $1\kpc$, we only weakly constrain the radial profile
of the dark matter, but we nevertheless obtain a first constraint:
$\alpha < 1.53$ at 95\,\% confidence for a model in which
$\rho_{\mathrm{DM}} (r;\approx R_0) \propto 1 / r^{\alpha}$.

The outline of this paper is as follows. We describe the SEGUE G-dwarf
data in \sectionname~\ref{sec:data}. In \sectionname~\ref{sec:method}
we explain in detail the various steps involved in the dynamical
modeling of abundance-selected populations of stars using a DF
model. We present the results from this modeling in
\sectionname~\ref{sec:indivresults} and discuss how we turn these
results into a measurement of the surface density to $1.1\kpc$ at a
single radius for each \map. We show the resulting surface-density and
vertical-force profiles in \sectionname~\ref{sec:surfresults} and
discuss systematics associated with this measurement in
\sectionname~\ref{sec:systematics}. In
\sectionname~\ref{sec:potentialfit}, we use these new measurements
together with additional dynamical data to fit mass models of the
inner Milky Way. We discuss the implications of these new measurements
in \sectionname~\ref{sec:discuss}. We recapitulate our main
conclusions in \sectionname~\ref{sec:conclusion} and look ahead to the
dynamical analysis of Gaia data. \appendixname~\ref{sec:normint} gives
technical details of the calculation of the effective survey volume
required in the dynamical modeling, and
\appendixname~\ref{sec:datamodel} presents various detailed
comparisons between the spatial and kinematical distribution of the
data and that of our best-fit dynamical models. We recommend that
readers not interested in the technical details of the dynamical
models read the introduction of \sectionname~\ref{sec:method}, which
briefly summarizes the main ingredients of the modeling, and then
\sectionname~\ref{sec:surfresults} for the measurement of the
vertical-force profile, and \sectionname~\ref{sec:potentialfit} and
following Sections for the implications of these new measurements for
the mass distribution in the inner Milky Way. In this paper, we use
\bo\ to collectively refer to the set of
\citet{BovyNoThickDisk,BovyMAPkinematics,BovyMAPstructure} papers,
appending b,c, or d to indicate specific papers in that series.


\section{SDSS/SEGUE data}\label{sec:data}

The data used in this paper are largely the same as the SEGUE G-dwarf
data used in \bo. We summarize the main properties of these data with
a focus on what is new. We refer the reader to \bo cd for a full
description of the G-star data used here.

\subsection{G-dwarf data}\label{sec:Gdata}

The G-star sample used in this analysis is identical to the one used
in \bo d, but we now also use the stars' line-of-sight velocities and
proper motions explicitly to calculate the vertical velocities and
errors, as in \bo c.  For a detailed description of the sample, we
refer to \bo d, \bo c, \citet{Yanny09a}, and \citet{Lee11b}. The
G-dwarf sample is the largest of the systematically targeted
sub-samples in \segue\ to explore the Galactic disk; they are the
brightest tracers whose main-sequence lifetime is larger than the
expected disk age ($\approx 10\Gyr$) for all but the most metal-rich
stars with $\feh \gtrsim 0.2\dex$. Their rich metal-line spectrum
affords velocity determinations good to $\sim\!  5$ to 10~\kms
\citep{Yanny09a}, as well as good abundance ($\afe , \feh$)
determinations ($\delta_{\afe}\sim 0.1$ dex, $\delta_{\feh}\sim 0.2$
dex\footnote{In this paper, we use $\delta$ to indicate observational
  uncertainties and reserve $\sigma$ for the velocity dispersion of
  stellar populations.},
\citealt{Lee08a,Lee08b,AllendePrieto08a,Schlesinger10a,Lee11a,Smolinski11a};\bo
c). The distances to the sample stars range from $0.6$~kpc to nearly
4~kpc, with stars somewhat closer to the disk plane being sampled by
the lines of sight at lower Galactic latitudes. The effective minimal
distance limit of the stars (600~pc) implies that the vertical heights
below one scale height ($|Z| < h_z$) of the thinner disk components is
not well sampled by the G-dwarf data. As in \bo, we employ a
signal-to-noise ratio cut of S/N $> 15$.

We transform distances and velocities to the Galactocentric rest-frame
by assuming that the Sun's displacement from the midplane is 25 pc
toward the north Galactic pole \citep{Chen01a,Juric08a}, that the Sun
is located at 8 kpc from the Galactic center
\citep[\eg,][]{Ghez08a,Gillessen09a,Bovy09b}, and that the Sun's
vertical peculiar velocity with respect to the Galactic center is
$7.25\,\kms$ \citep{Schoenrich10a}. In particular, we note that none
of the results in this paper are affected by the details of the Sun's
peculiar motion in the plane of the Galaxy. We use the uncertainties
on the distances, line-of-sight velocities, and proper motions to
calculate the uncertainty in the vertical velocities, but otherwise we
assume that distance uncertainties are negligible (which is a good
approximation because the typical distance uncertainty of
$\sim\!10\,\%$ is much smaller than any Galactic spatial
gradient). Uncertainties are typically 10\,\% in distance,
$\sim\!3.5$~mas yr\inv\ in proper motion \citep{Munn04a}, and $5$ to
$10\kms$ in line-of-sight velocity. The distance errors are obtained
by marginalizing over the color and apparent-magnitude errors that
enter the photometric distance relation (Equation~(A7) of
\citealt{Ivezic08a}, \iv\ hereafter, see also \bo d), which is assumed
to have an intrinsic scatter of 0.1 mag in the distance modulus. These
uncertainties lead to tangential velocity uncertainties that are
$\sim\!50\kms$ at $3\kpc$ and a smaller contribution than that to the
vertical velocity uncertainty (depending on Galactic latitude). Stars
in our sample at such large distances are either far from the plane as
part of a population with a large velocity dispersion, or closer to
the Galactic center, where the velocity dispersion is larger by
$\sim\!50\,\%$ because it rises exponentially with a scale length of
$\sim\!7\kpc$ (\bo c). Therefore, the intrinsic dispersion is larger
than the typical velocity-measurement uncertainty for all populations
analyzed below.

\begin{figure}[t!]
  \includegraphics[width=0.5\textwidth,clip=]{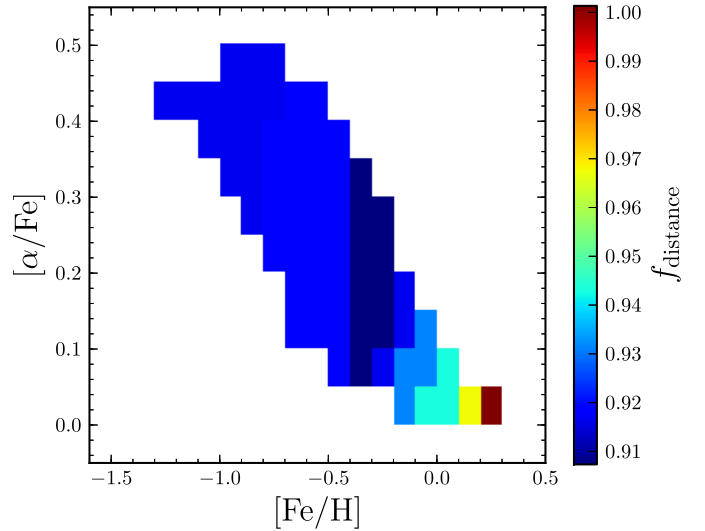}
  \caption{Distance factors applied to the \citet{Ivezic08a}
    photometric distance relation to put it on the distance scale of
    \citet{An09a}. These factors are primarily a function of \feh, but
    are calculated based on the mean \feh\ of each
    \map.}\label{fig:andistance}
\end{figure}

Motivated by the analysis in \bo\ and \citet{Ting13a}, we treat in
what follows \map s as independent tracer populations of the Galactic
potential, whose fits provide independent constraints on the
potential. Operationally \map s are defined here as stars whose
\feh\ and \afe\ lie within bins $0.1\,\dex$ and $0.05\,\dex$ wide, as
in \bo.

One significant change to the distances that we made for the current
analysis is that we have placed the distances on the scale of
\citet{An09a}, \an\ hereafter, which was used in many other analyses
of the SEGUE G and K-dwarf samples
\citep[\eg,][]{Schlesinger12a,Zhang13a}, and which appears to 
provide more accurate distances from the kinematical tests of
\citet{Schoenrich12a}. These distances are typically $\sim\!7\,\%$
smaller than the distances calculated using the \iv\ relation. Because
our method for correcting for color and metallicity biases in the
sample selection requires a simple photometric distance relation (see
\sectionname~\ref{sec:method} and \appendixname~\ref{sec:normint}
below), we apply a single distance factor to the stars in each
\map\ calculated as the mean distance modulus offset between the
\iv\ and \an\ relations for the stars in the \map, taking into account
the \map's $g-r$ and \feh\ distributions (see Figure 1 of \bo d for
the difference between the \iv\ and \an\ distance scales as a function
of $g-r$ and \feh). These distance factors are shown in
\figurename~\ref{fig:andistance}. As discussed below, we have checked
that our measurements of the surface density as a function of radius
are affected by only a few percent by the difference between these two
distance scales and are therefore much less sensitive to distance
systematics than might be assumed (but we stress that the
\an\ distances are more accurate and therefore the correct distances
to use).

Because dynamical modeling such as that performed in this paper
essentially succeeds by finding the gravitational potential that makes
the observed velocities consistent with their observed spatial
distribution in a dynamical steady state, the detailed understanding
of the data sampling function that was crucial for \bo b and \bo d is
again an important ingredient for the present analysis. We refer the
reader to Appendix A of \bo d for a full description of the
\segue\ G-star selection function.

We begin with a \segue\ G-dwarf sample that has about 28,000 stars
with acceptably well-determined measurements, but here we only use
those 23,767 stars that fall within well-populated `mono-abundance'
bins in the (\feh,\afe) plane (\bo d; a bin is well-populated when it
contains more than 100 stars; the maximum number of stars in a bin is
789). Furthermore, we exclude \map s whose best-fit scale length in
\bo d was larger than $4.85\kpc$. We exclude these \map s because
their radial profile cannot be well-fit with the simple DF ansatz that
we make below. These \map s are all \feh-poor and have \afe\ close to
solar; they are primarily associated with the outer disk (see Figure 7
in \bo d). This removes 17 \map s and leaves 16,653 stars in 45 \map
s. For reasons discussed below, we remove 2 \map s from this sample,
such that the final data sample consists of 16,269 stars in 43 \map s.

\section{Analysis method}\label{sec:method}

While stellar dynamics overall is a long- and well-established field
\citep{binneytremaine}, the best ways to model large discrete data
sets with full phase-space information are still being developed
(cf. \citealt{McMillan12a,Rix13a}). In what follows, we analyze each
\map\ separately and independently with the end goal of constraining
the total surface density to $Z= 1.1\kpc$---$\Sigma(R,|Z| \leq
1.1\kpc) \equiv \int_{-1.1\kpc}^{1.1\kpc}\dd Z \rho(R,Z)$, abbreviated
as $\sigoneone(R)$---at a \emph{single} Galactocentric radius for each
\map. In this and the following Section we describe all of the steps
involved in fitting each \map\ as a population in a dynamical steady
state in order to measure the gravitational potential and how we
translate this into a single, robust measurement of \sigoneone\ at a
single radius for each \map. Our measurements of $\sigoneone(R)$
assume that the rotation curve is flat; if this is not the case, then
we show in \sectionname~\ref{sec:systematics} that we still robustly
measure the vertical force at $1.1\kpc$.

The basic ingredients of the fitting procedure are:
\begin{itemize}

\item Each \map\ is fit as a population of stars drawn from a
  quasi-isothermal DF (\qdf;
  \citealt{Binney10a,Binney11a,Ting13a}). This form for the DF has a
  number of free parameters related to the radial density profile of
  the tracer population and the radial and vertical velocity
  dispersions and how they change with $R$. The qDF is described in
  detail in \sectionname~\ref{sec:qdf}.

\item The \qdf\ is a function of the orbital actions $\vecj=(\jr,
  \jphi,\jz) \equiv (\jr,\lz,\jz)$. For the fit we only need the
  $(\vecx,\vecv) \rightarrow \vecj$ transformation, which for the disk
  orbits in our sample can be efficiently, accurately, and precisely
  computed using a St\"{a}ckel approximation of the potential for any
  reasonable form of the potential near the Galactic plane
  \citep{Binney12a}. This procedure is briefly summarized
  in \sectionname~\ref{sec:qdf}.

\item To calculate the orbital actions we require a full 3D model for
  the Milky Way's potential. We use a bulge + stellar disk + gas disk
  + dark halo model that a priori has many free parameters. For the
  present application we fix at fiducial values all but two of these
  parameters, the stellar disk scale length and the relative
  halo-to-disk contribution to the radial force (= circular velocity
  squared) at $R_0$. These two parameters are varied to scan through a
  wide range of $\sigoneone(R)$, including all a priori reasonable
  values for the range of Galactocentric radii that we are interested
  in ($4\kpc < R < 10\kpc$). In \sectionname~\ref{sec:systematics} we
  show explicitly that changing the parameters that are held fixed in
  the fiducial fit does not significantly impact our measurement of
  $\sigoneone(R)$ and our measurement of the vertical force at
  $1.1\kpc$ at a single Galactocentric radius is particularly free
  from systematics. Our model for the Milky Way's potential is
  described in detail in \sectionname~\ref{sec:potential}.

\item We determine the probability of the position--velocity data (the
  likelihood) for each set of DF + potential parameters using an
  expanded version of the methodology described in \sectionname~3.1 of
  \bo d, carefully accounting for the sample selection. The
  calculation of the effective survey volume consists of a
  nine-dimensional integral, whose fast computation is discussed in
  \appendixname~\ref{sec:normint}. We only use the vertical velocity
  component of the data and marginalize the six-dimensional qDF over
  the velocity components in the plane, because the planar motions do
  not constrain the vertical force and they may be affected by
  non-axisymmetric streaming motions. The fitting procedure is
  discussed in \sectionname~\ref{sec:fit}.

\item For each \map\ we determine the full joint PDF for the DF and
  the potential parameters. By marginalizing this over the DF
  parameters we obtain constraints on the potential that include
  uncertainties in the determination of the correct DF. The details on
  how we explore the PDF are given in
  \sectionname~\ref{sec:details}. From the properties of the DF we
  determine the Galactocentric radius $R$ at which $\sigoneone(R)$ is
  best constrained for each \map\ and we calculate $\sigoneone(R)$ and
  its uncertainty $\delta \sigoneone(R)$ from the PDF. This is
  described in \sectionname~\ref{sec:indivresults}.

\end{itemize}

We now describe these various steps in more detail.

\subsection{Distribution function model}\label{sec:qdf}

We use the quasi-isothermal DF first proposed by \citet{Binney10a} and
later refined by \citet{Binney11a} to fit the dynamics of a single
\map. \bo\ found that the radial and vertical density profiles of
individual \map s are well represented by single exponentials and that
the vertical velocity distribution is isothermal, \ie, the vertical
velocity dispersion is constant with height above the plane. Further
investigation of the Galactocentric radial velocities showed that the
radial velocity dispersion is also isothermal (unpublished). As shown
in \citet{Ting13a}, these properties of \map s can be reproduced by
assuming that the DF is a quasi-isothermal DF of the form of
\citet{Binney11a}. Therefore, in what follows we fit each \map\ as a
single qDF. This is a simple way of including abundance information to
separate different populations into the general approach followed by
\citet{Binney12b}, where disk stars with any abundance were fit using
a superposition of a large number of qDFs. In the current application,
abundances are used to group stars that presumably can be fit using
just a single qDF. The requirement that a \map\ can be modeled using a
single qDF drives our fine-grained pixelization of the (\feh,\afe)
plane.

The form for the qDF as it is used here is that of \citet{Binney11a}
\begin{equation}\label{eq:qdf}
  \mathrm{qDF}(\vecx,\vecv) = f_{\sigma_R}(\jr,\lz) \times \frac{\nu}{2\pi\sigma_Z^2}\,\exp\left(-\frac{\nu \jz}{\sigma_Z^2(R_c)}\right)\,,
\end{equation}
where $f_{\sigma_R}$ is given by
\begin{equation}\label{eq:qdf2}
\begin{split}
  f_{\sigma_R}(\jr,\lz) & = \left.\frac{\Omega n(R_c)}{\pi\sigma_R(R_c)^2
    \kappa}\right|_{R_c}\times\left[1+\tanh(L_z/L_0)\right]\\
  & \qquad \times\,\exp\left(-\frac{\kappa
    \jr}{\sigma_R^2(R_c)}\right)\,.
\end{split}
\end{equation}
Here $\kappa$, $\Omega$, and $\nu$ are the epicycle, circular, and
vertical frequencies \citep{binneytremaine}; $R_c$ is the radius of
the circular orbit with angular momentum $\lz$. We include the factor
in \eqnname~(\ref{eq:qdf2}) containing the $\tanh$ to eliminate stars
on counter-rotating orbits following \citet{Binney11a}, but we also
explicitly set the DF to zero for counter-rotating orbits. $n$,
$\sigma_R$, and $\sigma_Z$ are free functions of $R_c$, which
indirectly determine the radial profiles of the tracer density, the
radial velocity dispersion, and the vertical dispersion. However, it
should be noted that these are merely \emph{scale} profiles, as
opposed to the actual, physical profiles that can be calculated by
taking the appropriate moments of the DF. In principle these three
functions can take any form, but based on observations of the Milky
Way and external galaxies (\eg, \citealt{Lewis89a}; \bo c) we assume
that each of these functions is an exponential
\begin{align}\label{eq:inputprofile}
  n(R_c) & \propto \exp\left(-R_c/h_R\right)\,,\\
  \sigma_R(R_c) & = \sigma_{R,0}\,\exp\left(-\left[R_c-R_0\right]/h_{\sigma_R}\right)\,,\\
  \sigma_Z(R_c) & = \sigma_{Z,0}\,\exp\left(-\left[R_c-R_0\right]/h_{\sigma_Z}\right)\,.\label{eq:inputprofile2}
\end{align}

To evaluate the qDF for a given position and velocity we need to
calculate the actions for that position and velocity, $(\vecx,\vecv)
\rightarrow \vecj$, given $\Phi(R,Z)$. We calculate the actions using
the approximate algorithm of \citet{Binney12a}, where the actions are
calculated using an (implicit) approximation of the potential as a
St\"{a}ckel potential. For the sake of a self-contained description,
we summarize the relevant parts of the algorithm here. This algorithm
proceeds by introducing prolate confocal coordinates $(u,v)$
\begin{equation}
  R = \Delta \,\sinh u\,\sin v; \qquad z = \Delta \,\cosh u \cos v\,.
\end{equation}
\citet{Binney12a} discusses how the choice of $\Delta$ influences the
precision with which the actions are calculated. Based on this
discussion, we fix $\Delta$ to $0.45\,R_0$ for all potentials. The
momenta in this new coordinate system are given by
\begin{align}
  p_u & = \Delta \,(p_R\,\cosh u \,\sin v + p_z\,\sinh u \cos v)\,,\\
  p_v & = \Delta \,(p_R\,\sinh u \,\cos v - p_z\,\cosh u \sin v)\,.
\end{align}
If the potential is of the St\"{a}ckel form $\Phi_S(u,v) =
(U(u)-V(v))/(\sinh^2 u + \sin^2 v)$ then these momenta are functions
of their associated coordinate only
\begin{align}\label{eq:pupv}
  \frac{p_u^2}{2\,\Delta^2} & = E\,\sinh^2 u - I_3 - U(u) - \frac{\lz^2}{2\,\Delta^2\sinh^2 u}\,,\\
  \frac{p_v^2}{2\,\Delta^2} & = E\,\sin^2 v + I_3 + V(v) - \frac{\lz^2}{2\,\Delta^2\sin^2 v}\,,\label{eq:pupv2}
\end{align}
where $E$ is the energy and $I_3$ is a constant of the motion (the
third integral). This allows one to calculate the actions through a
single integration
\begin{equation}\label{eq:jrjzint}
  \jr = \frac{1}{\pi} \int_{u_\mathrm{min}}^{u_{\mathrm{max}}} \dd u \,p_u(u)\,,\qquad 
  \jz = \frac{2}{\pi} \int_{v_\mathrm{min}}^{\pi} \dd v \,p_v(v)\,.
\end{equation}

The crucial ingredient of \citet{Binney12a}'s algorithm is that
although it appears from \equationname
s~(\ref{eq:pupv}-\ref{eq:pupv2}) that we require an explicit form for
$U(u)$ and $V(v)$---and thus an explicit representation of the
potential in St\"{a}ckel form, as in \citet{Sanders12a}---if the
potential is close to a St\"{a}ckel potential then $U(u)$ can be
re-written as $U(u_0) + \delta U(u)$, where $\delta U(u) \equiv U(u) -
U(u_0)$ and $u_0$ is a reference value for $u$ that does not affect
the calculated actions, and similarly $V(v)$ can be re-written as
$V(\pi/2) + \delta V(v)$. These $\delta U(u)$ and $\delta V(v)$ can be
calculated directly from the potential as
\begin{align}
\delta U(u) & = (\sinh^2 u+\sin^2 v)\Phi(u, v)\\
& \qquad- (\sinh^2 u_0 +\sin^2 v)\Phi(u_0, v)\,,\\
\delta V(v) & = \cosh^2 u \Phi(u,\pi/2)-(\sinh^2 u+\sin^2 v)\Phi(u, v)\,,
\end{align}
since $(\sinh^2 u+\sin^2 v)\Phi(u, v) = U(u)-V(v)$ for a St\"{a}ckel
potential. If the potential is close to a St\"{a}ckel potential, then
the dependence of $\delta U(u)$ on $v$ and that of $\delta V(v)$ on
$u$ is small, so they can be calculated for a chosen reference value
of $v$ and $u$, respectively.

While we cannot calculate $U(u_0)$ and $V(\pi/2)$ directly without an
explicit representation of the potential in St\"{a}ckel form, we can
compute the combinations $I_3+U(u_0)$ and $I_3 + V(\pi/2)$ from the
current position and velocity of the orbit (the position and velocity
at which the actions are required) if we set $u_0$ to the value
corresponding to the given position
\begin{align}
  I_3 + U(u_0) & = E \sinh^2 u - \frac{p_u^2}{2\,\Delta^2} - \frac{\lz^2}{2\,\Delta \sinh^2 u}\,,\\
  I_3 + V(\pi/2) & = -E \sin^2 v + \frac{p_v^2}{2\,\Delta^2} + \frac{\lz^2}{2\,\Delta \sin^2 v}\,.
\end{align}

\begin{figure}[t!]
  \includegraphics[width=0.4\textwidth,clip=]{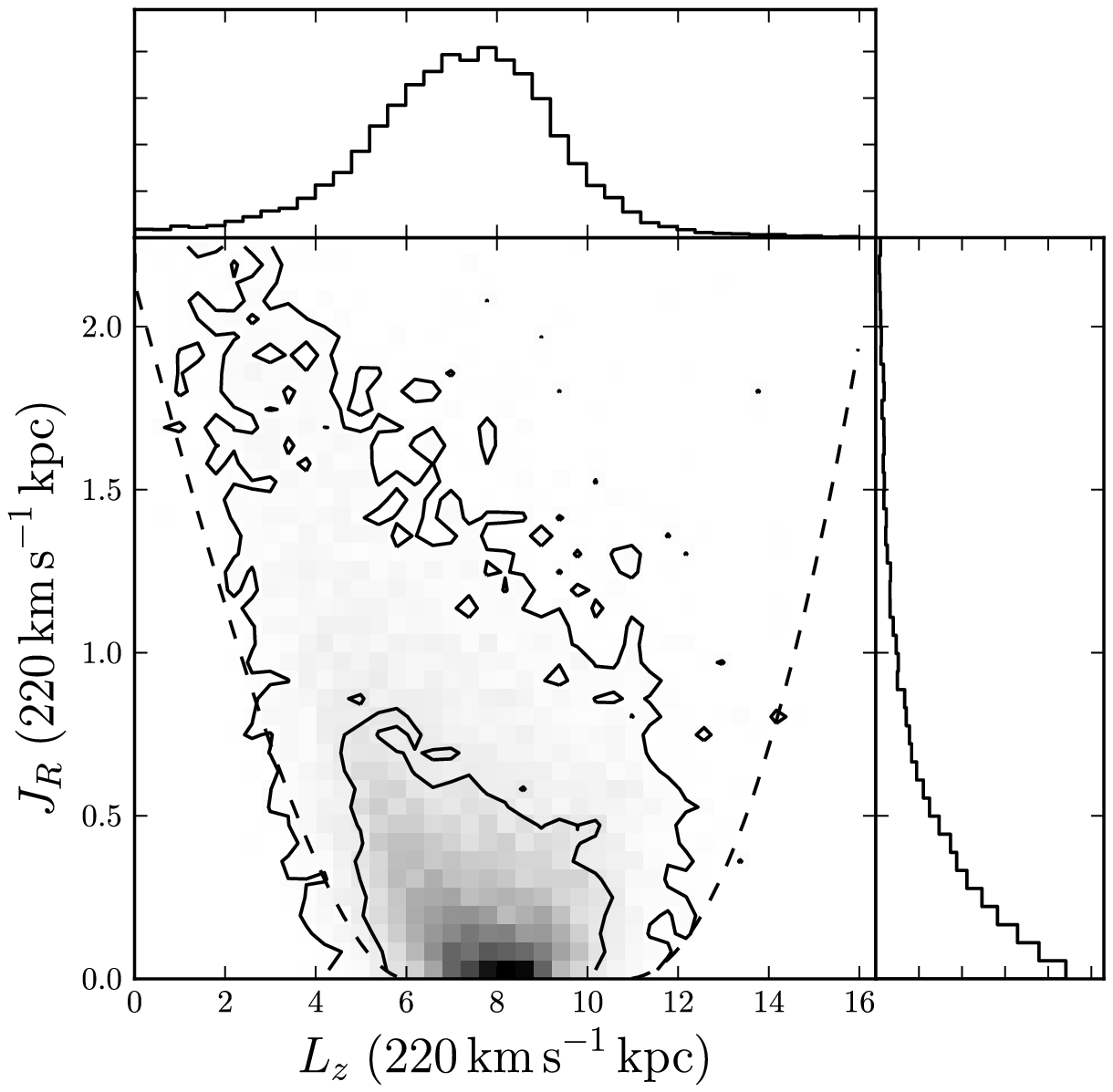}\\
  \includegraphics[width=0.4\textwidth,clip=]{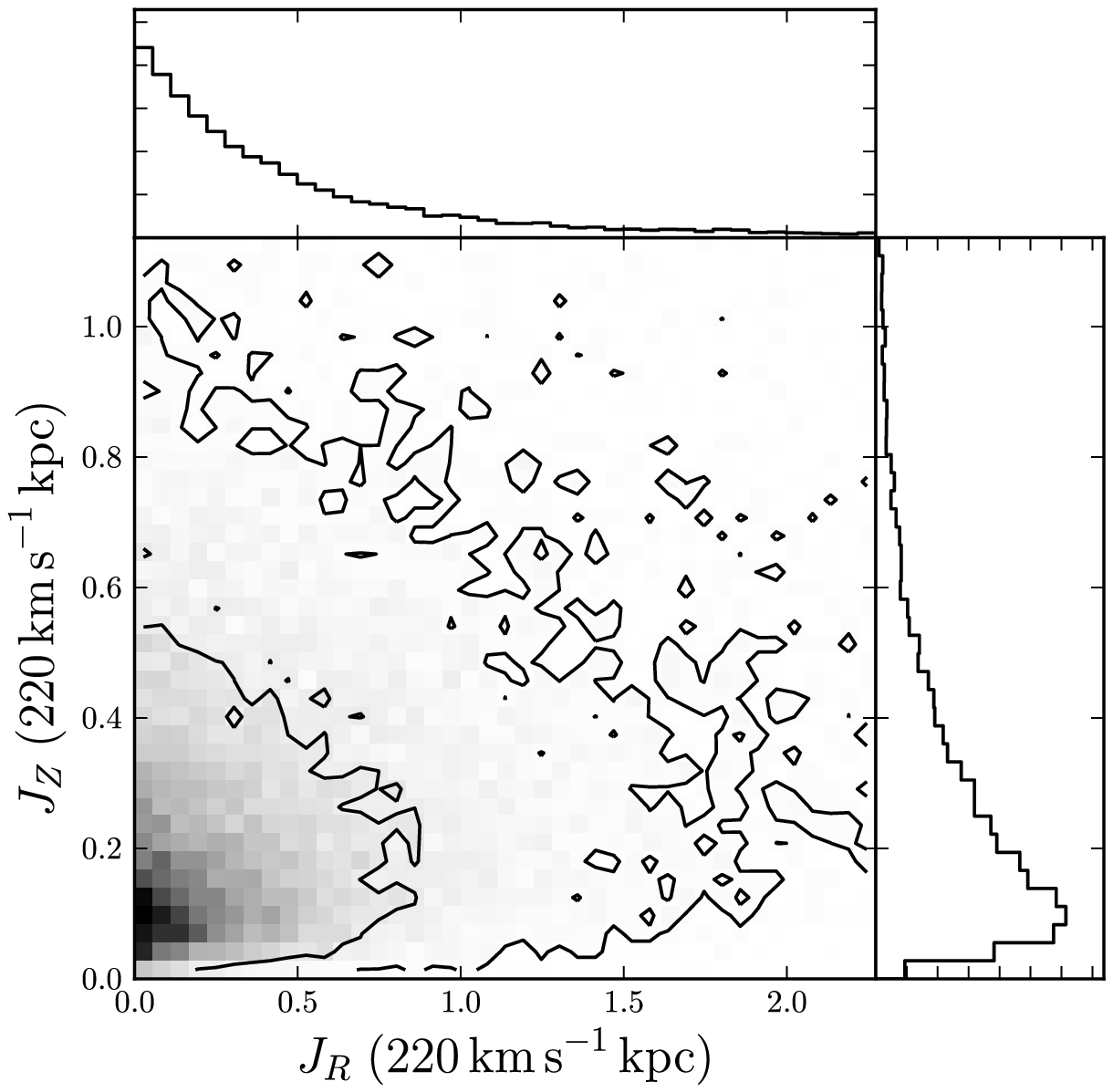}
  \caption{Distribution of the SEGUE G-type dwarfs in action
    space. Actions are calculated in the best-fitting potential of
    \sectionname~\ref{sec:potentialfit}. The dashed lines in the top
    panel show the loci of orbits in the plane with a pericenter of
    $6\kpc$ (left) and an apocenter of $11\kpc$
    (right).}\label{fig:data_jrlzjz}
\end{figure}

In what follows we always use this algorithm to calculate the actions
rather than constructing an interpolation look-up table for a grid of
integrals of the motion. The range of the integrals in \equationname
s~(\ref{eq:jrjzint}) is determined using \citet{Brent73a}'s method for
finding the zeros of the integrands and the integrals are then
calculated using 10-th order Gauss-Legendre quadrature.

The distribution of the actions of the 16,269 sample stars calculated
in a fiducial potential is shown in
\figurename~\ref{fig:data_jrlzjz}. The diagonal edges in the
distribution of \jr\ and \lz\ (the radial action and the angular
momentum, see below) are due to the fact that circular orbits far from
the solar neighborhood (\ie, outside of the bounds of our sample) are
not represented in our data set. Similarly, in the distribution of the
vertical action \jz, stars with small vertical actions ($\sim$small
vertical energies) are missing in our sample because of the effective
lower limit on the vertical height of the stars in our sample; stars
with low vertical action do not reach high enough above the plane of
the disk to be included in our sample.

\begin{figure}[t!]
  \includegraphics[width=0.38\textwidth,clip=]{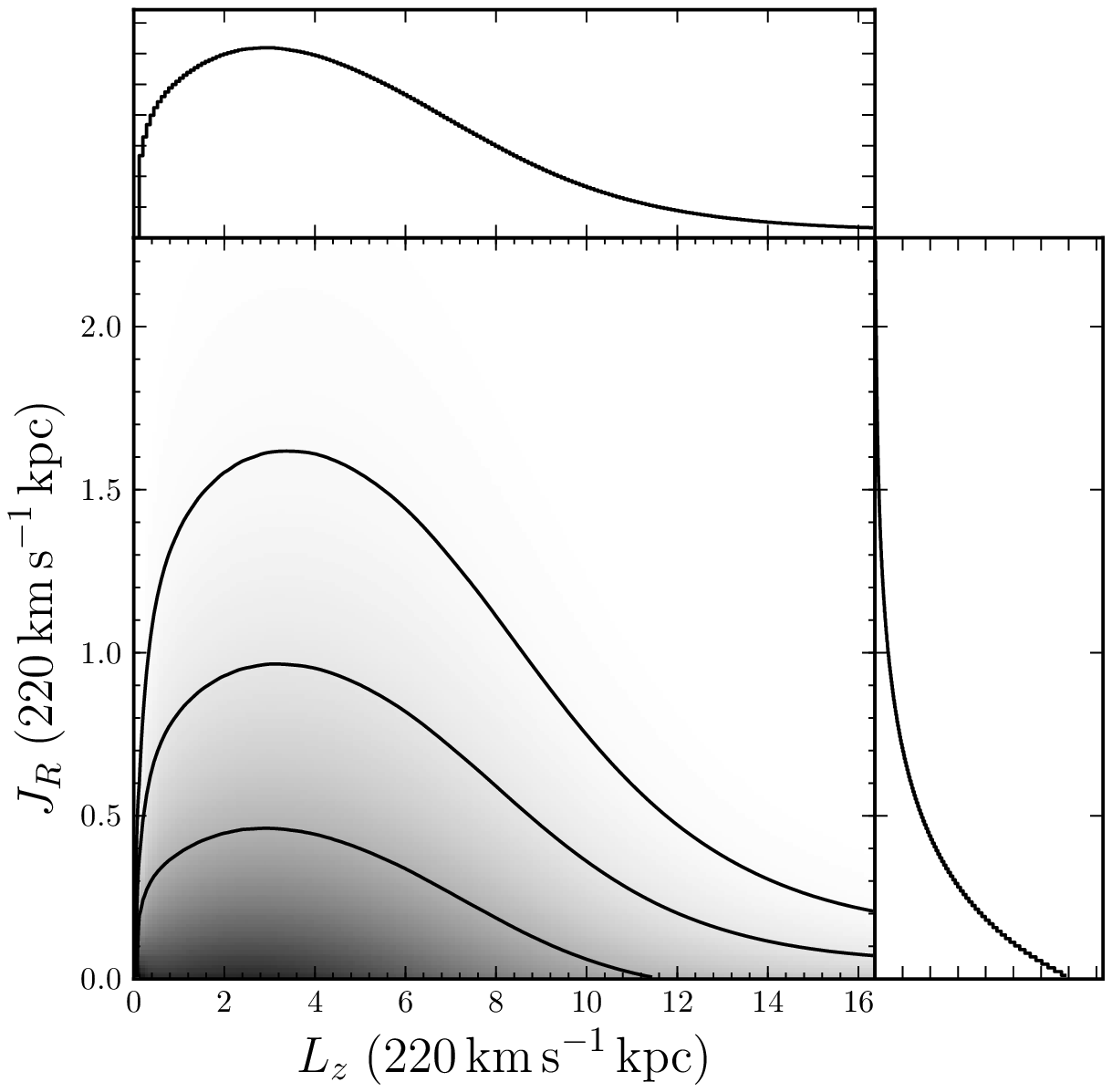}\\
  \includegraphics[width=0.38\textwidth,clip=]{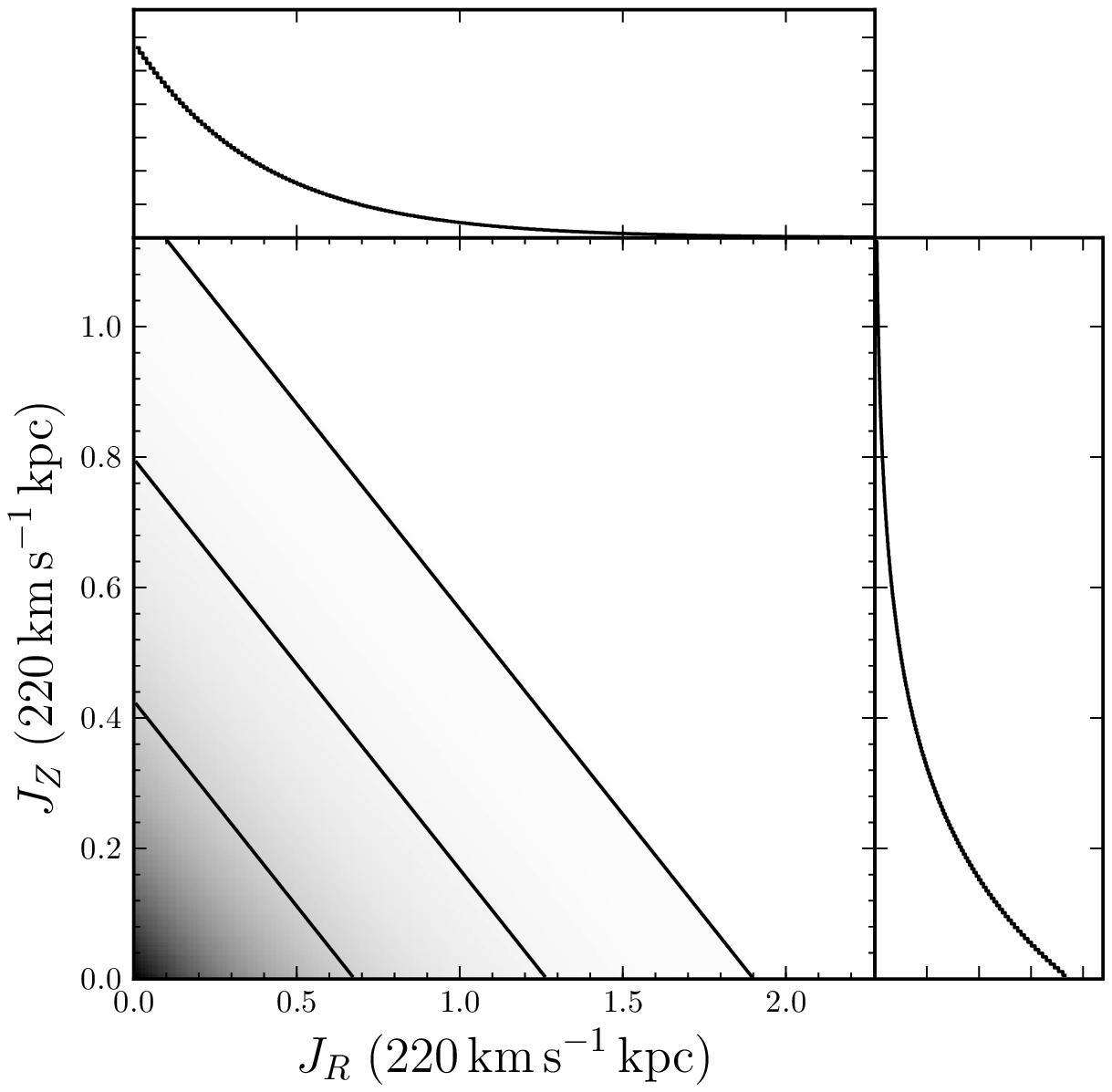}
  \caption{The quasi-isothermal DF as a function of the actions. The
    qDF of \equationname~(\ref{eq:qdf}) is shown as a function of
    \lz\ and \jr\ for $\jz = 0$ and as a function of \jr\ and \jz\ for
    $\lz / 8\kpc= 200\kms$. The qDF parameters are chosen to mimic
    those of one of the \aenhanced\ \map s considered in this paper:
    $h_R = 2\kpc$, $\sigma_Z(R_0) = 66\kms$, $\sigma_R(R_0) = 55\kms$,
    $h_{\sigma_Z} = 7\kpc$, and $h_{\sigma_R} = 8\kpc$ (see
    \figurename~\ref{fig:qdf_invsout} for how these scale parameters
    relate to the physical scale length and dispersion
    parameters). The potential is the same as that used in
    \figurename~\ref{fig:data_jrlzjz}.}\label{fig:fidDF_jrlzjz}
\end{figure}

\begin{figure*}[t!]
  \includegraphics[width=0.25\textwidth,clip=]{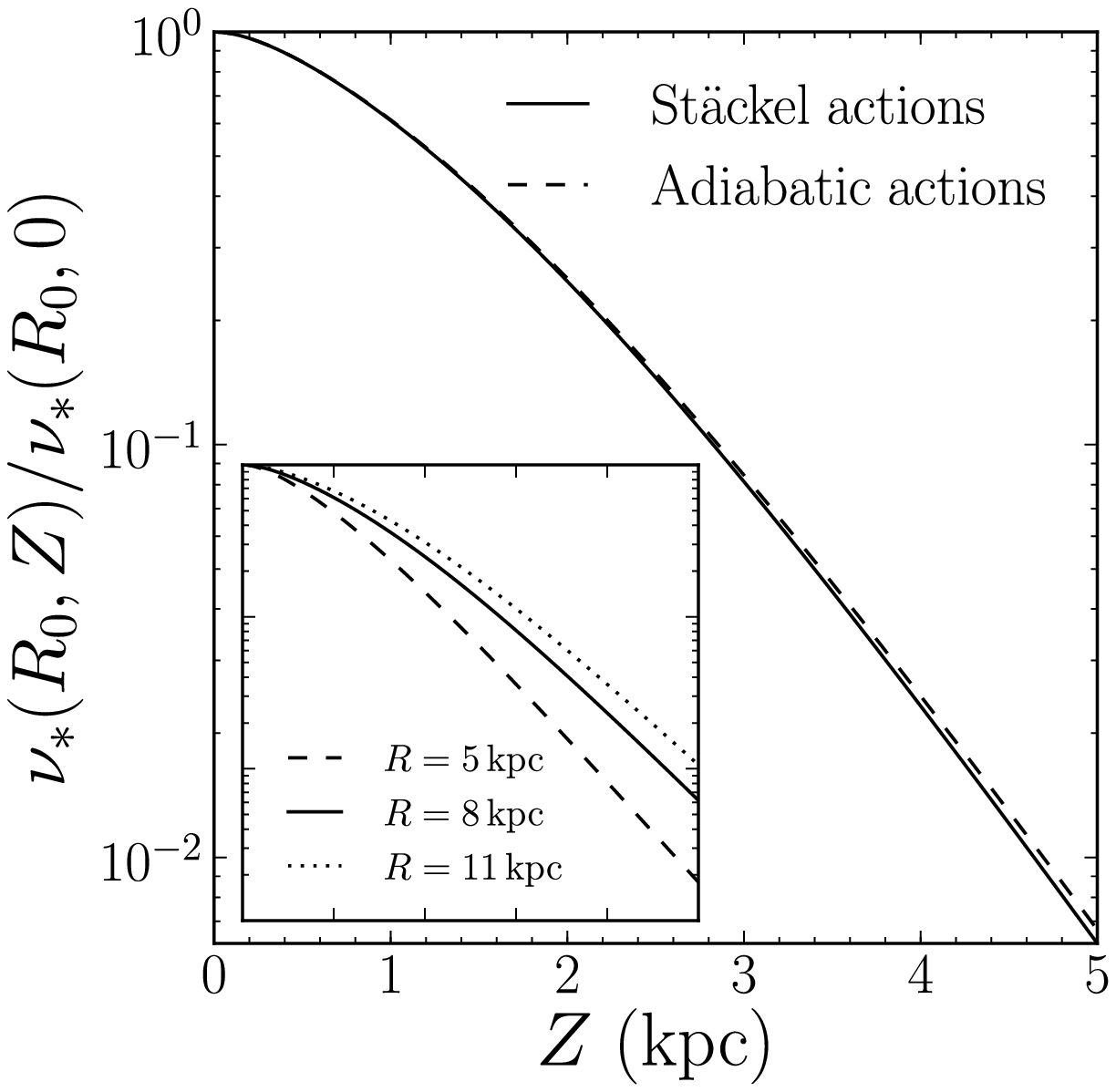}
  \includegraphics[width=0.25\textwidth,clip=]{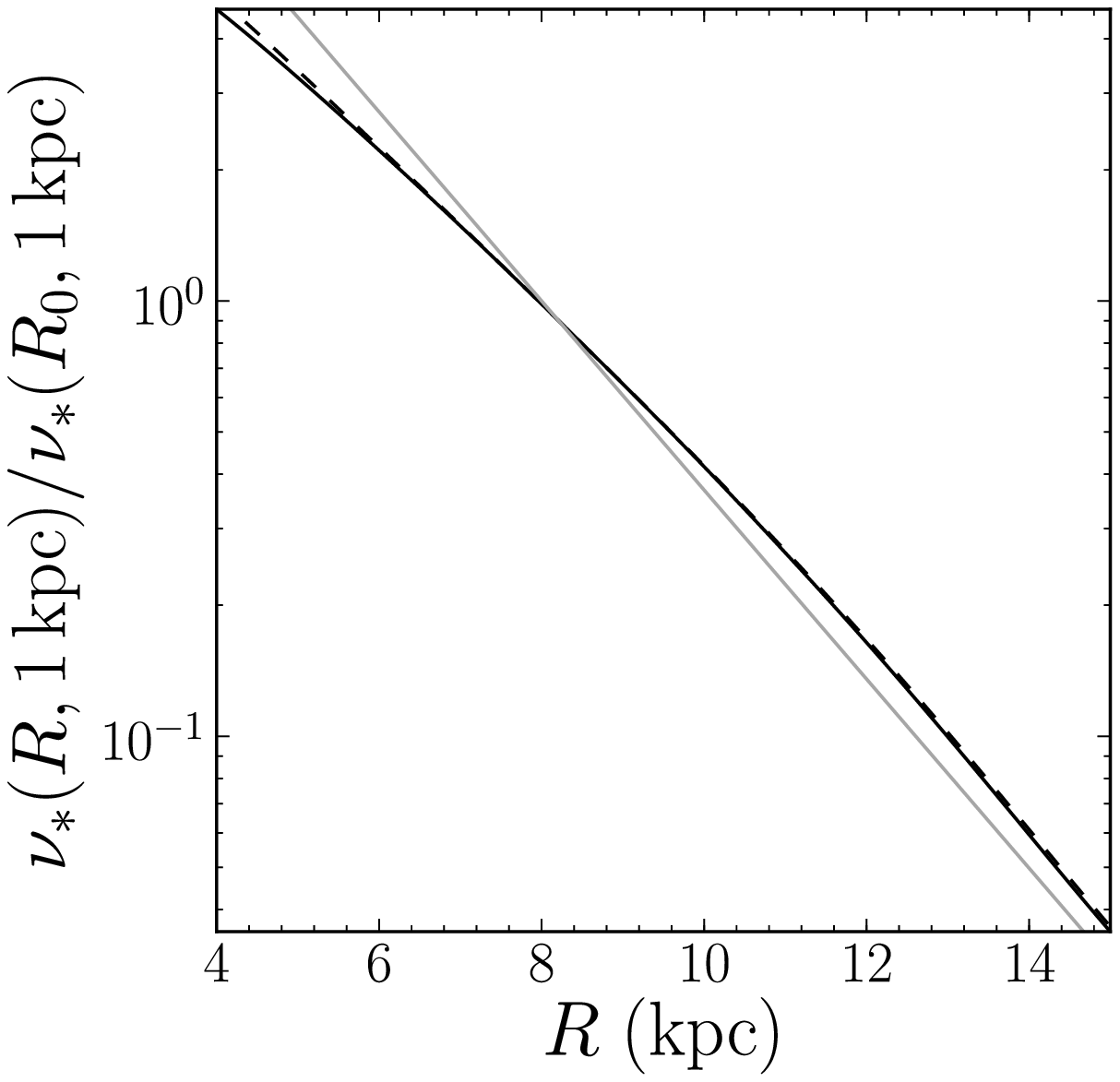}
  \includegraphics[width=0.24\textwidth,clip=]{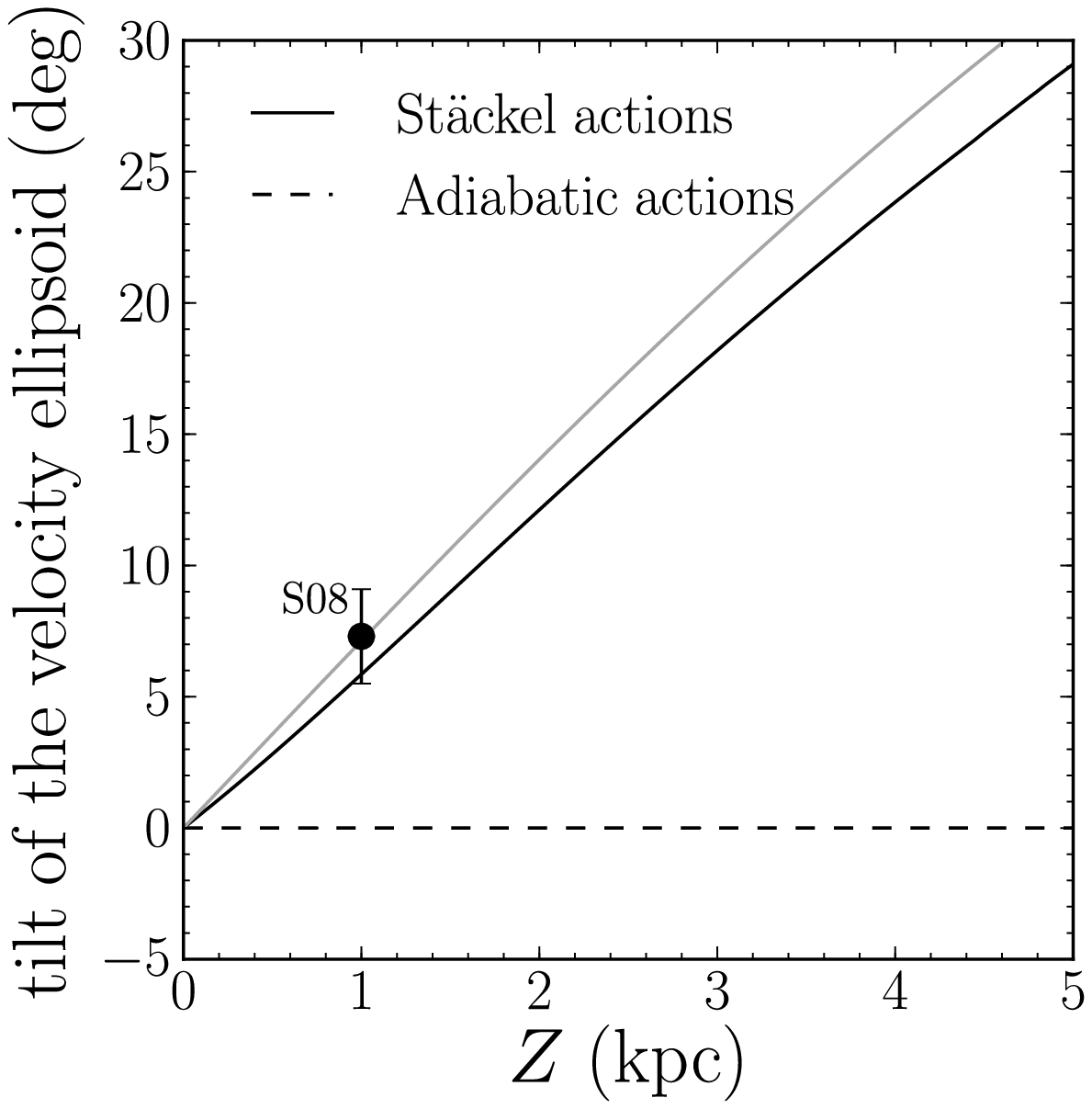}
  \includegraphics[width=0.24\textwidth,clip=]{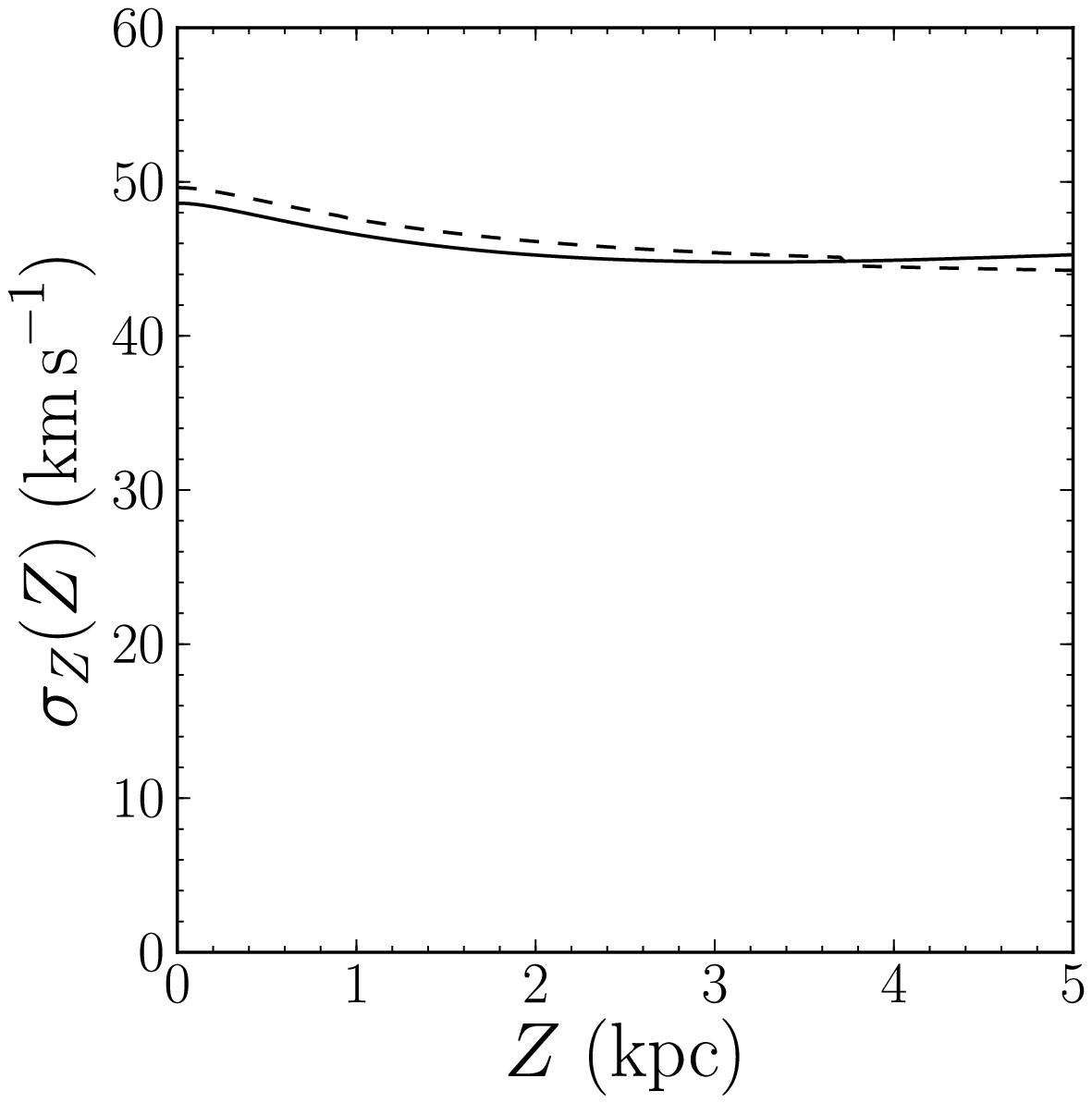}
  \caption{Same as \figurename~\ref{fig:fidDF_jrlzjz}, but now the qDF
    is projected into position and velocity space. The leftmost panel
    shows the vertical density profile; the inset shows that this qDF
    disk population flares slightly (these profiles are calculated
    using the St\"{a}ckel-based approximation of the actions). The
    second panel shows the radial density profile at $1\kpc$ from the
    plane; an exponential with a scale length of $2\kpc$ is shown for
    comparison. The third and fourth panels shows the tilt of the
    velocity ellipsoid and the vertical velocity dispersion at $R_0$
    as a function of height.  All four profiles are computed both
    using the adiabatic approximation for the calculation of the
    actions and for the St\"{a}ckel-based approximation. The position
    and velocity dependence of the qDF is very similar when using
    these two approximations, except for the tilt of the velocity
    ellipsoid. This tilt is zero in the adiabatic approximation and
    close to pointing to the Galactic center (gray line in the third
    panel) when using the St\"{a}ckel approximation. The S08
    measurement of the tilt is from \citet{Siebert08a}. The dynamical
    analysis in this paper is performed using the St\"{a}ckel
    approximation and therefore includes a realistic
    tilt.}\label{fig:fidDF_posvel}
\end{figure*}

The DF as a function of the actions for a fiducial set of parameters
is shown in \figurename~\ref{fig:fidDF_jrlzjz}. A comparison between
this distribution and the distribution of the data actions in
\figurename~\ref{fig:data_jrlzjz} shows that the data's action
distribution is heavily affected by selection biases related to the
spatial sampling of the data.

The density, mean velocity, velocity dispersions, etc. of the qDF can
be calculated as velocity moments of the 6D qDF:
\begin{align}\label{eq:outputprofile}
\nu_*(R,Z) & = \int \dd \vecv \,\mathrm{qDF}(R,Z,\vecv)\,,\\
\nu_*(R,Z)\,\langle \vecv \rangle(R,Z)  & = \int \dd \vecv \,\vecv\,\mathrm{qDF}(R,Z,\vecv)\,,\\
\nu_*(R,Z)\,\langle \vecv^2 \rangle(R,Z)  & = \int \dd \vecv \,\vecv^2\,\mathrm{qDF}(R,Z,\vecv)\,;
\end{align}
these can be straightforwardly transformed into the velocity
dispersions. In what follows these velocity moments are calculated
using 20-th order Gauss-Legendre quadrature between $-4\,\sigma_R(R)$
and $4\,\sigma_R(R)$ in the radial velocity $V_R$, $-4\,\sigma_Z(R)$
and $4\,\sigma_Z(R)$ in the vertical velocity $V_Z$, and $0$ and
$3\,V_c(R_0)/2$ in the rotational velocity $V_T$; the dispersions
$\sigma_R(R)$ and $\sigma_Z(R)$ in these expressions for the ranges
are calculated using the \emph{scale} dispersion profiles from
\equationname s~(\ref{eq:inputprofile}-\ref{eq:inputprofile2}). We use
actions calculated using the adiabatic approximation in a few
instances below; in this case we use 20-th order Gauss-Legendre
quadrature between $0$ and $4\sigma_{R,Z}(R)$, because in the
adiabatic approximation the qDF is perfectly symmetric in $V_R$ and
$V_Z$ separately, which is not the case when using the more correct
St\"{a}ckel actions (see below).

In \figurename~\ref{fig:fidDF_posvel} we show the vertical and radial
tracer density profile for a fiducial set of qDF parameters, as well
as the tilt of the velocity profile and the vertical velocity
dispersion as a function of height. We calculate these using the
St\"{a}ckel approximation for the actions as well as the simpler, but
less precise, adiabatic approximation \citep[\eg.][]{Binney10a}. The
vertical and radial density profiles are close to exponential, but the
vertical profile gradually flattens closer to the plane of the
disk. The inset in the leftmost panel shows that the tracer density
flares slightly. The right panel shows that the qDF is indeed close to
isothermal as the vertical velocity dispersion only increases by a few
\kms\ over $5\kpc$; the same holds for the radial velocity dispersion
\citep[see also][]{Ting13a}.

The difference between the St\"{a}ckel and adiabatic approximations is
small for most observables, except for the tilt of the velocity
ellipsoid. The tilt is equal to zero when using the adiabatic
approximation because this assumes that vertical and planar motions
are close to decoupled, implying that there is no correlation between
vertical and radial oscillations along the orbit. The St\"{a}ckel
approximation does not make this assumption and it correctly captures
the correlation between the radial and vertical oscillations for stars
that reach large heights above the plane ($\gtrsim500\,\mathrm{pc}$);
the gray line shows the tilt if the velocity ellipsoid is pointing
toward the Galactic center and the tilt included in the qDF falls only
slightly short of this. Also shown in the third panel of
\figurename~\ref{fig:fidDF_posvel} is the measurement of the tilt of
the velocity ellipsoid from \citet{Siebert08a}, which agrees well with
the tilt that results from the qDF if the actions are calculated in
the St\"{a}ckel approximation.

\begin{figure*}[t!]
  \includegraphics[width=0.32\textwidth,clip=]{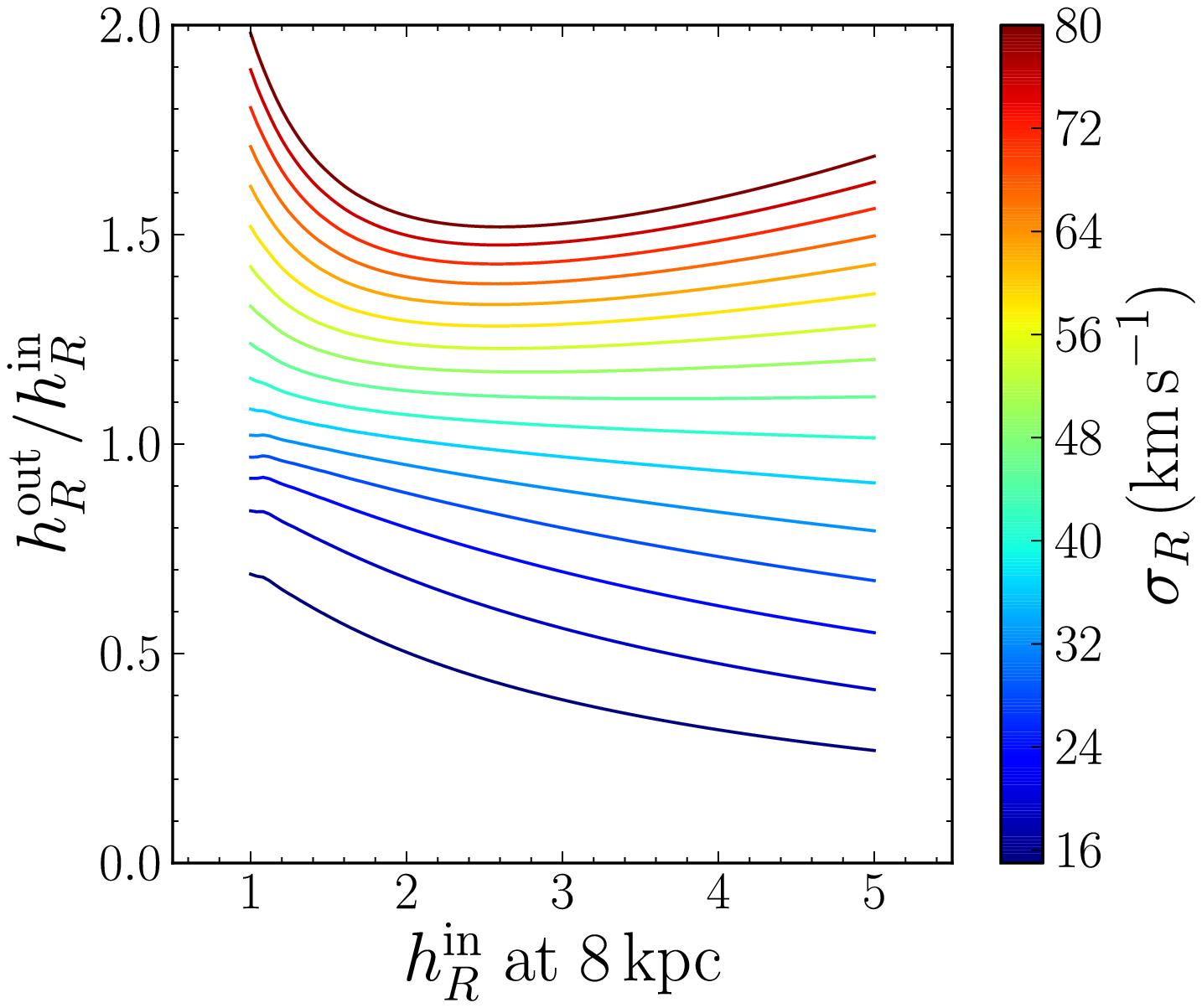}
  \includegraphics[width=0.32\textwidth,clip=]{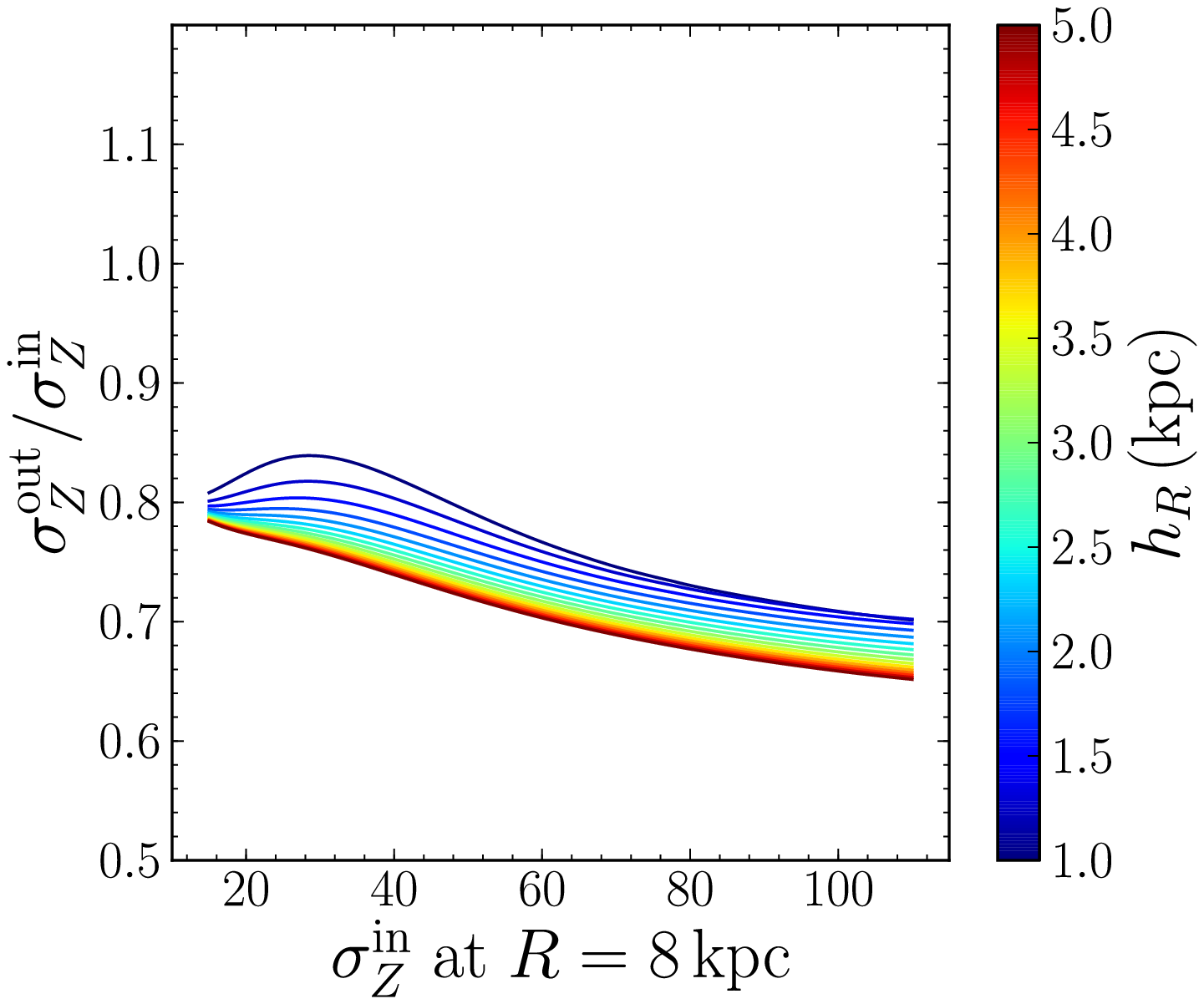}
  \includegraphics[width=0.32\textwidth,clip=]{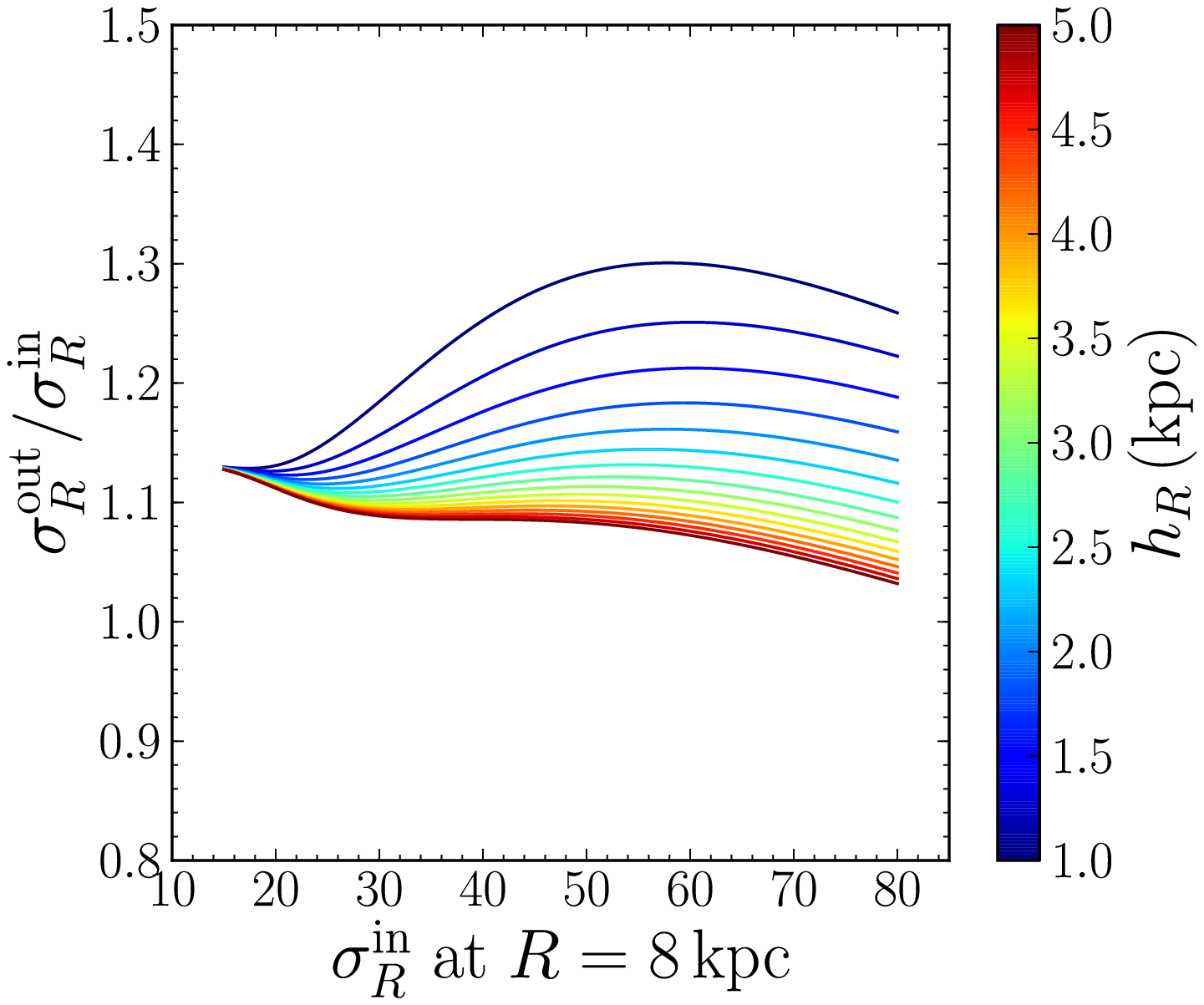}
  \caption{Physical properties of the quasi-isothermal DFs versus
    their scale parameters. The left panel shows the radial scale
    length measured over $\Delta R = R_0/3$ around $R_0$ versus the
    scale length scale parameter, for various values of the radial
    dispersion. The middle and right panels show the vertical and
    radial dispersion, respectively, at $(R,Z) = (R_0, 800\pc)$ versus
    the dispersion scale parameters for various values of the q radial
    scale length scale parameter. We assume that
    $\sigma^{\mathrm{in}}_R / \sigma^{\mathrm{in}}_Z = \sqrt{3}$,
    $h_{\sigma_Z} = 7\kpc$, and $h_{\sigma_R} = 8\kpc$ for all DF
    models in this Figure.}\label{fig:qdf_invsout}
\end{figure*}

\begin{figure*}[t!]
  \includegraphics[width=0.32\textwidth,clip=]{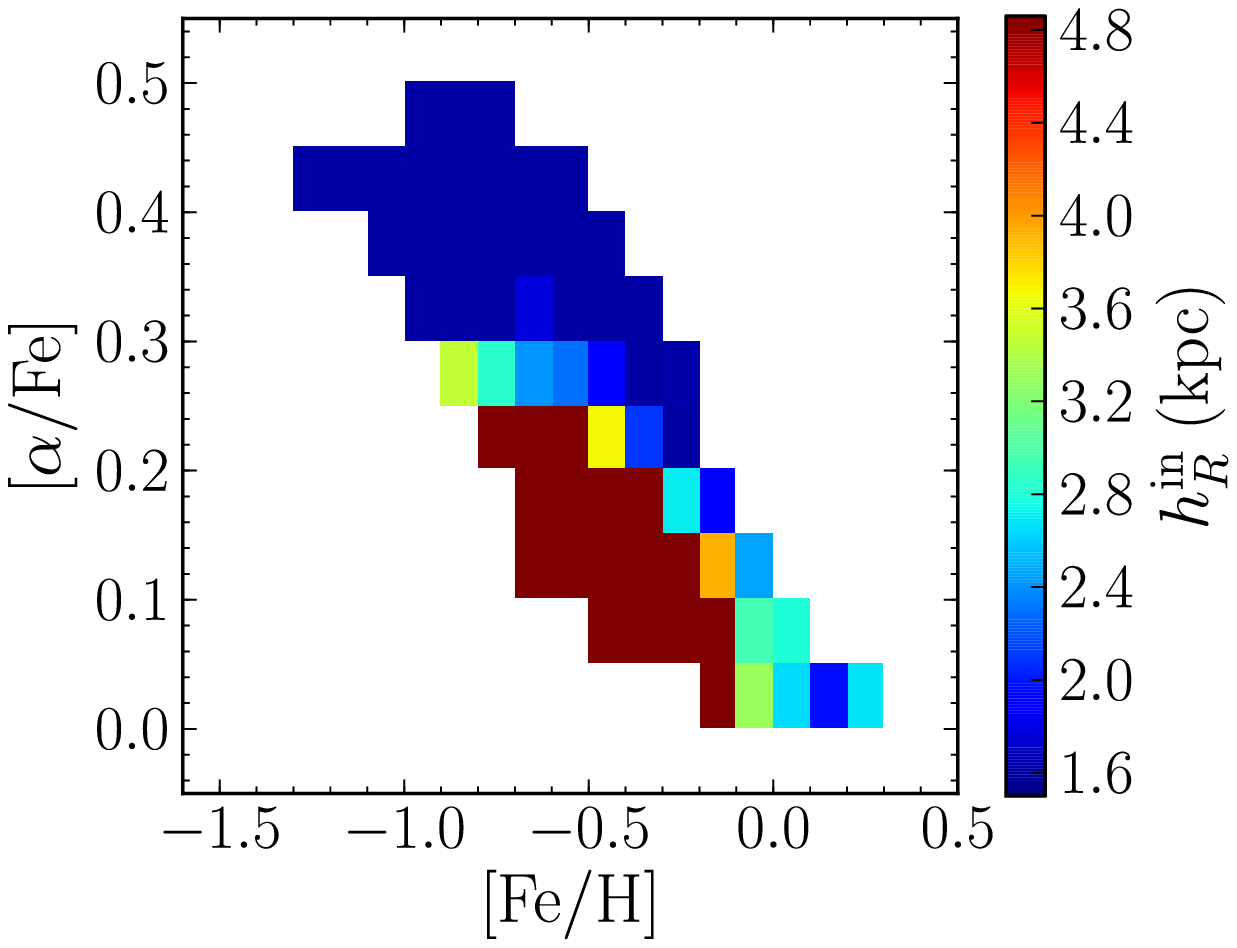}
  \includegraphics[width=0.32\textwidth,clip=]{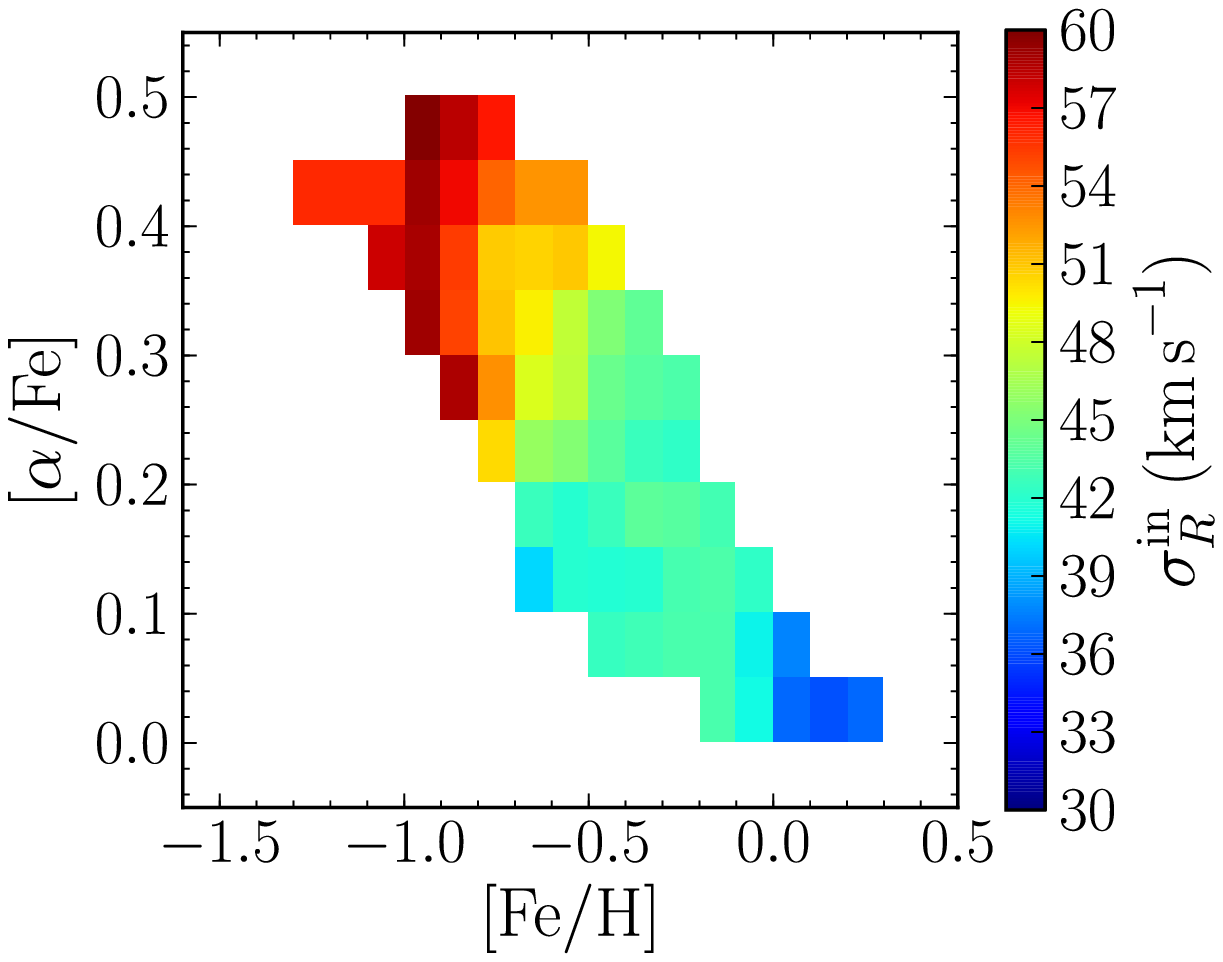}
  \includegraphics[width=0.32\textwidth,clip=]{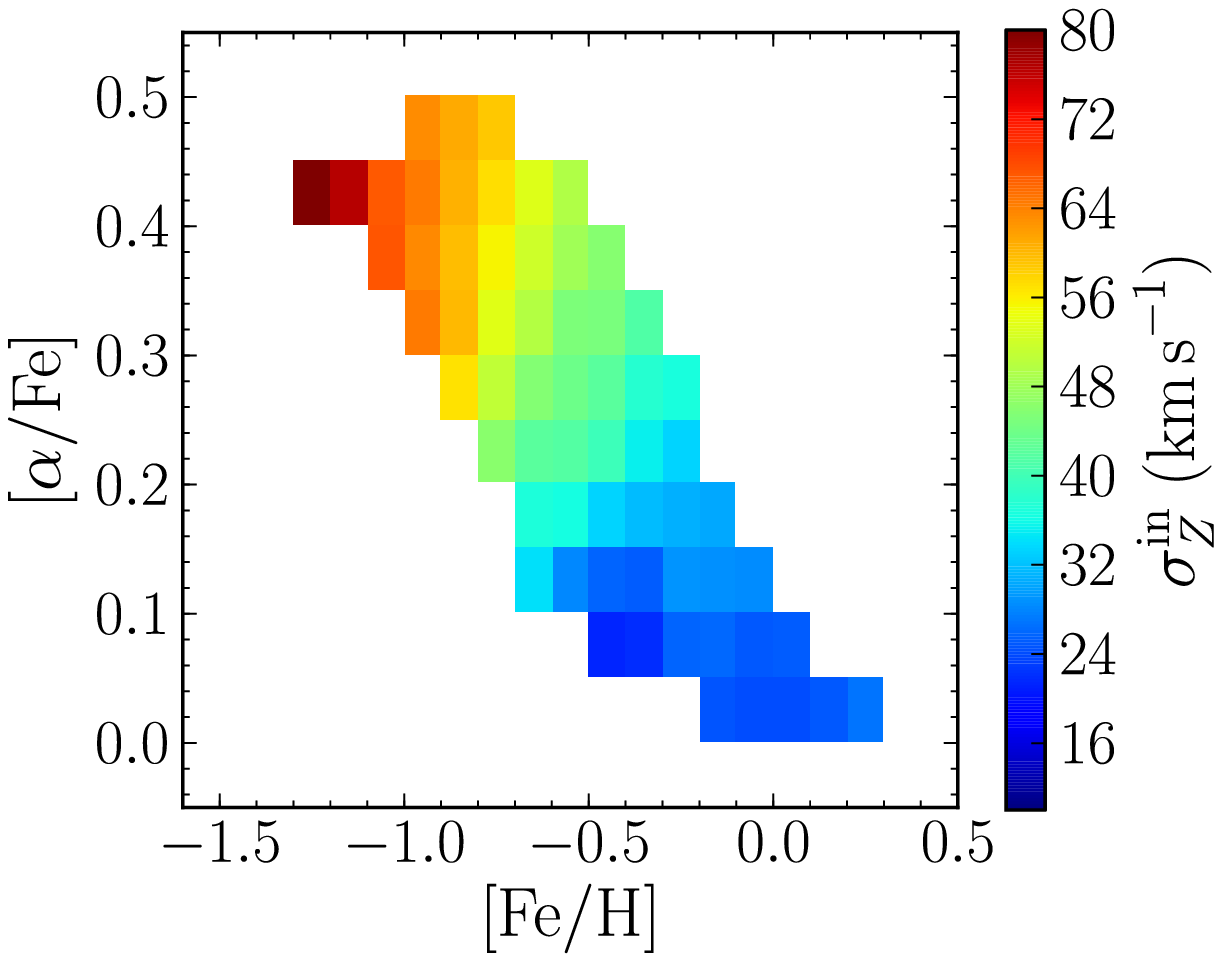}
  \caption{Estimates of the parameters of the quasi-isothermal DF that
    describes each MAP, based on the results from the previous,
    non-dynamical fits in \bo cd. The previous results for the
    dispersions were smoothed before estimating the DF parameters. The
    estimated parameters are shown for all \map s considered in \bo;
    only \map s with \bo d radial scale lengths $< 4.85\kpc$ are
    included in the dynamical modeling in this paper (that is, the
    \map s that are dark brown in the leftmost panel are
    excluded).}\label{fig:previousAsQdf}
\end{figure*}

As discussed before, the density and dispersion scale profiles in
\equationname s~(\ref{eq:inputprofile}-\ref{eq:inputprofile2}) are not
the physical profiles and the actual density and dispersion profiles
as calculated using \equationname s~(\ref{eq:outputprofile}) are
slightly different. In what follows we use the results from \bo\ for
the dispersions at $R_0$ to determine a likely range for the qDF
parameters for each \map\ used in a grid-based exploration of the
joint DF--potential PDF. Therefore, we need to determine how the scale
profiles relate to the physical profiles. In
\figurename~\ref{fig:qdf_invsout} we show these relations calculated
in a potential similar to potential II in \citet{binneytremaine}. We
use this relation to calculate the approximate scale profiles that
give physical profiles similar to the best-fit profiles found in
\bo\ for the radial density and the radial and vertical velocity
dispersion at $R_0$. These estimates are shown in
\figurename~\ref{fig:previousAsQdf}.

\subsection{Milky Way potential model}\label{sec:potential}

While for the present paper we are primarily interested in
$\sigoneone(R)$, the modeling of each \map\ using the qDF requires a
full 3D model for the Galactic potential. We use a four-component
model consisting of a Hernquist bulge with a mass of
$4\times10^9\,M_\odot$ and a scale radius of $600\pc$
\footnote{The details of the bulge model do not matter for the
  analysis that follows; we use a slightly lower bulge mass than
  typically found from dynamical considerations, \eg,
  \citet{McMillan11a}, because we use the Hernquist model rather than
  an exponentially cut-off model as is customary. The rotation curve
  at $R \gtrsim 4\kpc$, which is the only thing that matters in the
  analysis below, is similar for our bulge model and that of
  \citet{binneytremaine} and \citet{McMillan11a}.}, a stellar
exponential disk component characterized by a (single, effective)
scale height $z_h$ and scale length $R_d$, a spherical power-law dark
halo, and a gas disk modeled as an exponential disk with a local
surface density of $13\,M_\odot\,\pc^{-2}$, a scale height of
$130\pc$, and a scale length fixed to be twice that of the stellar
disk (following \citealt{binneytremaine}). We do not include a stellar
halo component, as the mass of the stellar halo is negligible compared
to the other components.

In the qDF fits to data of individual \map s, the circular velocity
curve, the epicycle, and the vertical frequencies are calculated for
each potential on a logarithmic grid in $R$ between $0.01\,R_0$ and
$20\,R_0$. The potential is calculated on the same radial grid and on
a linear grid in $Z$ between $0$ and $R_0$. These values are
interpolated using two and one-dimensional cubic B-splines and the
interpolating functions are used to first calculate the actions for a
given $(\vecx,\vecv)$ and then to evaluate the qDF.

The calculation of the bulge and halo potential and frequencies is
straightforward. The potential and forces of an exponential disk with
density $\rho(R,Z) = \rho_d\,e^{-R/R_d-|Z|/z_h}$ are calculated using
the expressions in Appendix A of \citet{Kuijken89a}, which we repeat
here for completeness
\begin{align}
  \Phi(R,Z) & = -4\pi\,G\rho_d\frac{z_h}{R_d} \int_0^\infty \dd k\,J_0(kR)\,(k^2+\frac{1}{R_d^2})^{-3/2}\nonumber\\
& \qquad \times \,\frac{e^{-k|Z|} - k z_h e^{-|Z|/z_h}}{1-(k\,z_h)^2}\,,\\
  F_R(R,Z) & = -4\pi\,G\rho_d\frac{z_h}{R_d} \int_0^\infty \dd k\,k\,J_1(kR)\,(k^2+\frac{1}{R_d^2})^{-3/2}\nonumber\\
  & \qquad \times\,\frac{e^{-k|Z|} - k z_h e^{-|Z|/z_h}}{1-(k\,z_h)^2}\,,\\
  F_Z(R,Z) & = -\mathrm{sign}(Z)\,4\pi\,G\rho_d\frac{z_h}{R_d} \int_0^\infty \dd k\,k\,J_0(kR)\nonumber\\
  & \qquad \times\,(k^2+\frac{1}{R_d^2})^{-3/2}\,\frac{e^{-k|Z|} - e^{-|Z|/z_h}}{1-(k\,z_h)^2}\,,
\end{align}
and similar expressions for the second derivative of the potential.
The integrals in these expressions contain strongly oscillating Bessel
functions $J_i(\cdot)$, especially when evaluating them near the
Galactic plane. We calculate these integrals using 10-point
Gauss-Legendre integration between each of the zeros of the relevant
Bessel function out to $k = 2/z_h$ ($4/z_h$ for the
radial force) for $R \geq R_0$ and $k = 2\,R_0/(R\,z_h)$
($4\,R_0/(R\,z_h)$ for the radial force) for $R < R_0$. At large
radii ($R > 6\,R_0$) the exponential disk potential is approximated as
Keplerian.

\begin{figure}[t!]
  \includegraphics[width=0.5\textwidth,clip=]{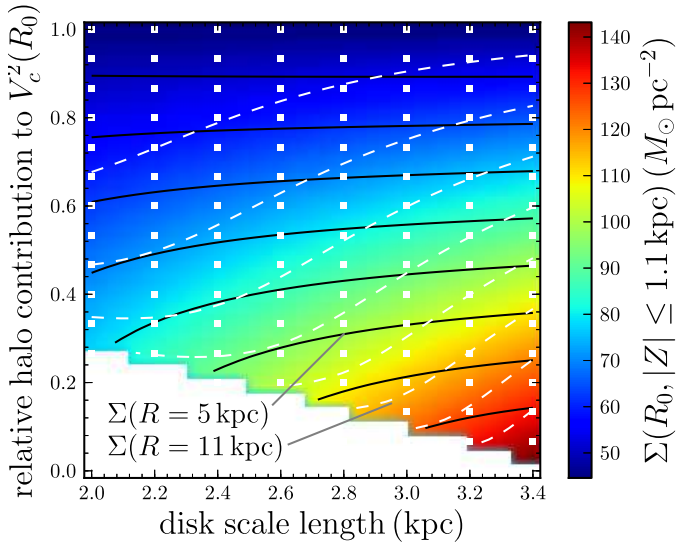}
  \caption{Range of surface densities spanned by our fiducial model
    for the Milky Way's potential near the plane of the disk, when
    varying the two free potential parameters (see
    \sectionname~\ref{sec:potential}): the stellar disk scale length
    and the relative contribution from the halo to the
    stellar-disk+halo contribution to the circular velocity at
    $R_0$. The colors represent surface densities at $R_0$, while the
    black, solid and white, dashed contours indicate lines of constant
    surface density at $5\kpc$ and $11\kpc$, respectively. The white
    squares show the actual grid points included in the fiducial model
    (see \tablename~\ref{table:potgrid}).}\label{fig:surfrdfh}
\end{figure}

The free parameters of the model for the gravitational potential are
parameterized using: (1 \& 2) the stellar disk mass scale length $R_d$
and mass scale height $z_h$, (3) the relative contribution of the halo
to the stellar-disk+halo contribution to the radial force at $R_0$,
(4) the circular velocity $V_c$ at $R_0$ (this sets the overall
amplitude of the potential), and (5) the logarithmic derivative of the
circular velocity at $R_0$ with respect to $R$, $\dd \ln V_c (R_0) /
\dd \ln R$ (\ie, the ``flatness'' of the rotation curve). All of the
other parameters of the potential can be calculated in terms of these
basic parameters. We perform all the calculations involving the
actions and qDF by first re-scaling the potential such that $V_c(R_0)
\equiv 1$ at $R_0 = 1$; this requires re-scaling the data positions by
$R_0$ and velocities by $V_c$ and re-scaling the input parameters of
the qDF similarly. The PDF (see below) then needs to be divided by
appropriate factors of $V_c(R_0)$ and $R_0$ to have the correct units.

\begin{figure}[t!]
  \includegraphics[width=0.5\textwidth,clip=]{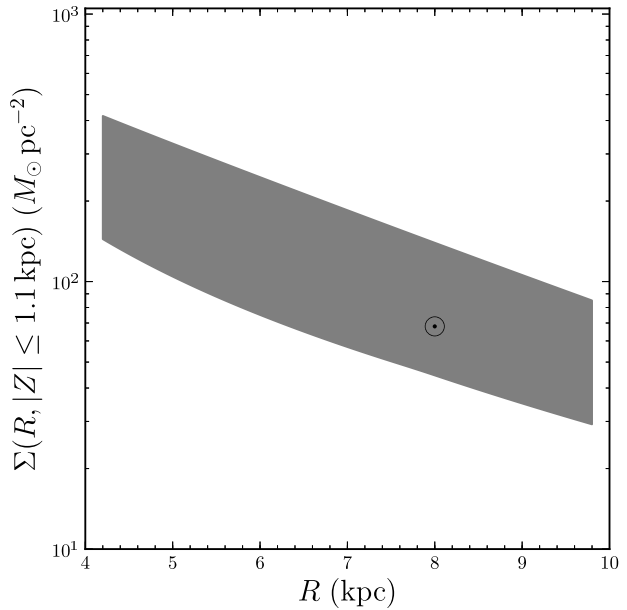}
  \caption{Range of surface densities spanned by our fiducial
    potential model as a function of radius, when varying the two
    potential parameters. This Figure shows the range of surface
    densities at every $R$ corresponding to the same models as shown
    in \figurename~\ref{fig:surfrdfh}. Our fiducial model includes the
    a priori reasonable range of surface densities at all $R$ between
    $4\kpc$ and $10\kpc$. A typical value (and the final measurement
    in \sectionname~\ref{sec:potentialfit}) of $\sigoneone(R_0)$ is
    indicated by the solar symbol.}\label{fig:prior-surface-range}
\end{figure}

In the fits of a qDF to individual \map s below we fix all of the five
basic parameters except for the stellar disk scale length and the
relative halo-to-disk contribution to the radial force at $R_0$. This
proved necessary because exploring the joint qDF+potential parameters
PDF for each \map\ is computationally expensive such that two
potential parameters is the maximum that can currently be efficiently
explored (see below for more discussion of this). In our fiducial
model we set the stellar disk mass scale height to $400\pc$ (this is
the mass-weighted mean scale height calculated based on the star-count
decomposition of \bo b, \ie, it is the mean of the histogram shown in
Figure 2 of \bo b), we fix the potential amplitude to be such that
$V_c = 230\kms$, and we fix the rotation curve to be flat at $R_0$,
\ie, $\dd \ln V_c(R_0) \dd \ln R = 0$. While these may appear to be
strong assumptions, we stress that the main goal of the \map\ fits is
to measure the surface density \sigoneone\ at $1.1\kpc$ at the
Galactocentric radius where this is best determined for each \map. As
such, the range of allowed \sigoneone\ is what really
matters. \figurename~\ref{fig:surfrdfh} shows the range of
$\sigoneone(R_0)$ in the fiducial model; it is clear that any
reasonable $\sigoneone(R_0)$ (cf. \citealt{BovyTremaine,Zhang13a}) is
included in the fiducial
model. \figurename~\ref{fig:prior-surface-range} shows the range of
$\sigoneone(R)$ between $4\kpc$ and $10\kpc$ included in the fiducial
model, and this is wide enough to encompass all a priori reasonable
$\sigoneone(R)$. We illustrate below that changing the potential
parameters that we keep fixed in the fiducial model does not
significantly change our results.

\subsection{Fitting procedure}\label{sec:fit}

In \sectionname~3.1 of \bo d we described a method for fitting the
density profiles of a sample of stars using the individual positions
of the stars in a likelihood-based approach and correcting for the
effects of various selection biases. Here we expand this methodology
to fitting both the positions and velocities to a 6D DF. As described
in \sectionname~\ref{sec:qdf}, we model the DF with a dynamical
equilibrium model specified by the qDF, which is a function of the
actions \vecj\ only, that is $\df(\vecx,\vecv) \equiv
\df(\vecj[\vecx,\vecv])$. We work in the approximation that the
selection function has only a spatial (or magnitude) dependence, which
is the case for the G-dwarf sample used in this paper.

We fit a qDF to the various \map s by taking into account the fact
that the observed stellar number counts do not reflect the underlying
(volume-complete) stellar distribution, but are instead strongly
shaped by (a) the strongly position-dependent selection fraction of
stars with spectra (see \figurename~11 in \bo d), (b) the need to use
photometric distances to relate the underlying phase-space model to
the observed magnitude distribution (as the magnitude-limited
\segue\ sample corresponds to a color- and metallicity-dependent
distance-limited sample), and (c) the pencil-beam nature of the
survey. As in \bo d, we model the observed distribution of stars as a
Poisson sampling of the underlying population. The details of this
method are the same as that described in \sectionname~3.1 of \bo d and
we refer the reader to that paper for a detailed discussion.

Since the photometric distance estimates $D$ depend on the $g-r$
color, metallicity \feh, and apparent $r$-band magnitude, and because
the selection function is a function of position, $r$, and $g-r$, we
need to model the observed density of stars in
color--magnitude--metallicity--phase (position--velocity) space,
$\lambda(l,b,d,\vecv,r,g-r,\feh)$. This density of stars can be
written as
\begin{equation}\label{eq:rate}
\begin{split}
\lambda(l,b,D,\vecv,r,&g-r,\feh)= \\ &
\rho(r,g-r,\feh|R,Z,\phi)\times\qdf(\vecx,\vecv) \\
& \ \times
|J(R,Z,\phi;l,b,D)| \times S(\mathrm{plate},r,g-r)\,.
\end{split}
\end{equation}
Here, $(R,Z,\phi)$ are Galactocentric cylindrical coordinates
corresponding to rectangular coordinates $\vecx\equiv(X,Y,Z)$, which
can be calculated from $(l,b,D)$. The factor
$\rho(r,g-r,\feh|R,Z,\phi)$ is the number density in
magnitude--color--metallicity space as a function of position (see
further discussion in \appendixname~B of \bo d). The $|J(R,z;l,b,D)|$
is a Jacobian term because of the $(X,Y,Z) \rightarrow (l,b,D)$
coordinate transformation; the factor $S(\mathrm{plate},r,g-r)$ is the
selection function as given in \equationname~(A2) of \bo d. Finally,
$\qdf(\vecx,\vecv)$ is the underlying DF of the sample. While we model
this DF as being a function of the actions \vecj\ only, we stress that
the DF always remains a density in
$(\vecx,\vecv)$--space\footnote{This caveat may seem unnecessary as
  the Jacobian of the canonical $(\vecx,\vecv) \rightarrow
  (\vecj,\vec{\theta})$ transformation is unity. However, because we
  cannot calculate the actions exactly, the Jacobian will not be
  exactly equal to one.}. The parameters \paramspot\ of the
gravitational potential $\Phi$ enter the observed density through the
dependence of \vecj\ on $\Phi$. We denote the parameters of the DF
using $\paramsdf \equiv (h_R,\sigma_Z(R_0), h_{\sigma_Z},
\sigma_R(R_0), h_{\sigma_R})$.

The likelihood of a given model for the DF and potential $\Phi$ is
given by that of a Poisson process with rate parameter
$\lambda$. After marginalizing over the amplitude of the DF with a
logarithmically-flat prior (which effectively normalizes the DF), the
likelihood of a single data point $i$ given by
\begin{equation}\label{eq:dflike}
\begin{split}
  & \ln \mathcal{L}_{i,\mathrm{DF}} = \ln
    \qdf(\vecj[\vecx_i,\vecv_i]|\paramspot,\paramsdf) \\
    & \qquad \qquad -\ln \int \dd l
    \,\dd b\, \dd D\, \dd \vecv\,\dd r\, \dd (g-r) \,\dd\feh\\
    & \qquad \qquad \qquad \ \lambda(l,b,D,\vecv,r,g-r,\feh|\paramspot,\paramsdf)\,.
\end{split}
\end{equation}
The second term normalizes the density $\lambda$ over the observed
volume. In \appendixname~\ref{sec:normint} we describe how we
calculate the 9-dimensional normalization integral efficiently. For
comparison with observed velocities, we convolve the likelihood in
\eqnname~(\ref{eq:dflike}) with the observational velocity
uncertainties.

Because we are primarily interested in measuring the surface density
$\sigoneone(R)$, the dynamics of \map s in the plane of the Milky Way
is mostly irrelevant as it does not contain much information about the
vertical surface density. Moreover, non-axisymmetric perturbations to
the Galactic potential (\eg, those from spiral structure or the bar)
strongly affect the kinematics in the plane of the Galaxy, such that
non-axisymmetric planar kinematics ($V_R$ and $V_T$) could bias the
fit of an axisymmetric model to the \map\ data. Practically,
discarding the planar kinematics also allows us to fix the parameters
of the qDF related to the radial velocity dispersion, as they only
affect the distribution of velocities in the plane; this reduces the
number of qDF parameters from five to three, thus significantly
reducing the computations involved in evaluating the likelihood (see
\sectionname~\ref{sec:details} below). Therefore, we do not consider
the planar kinematics and marginalize the likelihood over $V_R$ and
$V_T$ for each star. Thus, we only need to convolve the likelihood
with the observational uncertainty in the vertical velocity (since we
assume that distance uncertainties are only important as far as the
velocity is concerned). We perform this convolution using 20-th order
Gauss-Legendre quadrature over the range
$(V^i_Z-4\delta\,V^i_Z,V^i_Z+4\delta\,V^i_Z)$, where $V^i_Z$ and
$\delta\,V^i_Z$ are the observed velocity and its uncertainty. Thus,
the likelihood for a single data point $i$ becomes
\begin{equation}\label{eq:dflike2}
\begin{split}
  & \ln \mathcal{L}_{i,\mathrm{DF}} = \\
  & \ \ln
    \int_{V^i_Z-4\,\delta\,V^i_Z}^{V^i_Z+4\,\delta\,V^i_Z}\dd V_Z\,
    \int \dd V_R \,\dd V_T\,
    \qdf(\vecj[\vecx_i,\vecv_i]|\paramspot,\paramsdf)\\
    & \qquad \ \   -\ln \int
    \dd l \,\dd b\, \dd D\, \dd \vecv\,\dd r\, \dd (g-r) \,\dd\feh\\
    & \qquad \ \ \qquad \,\,\lambda(l,b,D,\vecv,r,g-r,\feh|\paramspot,\paramsdf)\,.
\end{split}
\end{equation}

We also add an outlier model consisting of a constant spatial density
and a Gaussian velocity distribution with a dispersion of
$100\kms\,\exp\left(-(R-R_0)/6\kpc\right)$; for \map s with best-fit
scale heights from \bo d larger than 500 pc, this outlier velocity
dispersion is multiplied by two. This outlier model mimics a halo
population. The outlier fraction $p_{\mathrm{out}}$ is added as a free
parameter. In the analysis below, the outlier fraction is small for
most \map s; only the most \aenhanced, \feh-poor \map s have outlier
fractions of $\sim\!10\,\%$, because their intrinsic dispersions are
large. Most importantly, there is no correlation between the outlier
fraction and the inferred value of $\sigoneone(R)$, such that it does
not affect our results. The full likelihood is then given by
\begin{equation}\label{eq:dflike3}
\begin{split}
\ln \mathcal{L} & = \sum_i\ln\left[(1-p_{\mathrm{out}})\,\mathcal{L}_{i,\mathrm{DF}} + p_{\mathrm{out}}\,\mathcal{L}_{i,\mathrm{out}}\right]\,,
\end{split}
\end{equation}
where $\mathcal{L}_{i,\mathrm{out}}$ is the likelihood contribution
from the outlier model.

\subsection{Further implementation details}\label{sec:details}

\begin{figure}[h!]
  \includegraphics[width=0.482\textwidth,clip=]{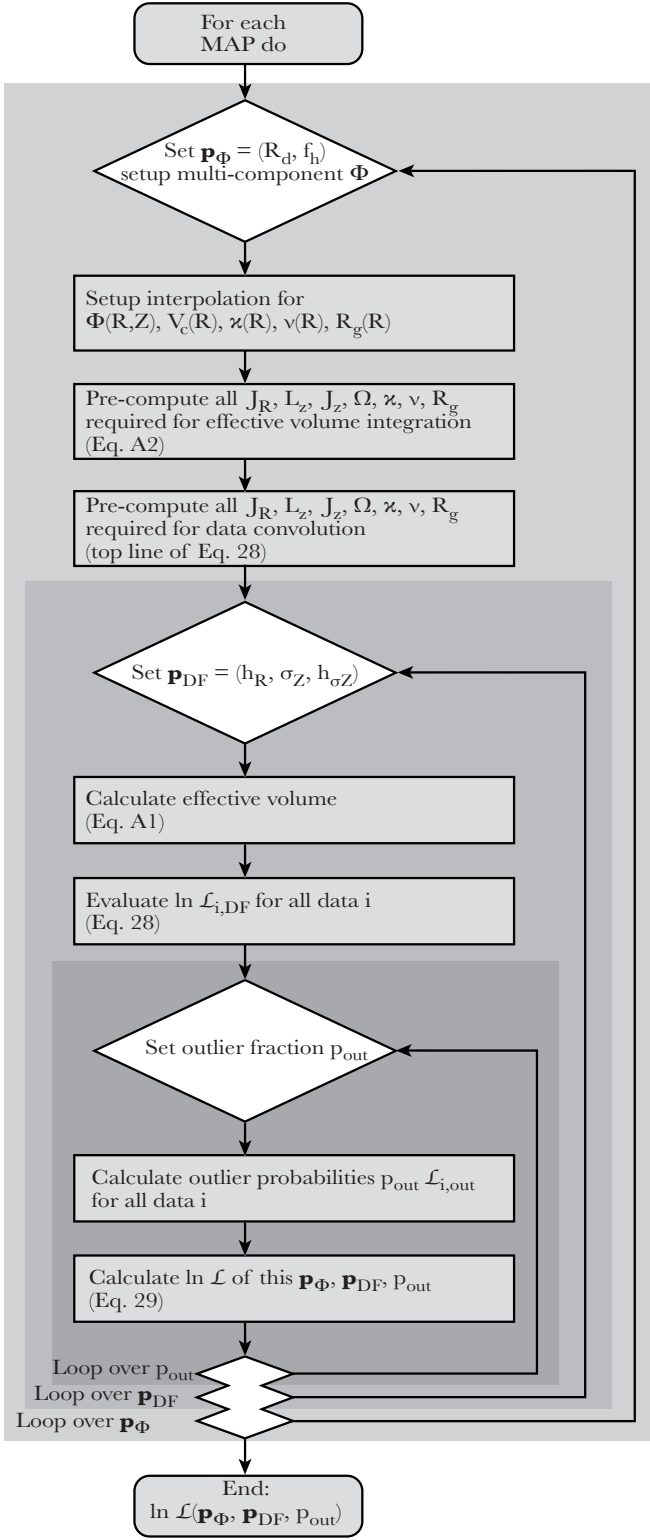}
  \caption{Graphical representation of the steps involved in
    determining the joint PDF of potential and qDF parameters for the
    data of each \map. These steps are described in detail in
    \sectionname~\ref{sec:method}, and the implementation in
    particular is discussed in \sectionname~\ref{sec:details}. The
    main loop steps through the different potential models of
    \tablename~\ref{table:potgrid}, for which all of the dynamical
    quantities involved in evaluating the qDF (actions and rotational,
    epicycle, and rotational frequencies) are pre-calculated. An inner
    loop then goes through the different qDF models of
    \tablename~\ref{table:qdfgrid} and outlier fractions to determine
    the likelihood of the potential+qDF models for the
    data.}\label{fig:flowchart}
\end{figure}

Using the methodology described in the previous Section, we explore
the joint PDF of the potential and qDF parameters for each
\map\ separately. We do this by calculating the PDF on a fixed grid
specified below. This Section gives the details of the specific
implementation choices made to explore the joint PDF of potential and
qDF parameters in a computationally efficient manner. The main steps
of the implementation are shown graphically in
\figurename~\ref{fig:flowchart}. 

\begin{deluxetable*}{lccccr}
\tablecaption{}
\tablecolumns{6}
\tablewidth{0pt}
\tabletypesize{\footnotesize}
\tablecaption{Potential models for which the analysis described in \sectionname~\ref{sec:method} to measure $\kzoneone(R)$ is performed}
\tablehead{\colhead{} & \colhead{$R_d$} & \colhead{$z_h$} & \colhead{$f_h$} & \colhead{$V_c$} & \colhead{$\frac{\dd \ln V_c(R_0)}{\dd \ln R}$}}

\startdata

fiducial model & 8 values from $2\kpc$ to $3.4\kpc$ & $400\pc$ & 16 values from $0$ to $1$ & $230\kms$ & $0.0$\\[10pt]

systematics: high $V_c$ & 8 values from $2\kpc$ to $3.4\kpc$ & $400\pc$ & 16 values from $0$ to $1$ & $250\kms$ & $0.0$\\
systematics: low $V_c$ & 8 values from $2\kpc$ to $3.4\kpc$ & $400\pc$ & 16 values from $0$ to $1$ & $210\kms$ & $0.0$\\
systematics: low $z_h$ & 8 values from $2\kpc$ to $3.4\kpc$ & $200\pc$ & 16 values from $0$ to $1$ & $230\kms$ & $0.0$\\
systematics: falling rot. curve & 8 values from $2\kpc$ to $3.4\kpc$ & $400\pc$ & 16 values from $0$ to $1$ & $230\kms$ & $-0.1$\\
systematics: rising rot. curve & 8 values from $2\kpc$ to $3.4\kpc$ & $400\pc$ & 16 values from $0$ to $1$ & $230\kms$ & $0.1$

\enddata\label{table:potgrid}
\end{deluxetable*}

The potential grid consists of 8 stellar disk scale lengths ranging
from $2\kpc$ to $3.4\kpc$ and 16 relative halo-to-disk contributions
to $V^2_c(R_0)$, ranging from $0$ to $1$. We do not consider models in
which the halo power-law is smaller than 0 or larger than 3; these
regions are blank in \figurename~\ref{fig:surfrdfh}. The list of
potential parameters, the range over which the two basic parameters
are varied, and the values at which the other potential parameters are
fixed are given in \tablename~\ref{table:potgrid} for all of the
potential models considered below.

\begin{deluxetable*}{lr}
\tablecaption{}
\tablecolumns{2}
\tablewidth{0pt}
\tabletypesize{\footnotesize}
\tablecaption{Parameters of the qDF and the range over which they are varied in the analysis}
\tablehead{\colhead{Parameter} & \colhead{Range considered}}

\startdata

$h_R / 8\kpc$ & 8 values log.-spaced between $-1.8$ and $0.9$ \\
$\sigma_Z(R_0)$ & 16 values log.-spaced over range of width 0.36 around estimate in \figurename~\ref{fig:previousAsQdf}\\
$h_{\sigma_Z}$ & 8 values log.-spaced between $\ln 4\kpc$ and $\ln16\kpc$ \\
$\sigma_R(R_0)$ & fixed at estimate in \figurename~\ref{fig:previousAsQdf} \\
$h_{\sigma_R}$ & fixed at $8\kpc$
\enddata\label{table:qdfgrid}
\end{deluxetable*}

The qDF as described in \sectionname~\ref{sec:qdf} has nominally five
free parameters. However, as discussed in the previous Section, we
marginalize the likelihood over the radial and azimuthal velocity
components $V_R$ and $V_T$, such that the parameters describing the
radial dispersion profile---$\sigma_R(R_0)$ and $h_{\sigma_R}$---are
not important for the data modeling. Therefore, we fix $\sigma_R(R_0)$
at the values estimated based on fits of the radial dispersion similar
to those performed in \bo c (see \figurename~\ref{fig:previousAsQdf})
and we set $h_{\sigma_R} = 8\kpc$. This leaves three free qDF
parameters for each \map: $h_R$, $\sigma_Z(R_0)$, and
$h_{\sigma_Z}$. \tablename~\ref{table:qdfgrid} lists all five qDF
parameters and the range over which they are varied in the analysis,
as described below.

For each \map\ we consider the same range of $\ln h_R / 8\kpc$: 8
values from $-1.86$ to $0.9$. The range of the vertical dispersion
scale length is also the same for each \map: 8 values logarithmically
spaced between $4\kpc$ and $16\kpc$. The range of $\sigma_Z(R_0)$ is
set individually for each \map\ based on the estimates for
$\sigma_Z(R_0)$ from \figurename~\ref{fig:previousAsQdf}. 16 values
are logarithmically spaced over a range of width $0.36$ around the
estimated $\ln \sigma_Z(R_0)$. The central values were slightly
adjusted based on visual inspection of the PDFs such that the best-fit
does not occur at the edge of the parameter range; these adjustments
are typically less than $20\,\%$.

In practice, for computational speed reasons, the likelihood
calculation is structured as shown in
\figurename~\ref{fig:flowchart}. We first setup an interpolation grid
for the potential and the rotational, epicycle, and vertical
frequencies as well as the guiding-star radii. These grids are then
used to quickly evaluate the gravitational potential when calculating
the actions and the interpolated frequencies and guiding-star radii
are used in the fast evaluation of the qDF (see below). This step is
necessary because directly evaluating the contribution to the
potential coming from the stellar and gas exponential disks would be
prohibitively computationally expensive.  The explored parameters for
each \map\ are then separated in three classes: (1) the potential
parameters \paramspot, (2) the 3 qDF parameters, and (3) the outlier
fraction.

Once the parameters of the potential are specified, all of the actions
that are involved in evaluating the likelihood in
\equationname~(\ref{eq:dflike2}) are pre-computed. This includes all
of the actions involved in the calculation of the effective survey
volume (\equationname~[\ref{eq:normintdens}]) and the actions for the
data that are used to convolve the likelihood with the vertical
velocity uncertainty (\equationname~[\ref{eq:dflike2}]). As discussed
above, these integrals over the velocities are performed over a range
specified by the model dispersions (those that are explicit parameters
of the qDF). In order to be able to re-use the actions for multiple
qDF parameter sets, we calculate the integrals over the velocity,
always using ranges based on the estimated dispersions from
\figurename~\ref{fig:previousAsQdf}. As the dispersions that are
considered during the PDF exploration are typically within $\lesssim
30\,\%$ of these estimates and we integrate out to $4\,\sigma$, this
procedure works well for all of the considered qDF parameters. It also
makes sure that there is no stochastic noise in the likelihood
evaluation, which could be large if the velocity and/or spatial
integrals are performed using Monte Carlo integration (see discussion
in \citealt{McMillan13a} and \citealt{Ting13a}). We also pre-calculate
the rotational, epicycle, and vertical frequencies as well as the
guiding star radii associated with all of the points at which the qDF
has to be evaluated in the course of a single likelihood evaluation
(cf. the definition of the qDF in \equationname~[\ref{eq:qdf}]). Once
the actions, frequencies, and guiding star radii are calculated, the
likelihood can be quickly evaluated for different sets of qDF
parameters, by evaluating the qDF using the pre-calculated values and
re-summing the integrals.

With all of the actions etc. pre-calculated we then vary the qDF
parameters (second loop in \figurename~\ref{fig:flowchart}). For any
given qDF parameter set we calculate the effective survey volume and
evaluate the likelihood related to the qDF for all of the data
points. The outlier fraction can then be varied very quickly and we
use a grid of 25 outlier fractions ranging from $0\,\%$ to $50\,\%$
(innermost loop in \figurename~\ref{fig:flowchart}).

We find that by using Gauss-Legendre quadrature to perform the
velocity integrals, we can evaluate the effective survey volume
accurately enough (that is, to an accuracy $\ll 0.001\,\%$,
sufficiently small such that this is not a source of noise in the
likelihood) using $\approx2\times10^6$ action evaluations. This is
less than the $\approx10^7$ evaluations required by
\citet{McMillan13a} and this leads to a significant reduction in the
time required for a single likelihood evaluation, as setting up the
grid of actions takes up a significant amount of time. The reason
Gauss-Legendre quadrature works well is that the qDF is a very smooth
function and it is easy to estimate the range over which it is
non-zero. Similarly, we have found that Gauss-Legendre quadrature is
far superior to Monte Carlo sampling for convolving the likelihood
with the velocity uncertainty: we can accurately calculate the
convolved likelihood with 20-th order Gauss-Legendre quadrature,
whereas when even using thousands of Monte Carlo samples the
convolution was not well-behaved (even though we used the \emph{same}
Monte Carlo samples for all different parameter settings, to remove
numerical noise as advocated in \citealt{Ting13a}). Future analyses of
Gaia data will still be mostly in the regime where distance
uncertainties are only important through their influence on the
velocity uncertainties. As 3D Gauss-Legendre quadrature could be
adequately performed using $\sim\!10^5$ evaluations of the integrand,
Gauss-Legendre quadrature will likely still be superior to Monte Carlo
integration.

The computational time is dominated both by setting up the grid of
actions and frequencies for a given potential and calculating the
effective survey volume for a given set of qDF parameters. Using all
of the speed-ups described in this Section, the exploration of the PDF
still requires $\sim\!3\,\mathrm{days}$ using 8 cpus for a single
\map. Each \map's PDF can be explored independently and all of the
steps involved in the likelihood evaluation can be easily
parallelized such that the availability of sufficient computational
resources is the main limitation for exploring a wider range of
potential models. The combined computational cost for all steps of
the analysis described in this paper was $20,000$ cpu-hours, or
$\approx\!1$ cpu-hour per datapoint.

\section{Results}\label{sec:results}

In this Section we present the results of fitting each \map\ with a
qDF model to measure \sigoneone\ at the Galactocentric radius $R$ at
which this quantity is best constrained for each \map. In
\sectionname~\ref{sec:indivresults} we discuss the joint constraints
on the dynamical parameters of the gravitational potential model and
on the qDF parameters for each \map\ individually. We then describe
how we determine the radius for each \map\ at which \sigoneone\ is
best constrained for that \map. In \sectionname~\ref{sec:surfresults}
we present the $\sigoneone(R)$ profile that results from combining the
individual fits to the different \map s. In
\sectionname~\ref{sec:systematics} we show that the $\sigoneone(R)$
profile that we measure using the fiducial model for the gravitational
potential does not significantly change if we change the parameters of
the fiducial potential. We also demonstrate that the impact of
systematic distance uncertainties is small.

\subsection{Results for individual \map s and determination of $\Sigma_{1.1}(R)$}\label{sec:indivresults}

As discussed in \sectionname~\ref{sec:method}, we fit each set of
\map\ data using a qDF model and we determine the joint PDF of two
dynamical parameters + three qDF parameters on a grid in all of the
parameters. We thus obtain constraints on the basic dynamical
parameters, disk scale length and relative halo-to-disk contribution
to $V_c^2(R_0)$, after marginalizing over the parameters of the qDF
(which is a weak form of the ``summing over all possible DFs'' of
\citet{Magorrian06a}). We can calculate the PDF of derived dynamical
parameters, such as the surface density at a given radius, the density
of the dark halo, etc., by appropriate transformations of the PDF of
the basic dynamical parameters (with the understanding that all of the
different derived parameters will be strongly correlated because there
are only two free parameters). Similarly, we can marginalize over the
dynamical parameters to constrain the qDF parameters for the
individual \map s, but this is not the focus of this paper.

\begin{figure}[t!]
  \includegraphics[width=0.23\textwidth,clip=]{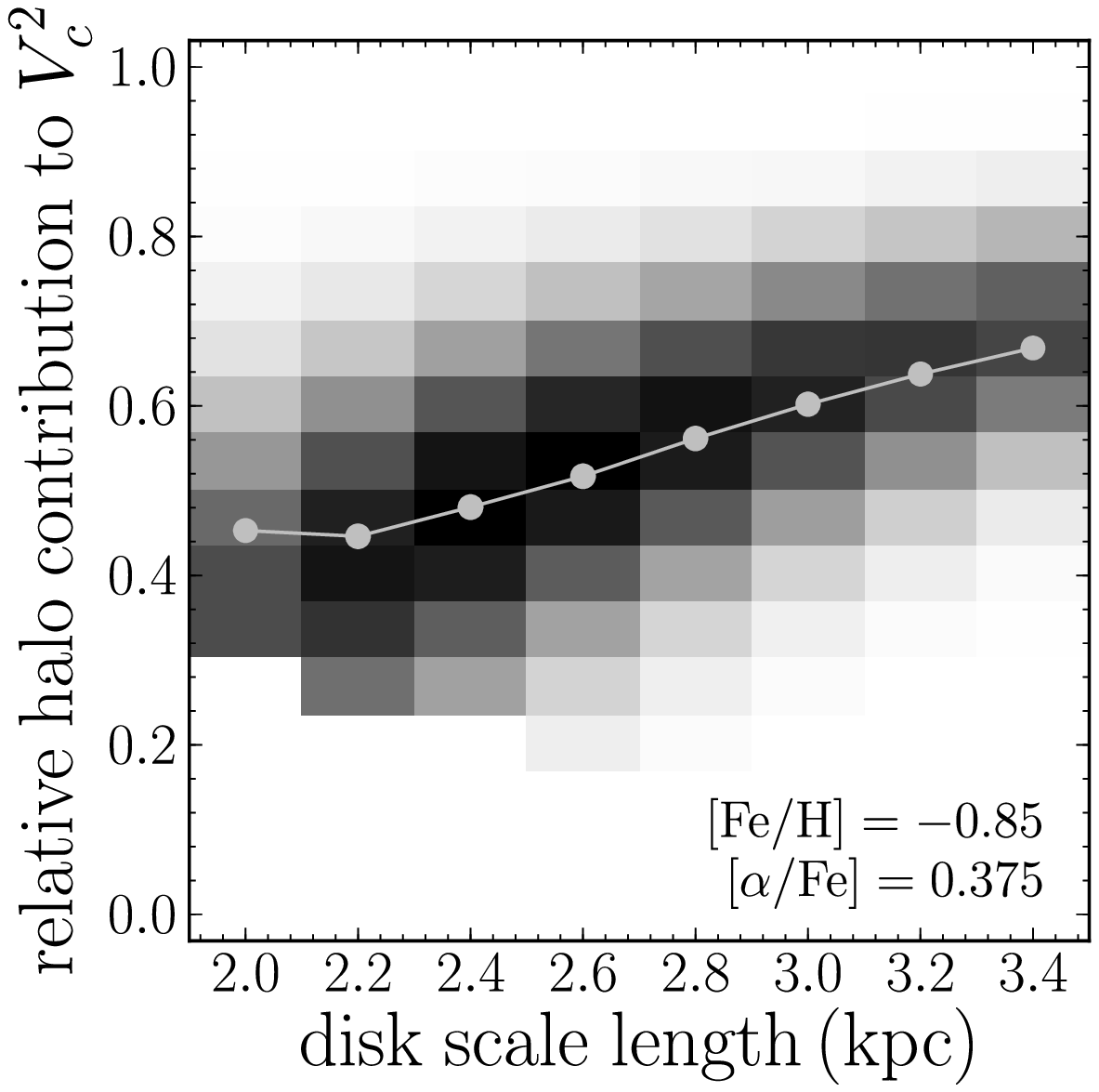}
\raisebox{-1.5pt}{\includegraphics[width=0.238\textwidth,clip=]{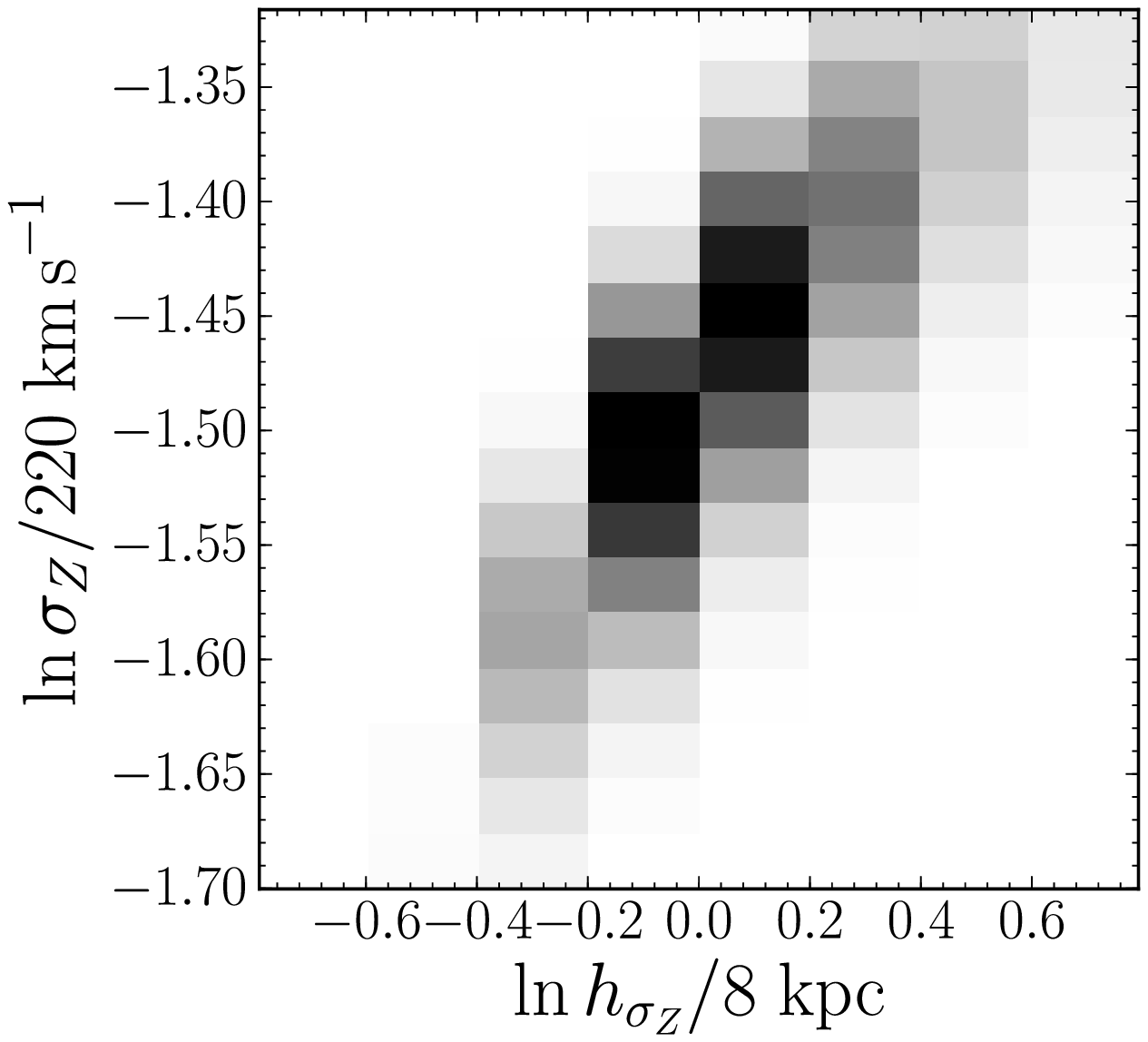}}\\
  \includegraphics[width=0.23\textwidth,clip=]{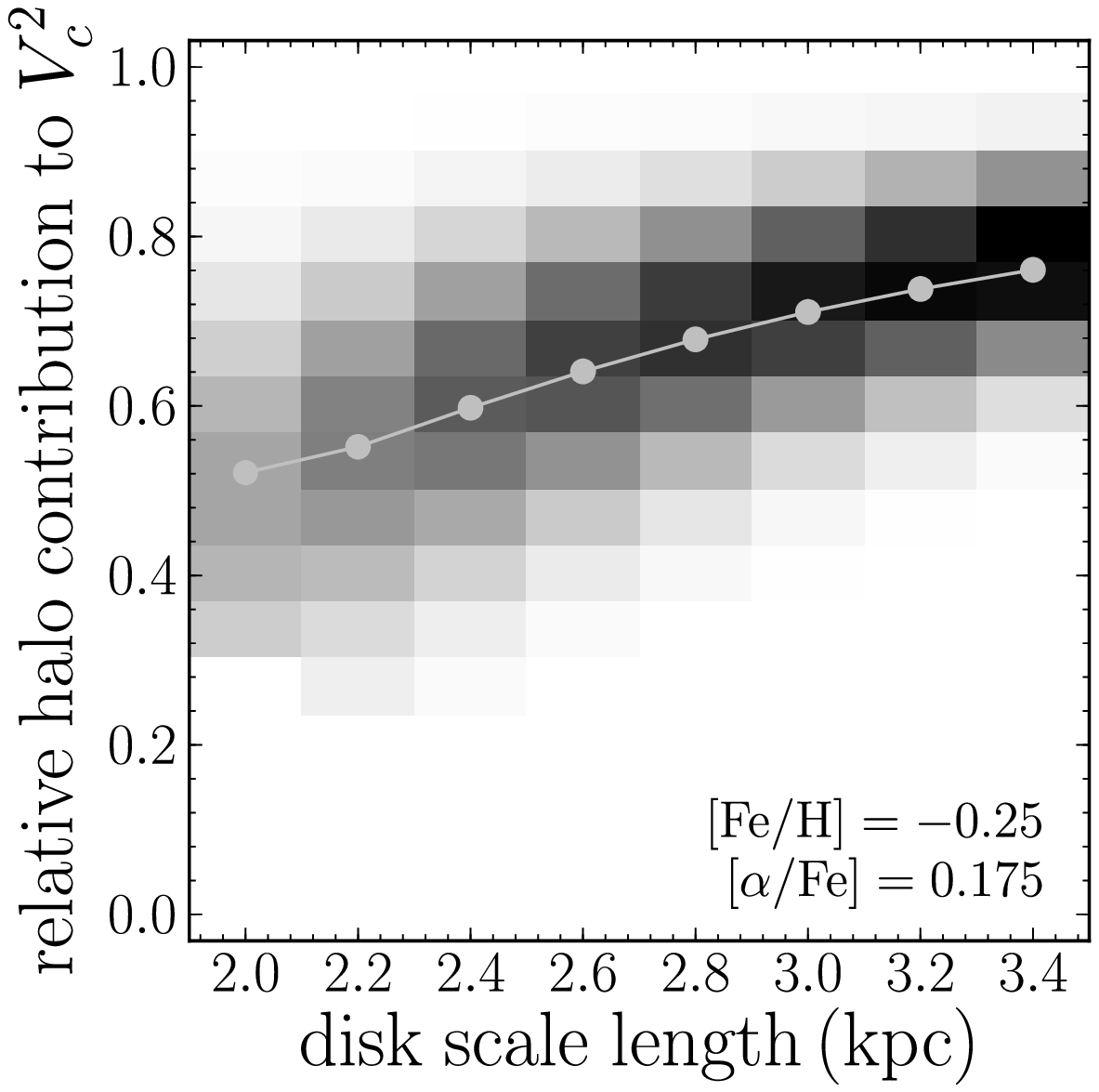}
  \raisebox{-1.5pt}{\includegraphics[width=0.238\textwidth,clip=]{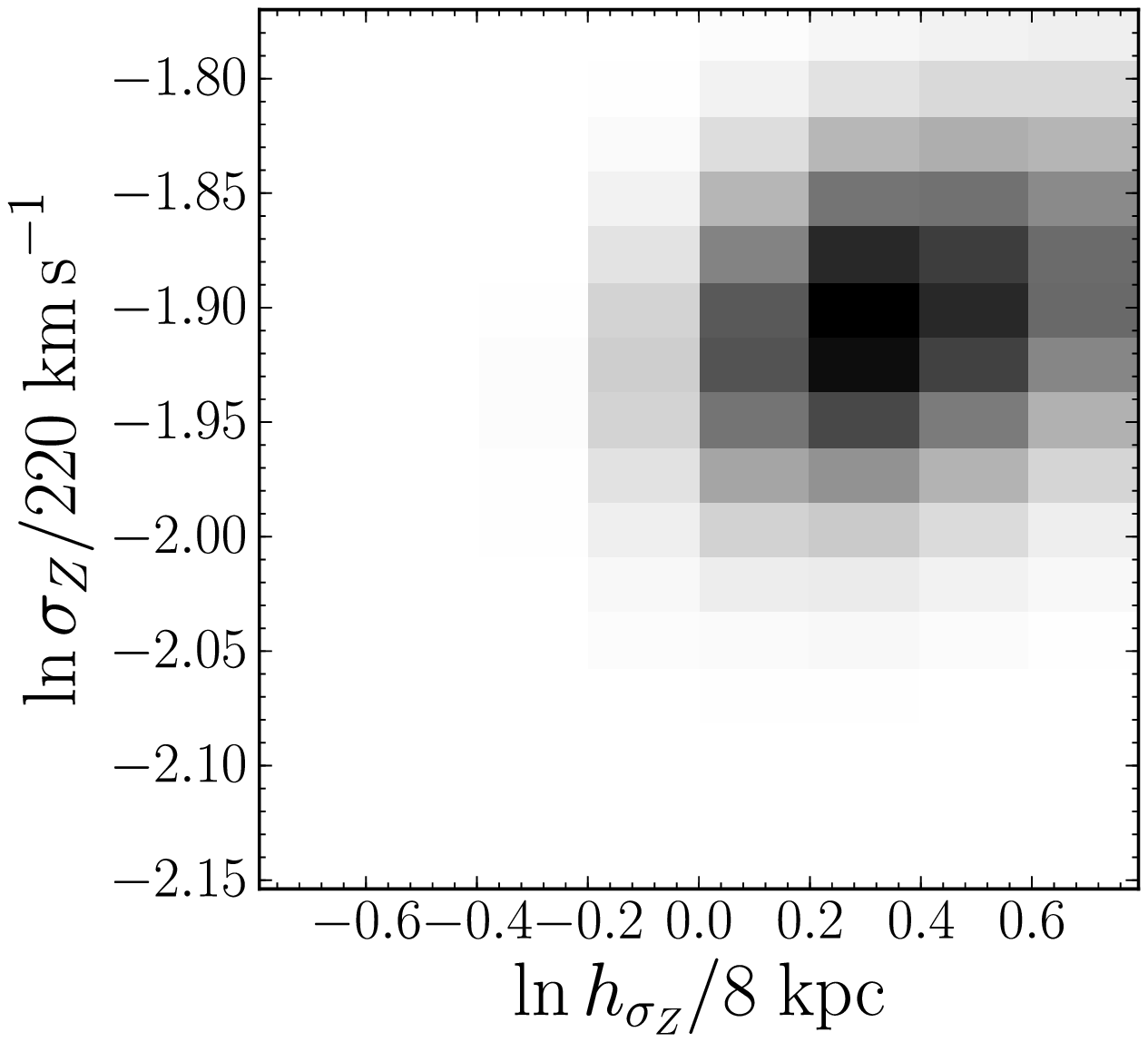}}
  \caption{Example PDFs for individual \map s. This Figure shows the
    2D PDF for the dynamical parameters (marginalized over the qDF
    parameters) and the 2D PDF for the qDF parameters governing the
    vertical velocity dispersion (marginalized over the dynamical
    parameters and the qDF parameter governing the radial density
    profile) for two example \map s. The gray points in the left
    panels are the mean of the PDF of the relative halo-to-disk
    contribution for a given scale length. The top panels are for a
    metal-poor, \aenhanced\ \map, while the bottom panels are for a
    more metal-rich \map. A quantitative comparison between the
    dynamical PDFs in the left panels and
    \figurename~\ref{fig:surfrdfh} shows that this metal-poor
    \map\ most robustly measures \sigoneone\ at $R = 6.6\kpc$, while
    this metal-rich \map\ measures \sigoneone\ best at $7.7\kpc$ (see
    \tablename~\ref{table:surf}). The optimal radius is determined as
    that for which the correlation between $\sigoneone$ and the disk
    scale length is minimized (see
    \figurename~\ref{fig:illustraterbest}); however, the dispersion
    PDFs in the right panels also show that the metal-poor
    \map\ measures the vertical dispersion closer to the Galactic
    center than the more metal-rich \map.}\label{fig:pdfs}
\end{figure}

\figurename~\ref{fig:pdfs} shows as examples the PDFs resulting from
the fits to two \map s for the basic dynamical parameters as well as
PDFs for the qDF parameters related to the vertical velocity
dispersion, $\sigma_Z(R_0)$ and $h_{\sigma_Z}$. We only show PDFs for
two \map s in \figurename~\ref{fig:pdfs} and we only show two
different 2D marginalized PDFs to give a sense of the PDFs and how
they are different for metal-poor and metal-rich \map s. We have
visually inspected all such PDFs as well as other 2D projections for
all individual \map s. We have done this for the fit of the fiducial
potential model as well as for all of the systematics checks in
\sectionname~\ref{sec:systematics}. All PDFs were found to be
sensible, although in a few cases the range of $\sigma_Z(R_0)$ had to
be changed and the PDF had to be re-computed since the best-fitting
value of $\sigma_Z(R_0)$ was found to lie on the boundary of the
$\sigma_Z(R_0)$ grid. Only for one of the 45 \map s in our initial
sample were we unable to adjust the range of $\sigma_Z(R_0)$ within an
acceptably small number of iterations and we dropped this bin from
further consideration: this \map\ is located at $\feh = -0.25$, $\afe
= 0.275$, and it contained only 129 data points.

To assess whether the qDF model is a good fit to the observed spatial
density and kinematics of the \map s we have performed extensive
direct comparisons between the data and the best-fitting model for
each \map. We do this by projecting the best-fitting model into the
space of observables taking into account the various selection biases
affecting the data. These data--model comparisons are described and
discussed in detail in \appendixname~\ref{sec:datamodel}. Overall, we
find that the spatial distribution and the vertical kinematics of the
data are very well fit by the qDF models for individual \map
s. However, we remove from further consideration the \map\ at $\feh =
-0.05$ and $\afe = 0.125$, because the detailed comparisons between
the data and the model indicates that no good fit is obtained for this
\map; this \map\ contains 255 data points. All 43 other \map s,
containing 16,269 data points, are included in the results described
below.

In principle, the optimal way of constraining the mass distribution
using individual \map s would be to multiply together the individual
PDFs for the dynamical parameters. However, these PDFs of the
individual \map s are not mutually consistent, e.g., the joint PDF for
all \aenhanced\ \map s is inconsistent with the joint PDF for all
metal-rich, \apoor\ \map s. This is a reflection of the fact that our
fiducial model for the gravitational potential, with only two free
parameters, is too restrictive, \ie, it does not contain a model that
is consistent with all of the individual \map s. However, the
2-parameter potential model \emph{is} general enough to provide a
measurement of the gravitational force in the small range of radii for
each \map\ that most affects the \map's orbital distribution. 

The PDF of the dynamical parameters in \figurename~\ref{fig:pdfs} and
those of all other \map s are characterized by a strong degeneracy
between the two basic dynamical parameters. A comparison with
\figurename~\ref{fig:surfrdfh} shows that this degeneracy is along
lines of constant surface density, which is expected as the vertical
kinematics of an individual \map\ primarily constrains the total
surface density. A comparison with \figurename~\ref{fig:surfrdfh} also
makes it clear that the direction of the degeneracy is indicative of
the \emph{radius} at which the surface density is best constrained by
the data of an individual \map. We focus on the surface density at a
height of $1.1\kpc$, because that is close to the mode of the
distribution of distances-from-the-midplane of the data for each
\map, which is set through the combination of the intrinsic vertical
distribution and the SEGUE selection procedure.

\begin{figure}[t!]
  \includegraphics[width=0.5\textwidth,clip=]{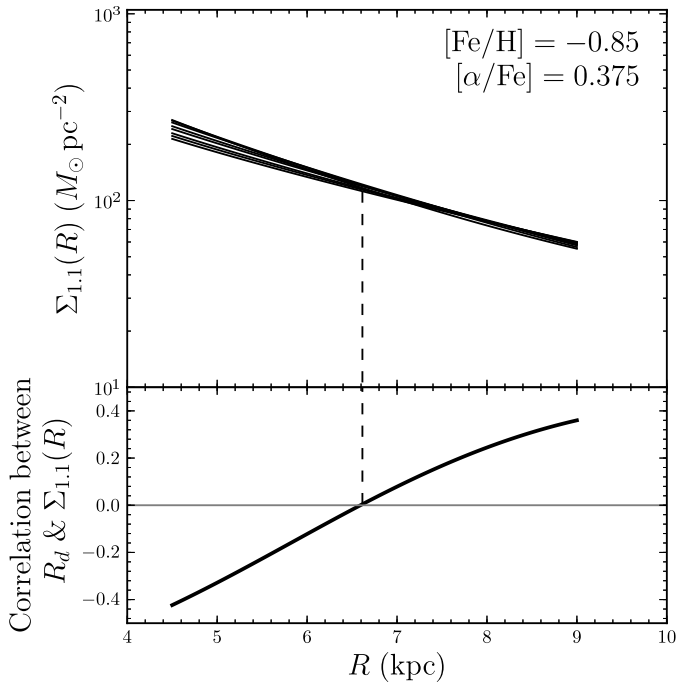}
  \caption{Illustration of the procedure used to determine the radius
    $R$ at which each \map\ best determines $\sigoneone(R)$, shown for
    the more metal-poor \map\ of \figurename~\ref{fig:pdfs}. The top
    panel shows the $\sigoneone(R)$ profile corresponding to the
    best-fit relative halo contribution to $V_c^2(R_0)$ for each of
    the 8 values of the disk scale length considered in the fiducial
    potential model. The bottom panel shows the posterior correlation
    between the disk scale length and $\sigoneone$ as a function of
    radius. The radius $R$ at which a given \map\ measures
    $\sigoneone$ is that for which the correlation between the disk
    scale length and $\sigoneone$ is minimized. The top panel shows
    that this is the radius around which the inferred $\sigoneone(R)$
    profile pivots when changing the assumed disk scale
    length.}\label{fig:illustraterbest}
\end{figure}

Formally, we can calculate the radius at which \sigoneone\ is best
determined by calculating the correlation between the disk scale
length and $\sigoneone$ as a function of $R$ and finding the radius at
which this correlation is minimized. This radius serves as the point
around which $\sigoneone$ pivots as the model's scale length is
changed. As such, it is the radius at which \sigoneone\ is most
robustly determined. From the PDF of the dynamical parameters, we can
then calculate the mean $\sigoneone(R)$ and its uncertainty at that
radius. \figurename~\ref{fig:illustraterbest} illustrates this
procedure.

\figurename~\ref{fig:pdfs} shows that the degeneracy in the PDF of the
dynamical parameters, and therefore the correlation between
\sigoneone\ and the disk scale length, is different for different \map
s. Thus, the radius at which \sigoneone\ is best determined is
different for the various \map s. This is plausible as stars of a
given \map, found in the Solar vicinity, are drawn from tracer
populations of very different scale lengths (\bo d). We expect that
\map s with abundances that are close to solar measure \sigoneone\ at
a radius that is close to the solar circle, because these stars
dominate the local population of stars and they spend much of their
orbits close to the solar circle. Likewise, we expect that the
\aenhanced\ \map s, which have short scale lengths and large velocity
dispersion (\bo), spend a much larger fraction of their orbit at
smaller radii and therefore measure \sigoneone\ closer to the Galactic
center.

\begin{figure}[t!]
  \includegraphics[width=0.5\textwidth,clip=]{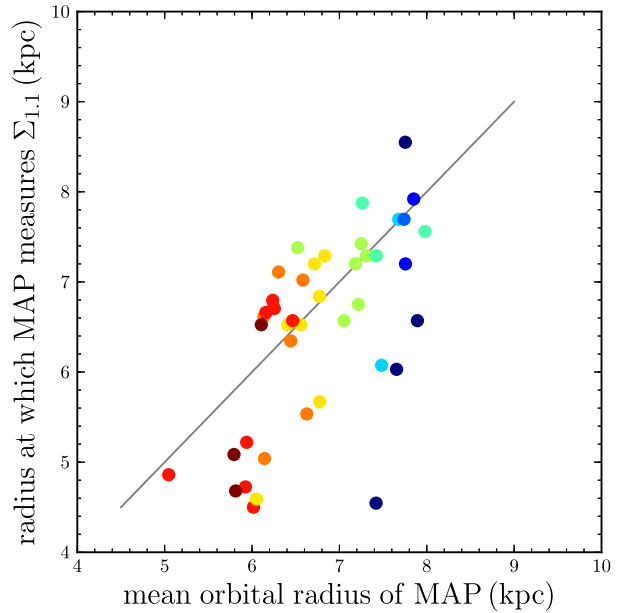}
  \caption{Comparison between the typical mean orbital radius of SEGUE
    stars in a MAP and the radius at which the correlation between
    $\sigoneone(R)$ and the model stellar disk scale length is
    minimized for each \map\ (see
    \figurename~\ref{fig:illustraterbest}). The typical mean radius is
    defined as the median of the midplane density weighted mean radii
    calculated in a simple model for the Milky Way's gravitational
    potential (see text). Points are color-coded by the \afe\ of the
    \map\ that they represent (as in \figurename~\ref{fig:surf}). \map
    s that are \aenhanced\ have small typical mean radii and measure
    \sigoneone\ at small radii, while \map s that are \apoor\ have
    mean radii close to $R_0$ and measure the surface density at
    $\sim\!7$ to $8\kpc$. The two radii and \afe\ are all strongly
    correlated, as shown by this figure.}\label{fig:rbest_vs_rmean}
\end{figure}

To test whether this expectation is borne out in practice, and
therefore, whether the procedure of determining the radius at which
\sigoneone\ is best measured makes sense, we have calculated the
median of the mean orbital radii of the data in each \map. We use a
simplified model for the Milky Way's gravitational potential that is
the same as that used in Sec. 5.4 and Figure 7 of \bo d, but the
details of the mass model for the Milky Way used are unimportant for
the calculation of the mean orbital radii. As we are interested in the
vertical dynamics, we time-average over the orbits weighting by the
midplane density ($=$ vertical oscillation frequency squared). These
median mean-orbital-radii for each \map\ are plotted against the radii
with the most robust \sigoneone\ determination in
\figurename~\ref{fig:rbest_vs_rmean}, color-coded by \afe.

It is clear from \figurename~\ref{fig:rbest_vs_rmean} that there is a
strong correlation between the radii calculated in these two entirely
different manners (apart from a single outlier in the lower-right
corner). Therefore, we can be confident that the procedure for
calculating the radius at which each \map\ best measures
\sigoneone\ provides a good definition for this radius. The
color-coding confirms the expectation that \aenhanced\ \map s
constrain $\sigoneone$ at radii much smaller than $R_0$, while
\apoor\ \map s measure \sigoneone\ at positions close to the
Sun's. \map s that are intermediate between these two measure
\sigoneone\ at intermediate radii. One of the more \apoor\ \map s in
\figurename~\ref{fig:rbest_vs_rmean} falls far from the relation
defined by the other \map s (best \sigoneone\ radius of $4.5\kpc$ for
a mean orbital radius of $7.5\kpc$). However, this single \map\ does
not influence any of the results obtained below. Below, we quote
results using the best radius as determined from the PDF; if we
instead require the measurements of the vertical force to be at the
median of the mean orbital radii of the stars in each \map, the
best-fit surface-density and vertical-force profiles are unchanged,
albeit with slightly increased uncertainties on the scale lengths.

\subsection{The surface density at 1.1\,kpc between 4\,kpc and 9\,kpc as measured by \map s}\label{sec:surfresults}

\begin{figure}[t!]
  \includegraphics[width=0.5\textwidth,clip=]{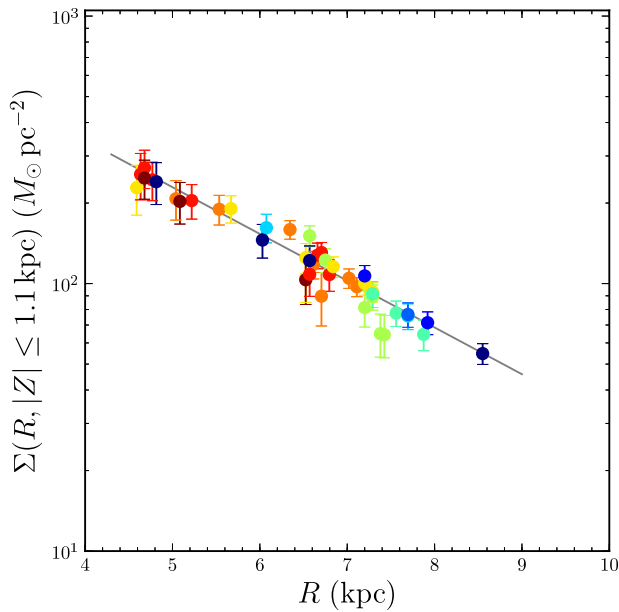}
  \caption{Surface density as a function of radius as measured by
    SEGUE G-dwarf \map s. Each point represents a measurement based on
    the data from 1 \map; each \map\ measures the surface density
    $\Sigma(R,|Z|\leq 1.1\,\mathrm{kpc})$ at the radius $R$ where the
    correlation between the model stellar disk scale length and
    $\Sigma(R_0,|Z|\leq 1.1\,\mathrm{kpc})$ is minimized in the \map
    's PDF (see \figurenames~\ref{fig:pdfs} and
    \ref{fig:illustraterbest}). The gray curve is an exponential fit
    with $\Sigma(R_0,|Z|\leq 1.1\,\mathrm{kpc}) =
    69\,M_\odot\,\mathrm{pc}^{-2}$ and a scale length of
    $2.5\kpc$. Points are color-coded by the \afe\ of the \map\ bin,
    from $\afe = 0.025$ (dark blue) to $\afe = 0.475$
    (red/brown).}\label{fig:surf}
\end{figure}

Even though the data from the different \map s were sampled by the
same survey centered on the solar radius (see Figure 7 in
\citealt{Rix13a}), they constrain $\sigoneone(R)$ at quite different
radii. As a consequence, we can measure the Galactic disk's surface
density profile. For each \map\ we calculate the mean $\sigoneone(R)$
and its uncertainty $\delta \sigoneone(R)$ at the best radius $R$ from
moments of the PDF of the dynamical parameters. Thus, we condense each
\map's dynamical information into the best single measurement of the
Milky Way's mass distribution for that \map. The surface densities
thus measured are shown in \figurename~\ref{fig:surf} and tabulated in
\tablename~\ref{table:surf}.

\figurename~\ref{fig:surf} shows that the \map s provide a highly
precise measurement of $\sigoneone(R)$ between $4.5\kpc$ and $9\kpc$
and that the results from all 43 \map s are in excellent agreement
with each other. From a purely phenomenological perspective this
measurement of $\sigoneone(R)$ can be well-characterized by a single
exponential
\begin{equation}\label{eq:bestfitsig}
  \sigoneone(R) = 69\,M_\odot\,\mathrm{pc}^{-2}\,\exp\left(-(R-R_0)/ 2.5\kpc\right)\,,
\end{equation}
with formal uncertainties of $\sim\!2.5\,\%$ in the normalization and
$4\,\%$ in the scale length. However, we have no reason to expect that
the real physical $\sigoneone(R)$ is a single exponential, as the mass
distribution is made up of various disk components and the halo. We
discuss this further in \sectionname~\ref{sec:potentialfit} below,
where we fit mass models.

\begin{figure}[t!]
  \includegraphics[width=0.5\textwidth,clip=]{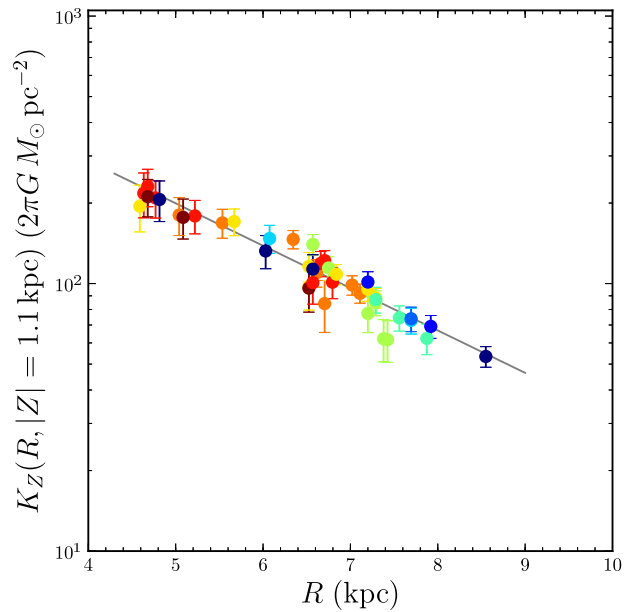}
  \caption{The vertical force at $1.1\kpc$ as a function of radius as
    measured by SEGUE G dwarf \map s. This Figure represents the same
    measurements as in \figurename~\ref{fig:surf}, but expressed as
    vertical forces. The gray curve is an
    exponential fit with $K_Z(R_0,|Z| = 1.1\,\mathrm{kpc}) / 2\pi G=
    67\,M_\odot\,\mathrm{pc}^{-2}$ and a scale length of
    $2.7\kpc$.}\label{fig:kz}
\end{figure}

An alternative way of presenting our results is as measurements of the
vertical force at $1.1\kpc$, which we denote by $\kzoneone$,
instead. These measurements are shown in \figurename~\ref{fig:kz} and
are also tabulated in \tablename~\ref{table:surf}. As with the
measurements of $\sigoneone(R)$, the radial dependence of the vertical
force is well-fit by an exponential as
\begin{equation}\label{eq:bestfitkz}
  \frac{\kzoneone(R)}{2\pi G} = 67\,M_\odot\,\mathrm{pc}^{-2}\,\exp\left(-(R-R_0)/ 2.7\kpc\right)\,,
\end{equation}
again with formal uncertainties of $\sim\!2.5\,\%$ in the
normalization and $4\,\%$ in the scale length. As expected based on
the vertical Jeans equation and shown explicitly in the following
Section, our \kzoneone\ measurements are more robust with respect to
changes in the assumptions about the derivative of the circular
velocity (\ie, the ``slope of the rotation curve''). Therefore, when
using our measurements to constrain dynamical models of the Milky Way,
it may be preferable to use the $\kzoneone(R)$ measurements rather
than the $\sigoneone$, especially when considering models where the
rotation curve is significantly non-flat. This is the approach we
follow in \sectionname~\ref{sec:potentialfit} below.

We have assumed $R_0 = 8\kpc$ throughout the analysis. The effect of
changing $R_0$ is to shift our measurements of $\sigoneone(R)$ and
$\kzoneone(R)$ such as to keep $R-R_0$ constant (\ie, our measurements
are at a fixed distance from $R_0$; see also
\sectionname~\ref{sec:systematics}). Therefore, our measurements can
be shifted to a different value of $R_0$ by keeping $R-R_0$ (see
\tablename~\ref{table:surf}) constant.

The measured $\sigoneone(R)$ or $\kzoneone(R)$ in \figurename
s~\ref{fig:surf} and \ref{fig:kz} and tabulated in
\tablename~\ref{table:surf} are 43 new, independent constraints on the
Milky Way's mass distribution and gravitational potential. They can be
combined with other constraints, such as measurements of the vertical
mass distribution as in \citet{BovyTremaine} and \citet{Zhang13a}, and
constraints on the rotation curve from the terminal velocity curve or
based on stellar tracers. This is the approach that we take in
\sectionname~\ref{sec:potentialfit}, where we show that the
\map\ measurements of $\sigoneone(R)$ quite directly constrain the
disk scale length and through combination with data on the rotation
curve can be used to disentangle the disk- and halo contributions to
the mass distribution in the inner Milky Way.

\begin{deluxetable*}{rrrrrrrr}
\tablecaption{}
\tablecolumns{6}
\tablewidth{0pt}
\tabletypesize{\footnotesize}
\tablecaption{Measured surface density and vertical force at different Galactocentric radii}
\tablehead{\colhead{\feh} & \colhead{\afe} & \colhead{$R$} & \colhead{$\Sigma_{1.1}(R)$} & \colhead{$\delta \Sigma_{1.1}(R)$} & \colhead{$R_0-R$} & \colhead{$K_{Z,1.1}(R)$} & \colhead{$\delta K_{Z,1.1}(R)$}\\
  \colhead{(dex)} & \colhead{(dex)} & \colhead{(kpc)} & \colhead{($M_\odot\,\mathrm{pc}^{-2}$)} & \colhead{($M_\odot\,\mathrm{pc}^{-2}$)} & \colhead{(kpc)} & \colhead{($2\pi G\,M_\odot\,\mathrm{pc}^{-2}$)} & \colhead{($2\pi G\,M_\odot\,\mathrm{pc}^{-2}$)}}

\startdata
-1.25 & 0.425 & 4.63 & 256.0 & 50.4 & 3.37 & 217.6 & 41.5\\
-1.15 & 0.425 & 4.68 & 270.9 & 44.3 & 3.32 & 230.6 & 36.8\\
-1.05 & 0.375 & 6.71 & 89.7 & 20.3 & 1.29 & 84.2 & 18.5\\
-1.05 & 0.425 & 4.77 & 244.2 & 40.1 & 3.23 & 209.0 & 33.2\\
-0.95 & 0.325 & 4.59 & 228.4 & 48.3 & 3.41 & 194.7 & 38.6\\
-0.95 & 0.375 & 5.04 & 207.6 & 34.9 & 2.96 & 180.5 & 29.2\\
-0.95 & 0.425 & 5.22 & 204.3 & 30.1 & 2.78 & 179.0 & 25.6\\
-0.95 & 0.475 & 4.68 & 247.9 & 41.3 & 3.32 & 211.4 & 33.7\\
-0.85 & 0.275 & 7.38 & 65.0 & 11.9 & 0.62 & 62.2 & 11.2\\
-0.85 & 0.325 & 6.53 & 104.9 & 19.8 & 1.47 & 97.5 & 18.0\\
-0.85 & 0.375 & 6.62 & 118.1 & 14.0 & 1.38 & 109.8 & 12.9\\
-0.85 & 0.425 & 6.66 & 127.6 & 14.0 & 1.34 & 119.0 & 12.9\\
-0.85 & 0.475 & 5.08 & 202.8 & 35.9 & 2.92 & 176.8 & 30.1\\
-0.75 & 0.275 & 7.42 & 64.4 & 11.6 & 0.58 & 61.7 & 11.0\\
-0.75 & 0.325 & 6.53 & 125.0 & 16.2 & 1.47 & 115.9 & 14.8\\
-0.75 & 0.375 & 7.11 & 97.3 & 8.0 & 0.89 & 92.0 & 7.5\\
-0.75 & 0.425 & 6.71 & 130.6 & 11.8 & 1.29 & 121.9 & 10.9\\
-0.75 & 0.475 & 6.53 & 103.3 & 19.8 & 1.47 & 96.1 & 18.0\\
-0.65 & 0.275 & 7.20 & 81.4 & 12.6 & 0.80 & 77.3 & 11.8\\
-0.65 & 0.325 & 7.20 & 100.2 & 9.9 & 0.80 & 95.2 & 9.3\\
-0.65 & 0.375 & 6.34 & 159.2 & 12.7 & 1.66 & 146.3 & 11.5\\
-0.65 & 0.425 & 6.79 & 108.0 & 14.5 & 1.21 & 101.2 & 13.4\\
-0.55 & 0.275 & 7.29 & 89.2 & 9.6 & 0.71 & 84.9 & 9.1\\
-0.55 & 0.325 & 7.29 & 93.1 & 7.8 & 0.71 & 88.5 & 7.4\\
-0.55 & 0.375 & 7.02 & 104.7 & 8.7 & 0.98 & 98.7 & 8.2\\
-0.55 & 0.425 & 6.57 & 108.3 & 18.7 & 1.43 & 100.8 & 17.1\\
-0.45 & 0.225 & 7.56 & 77.6 & 8.5 & 0.44 & 74.5 & 8.1\\
-0.45 & 0.275 & 6.75 & 122.4 & 12.6 & 1.25 & 114.3 & 11.7\\
-0.45 & 0.325 & 6.84 & 115.7 & 10.1 & 1.16 & 108.4 & 9.4\\
-0.45 & 0.375 & 5.54 & 189.6 & 24.1 & 2.46 & 168.6 & 20.9\\
-0.35 & 0.225 & 7.29 & 91.5 & 10.2 & 0.71 & 87.2 & 9.6\\
-0.35 & 0.275 & 6.57 & 150.9 & 13.5 & 1.43 & 140.1 & 12.4\\
-0.35 & 0.325 & 5.67 & 190.6 & 22.3 & 2.33 & 170.6 & 19.5\\
-0.25 & 0.175 & 7.70 & 75.6 & 8.4 & 0.30 & 72.8 & 8.0\\
-0.25 & 0.225 & 7.88 & 64.6 & 8.5 & 0.12 & 62.4 & 8.2\\
-0.15 & 0.125 & 7.70 & 76.7 & 8.1 & 0.30 & 73.9 & 7.8\\
-0.15 & 0.175 & 6.08 & 161.8 & 19.8 & 1.92 & 147.4 & 17.7\\
-0.05 & 0.025 & 6.57 & 121.9 & 16.4 & 1.43 & 113.2 & 14.9\\
-0.05 & 0.075 & 7.92 & 71.4 & 7.0 & 0.08 & 69.2 & 6.8\\
0.05 & 0.025 & 8.55 & 54.7 & 4.9 & -0.55 & 53.4 & 4.7\\
0.05 & 0.075 & 7.20 & 106.8 & 10.0 & 0.80 & 101.3 & 9.4\\
0.15 & 0.025 & 6.03 & 145.4 & 20.9 & 1.97 & 132.4 & 18.8\\
0.25 & 0.025 & 4.82 & 240.3 & 42.9 & 3.18 & 206.2 & 35.7\\
\enddata

\tablecomments{\protect{E}ach row gives the measurement of the surface
  density \sigoneone\ up to $|Z| = 1.1\,\mathrm{kpc}$ along with its
  uncertainty $\delta \sigoneone$ obtained from a single \map,
  specified by its central \feh\ and \afe. The radius is that where
  the correlation between the inferred surface density and the
  potential model's disk scale length is minimal; this is the radius
  at which the surface density is best measured by a \map\ (see
  \figurename~\ref{fig:illustraterbest}). The last two columns give
  the alternative measurement of the vertical force $K_{Z,1.1}$ at
  $1.1\kpc$ and its uncertainty $\delta \kzoneone$. The sixth column
  gives the difference between the radius at which \sigoneone\ or
  \kzoneone\ is measured and $R_0$ (to be held constant when using our
  measurements with a different value of $R_0$).}\label{table:surf}

\end{deluxetable*}

\subsection{Variations on the fiducial model}\label{sec:systematics}

The results in the previous Section are obtained in the fiducial model
in which we use the \an\ distance scale for the G-type SDSS dwarfs and
we fix the parameters of the gravitational potential model such that
$V_c(R_0) = 230\kms$, $z_h = 400\pc$, and $\dd \ln V_c(R_0) / \dd \ln
R = 0$. In this Section we determine the impact of these choices and
find them to be small and largely inconsequential for the measurement
of $\sigoneone(R)$ and especially $\kzoneone(R)$, and for the
dynamical analysis in \sectionname~\ref{sec:potentialfit} below.

We assess the impact of systematic distance uncertainties by
performing the analysis of the \map\ dynamics when using the distances
of \iv\ instead of the \an\ distances in our
fiducial model. As discussed in \sectionname~\ref{sec:data}, the
\iv\ distances are typically $7\,\%$ larger than the
\an\ distances to G-type dwarfs, with the detailed relative
distance factors given in \figurename~\ref{fig:andistance}. As
discussed in \bo d, the impact of unresolved binaries is such that
distances would be $\sim\!6\,\%$ larger, so the difference between the
results obtained using \iv\ distances and those derived
from \an\ is a good proxy for understanding the impact of
typical systematic distance uncertainties affecting the data.

We expect the impact of systematic distance uncertainties to be such
that $\sigoneone^{\mathrm{Ivezic}} / \sigoneone^{\mathrm{An}} =
(d^\mathrm{Ivezic} / d^{\mathrm{An}})^\alpha$, with $-1 \leq \alpha
\leq 1$ and similar for \kzoneone. This is because roughly $\sigoneone
\propto \sigma_Z^2 / h_Z$, such that if the measurement is dominated
by proper motions, such as when one observes the vertical motions of
disk stars far from the Sun, $\sigoneone \propto d$. Similarly, if the
measurement primarily depends on the line-of-sight velocities, which
is the case when measuring \sigoneone\ at $R_0$ by looking at the
motions of stars near the Galactic poles, $\sigoneone \propto
1/d$. Because the measurement of \sigoneone\ has contributions from
both proper motions and line-of-sight velocities, we expect
\sigoneone\ to scale with distance in a manner that is in between
these two extremes. We expect the scaling to be different for
different \map s, as \map s that measure \sigoneone\ at smaller $R$
rely more on proper motions than \map s that measure
\sigoneone\ closer to the solar circle.

\begin{figure}[t!]
  \includegraphics[width=0.5\textwidth,clip=]{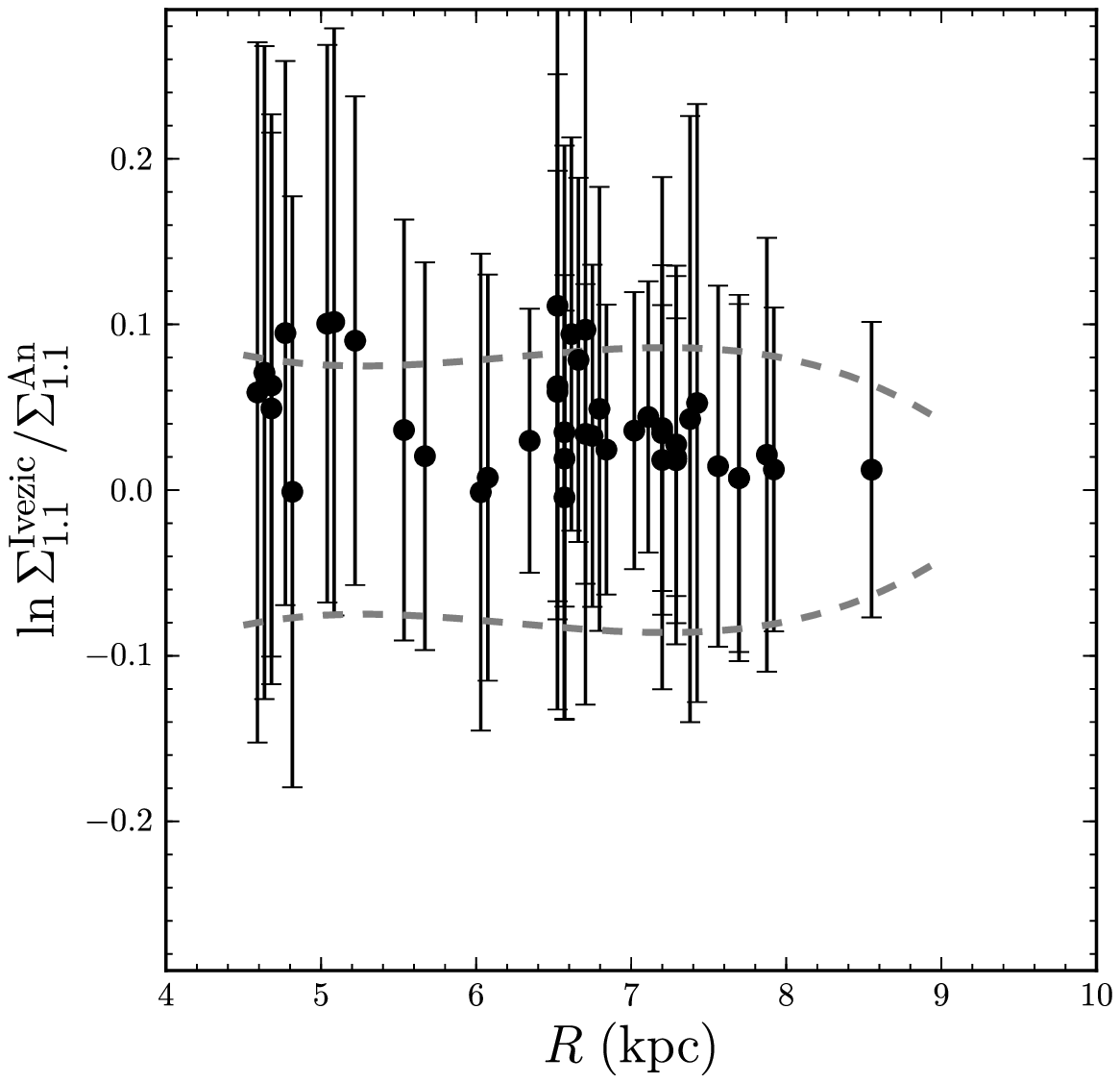}
  \caption{Impact of distance systematics: This Figure compares the
    measured \sigoneone\ for each \map\ when using distances on the
    \an\ scale with those when using the \iv\ distances. The
    differences are only a few percent for most \map s that measure
    \sigoneone\ at $6\kpc \lesssim R \lesssim 9\kpc$, while the
    differences are almost 10\,\% for \map s that measure \sigoneone\
    at $R \leq 5.5\kpc$. The dashed lines show the approximate locus
    where the impact of using the \iv\ distances results in a linear
    increase in \sigoneone\ ($\sigoneone^{\mathrm{Ivezic}} /
    \sigoneone^{\mathrm{An}} = (d^\mathrm{Ivezic} / d^{\mathrm{An}})$),
    as expected when the measurement is dominated by proper motions
    or in a linear decrease in \sigoneone\
    ($\sigoneone^{\mathrm{Ivezic}} / \sigoneone^{\mathrm{An}} =
    (d^\mathrm{Ivezic} / d^{\mathrm{An}})^{-1}$), as expected when the
    measurement is dominated by line-of-sight velocities. Overall,
    the impact of systematic distance uncertainties is much smaller
    than the statistical uncertainties, and the best-fitting
    exponential approximation to $\sigoneone(R)$ changes only by
    $2\,\%$ in the normalization and scale length (see
    text).}\label{fig:systematics_dist}
\end{figure}

We show the comparison between the measured \sigoneone\ obtained using
the \iv\ distances with those derived from the \an\ distances in
\figurename~\ref{fig:systematics_dist}. The difference in the measured
\sigoneone\ is shown for each \map\ at the radius at which the
\map\ best measures \sigoneone\ determined for the fiducial model. The
dashed lines show the approximate locus of the linear and inverse
scalings with distance discussed in the previous paragraph. This
Figure shows that the impact of the systematic distance shift between
the two distance scales is small for the majority of the \map s. \map
s that measure \sigoneone\ at $R < 5.5\kpc$ rely more on proper
motions and their measured \sigoneone\ therefore scales close to
linear with the distance scale. Most of the \map s that measure
\sigoneone\ at $R > 5.5\kpc$ are only affected at the few percent
level by the uncertainty in the distance scale. $\kzoneone$ is
affected in the same way as \sigoneone.

\begin{figure*}[t!]
  \begin{center}
  \includegraphics[width=0.7\textwidth,clip=]{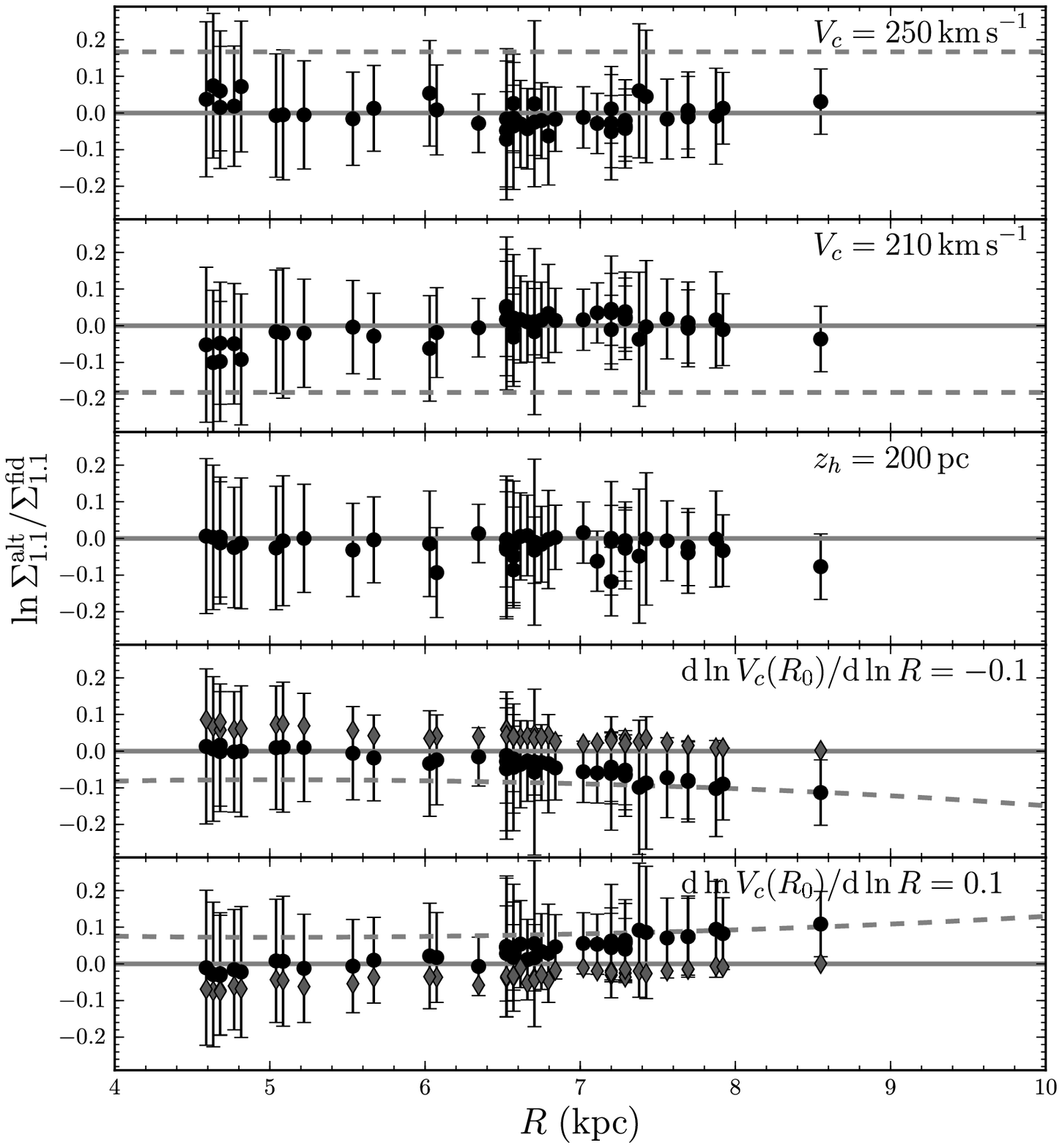}
  \caption{Influence of the most important fixed parameters of the
    fiducial potential model. This Figure shows the (logarithmic)
    difference between the surface-densities
    $\sigoneone^{\mathrm{fid}}$ inferred with the fiducial model for
    the potential ($V_c = 230\kms$, $z_h = 400\pc$, and $\dd \ln
    V_c(R_0) / \dd \ln R = 0$) and those inferred with a model where
    one of these three parameters is changed.  The top two panels show
    the influence of the normalization of the potential $V_c$, which
    is negligible and much smaller than the naive expectation that
    $\sigoneone^{\mathrm{alt}} / \sigoneone^{\mathrm{fid}} =
    (V_c^{\mathrm{alt}} / V_c^{\mathrm{fid}})^2$ if the surface
    densities simply scaled with $V_c^2$ (shown by the dashed
    line). The middle panel shows the outcome of changing the assumed
    disk scale height $z_h$. The bottom two panels explore the effect
    of changing the shape of the rotation curve (parameterized using
    the local logarithmic slope $\dd \ln V_c(R_0) / \dd \ln R$). The
    only significant systematic in measuring \sigoneone\ is that
    related to changing the shape of the rotation curve, which for
    measurements near $R_0$ is close to the naive expectation that the
    effect of changing the slope of the rotation curve is the addition
    of $|Z|\,(\dd V^2_c / \dd R)/2\pi G R$ (dashed curves). However, the
    change in \kzoneone, indicated by the gray diamond symbols, is
    still small (the change in \kzoneone\ is the same as that for
    \sigoneone\ for the upper three panels). Thus, the measurement of
    $\kzoneone(R)$ is unaffected by systematics, while the
    $\sigoneone(R)$ measurement is robust if the rotation curve is
    locally close to flat.}\label{fig:systematics1}
  \end{center}
\end{figure*}

The overall impact of the change in the distance scale between that of
\iv\ and \an\ on $\sigoneone(R)$ is small. It is clear from
\figurename~\ref{fig:systematics_dist} that the effect of using the
larger distances is to slightly steepen the $\sigoneone(R)$
profile. The best-fit exponential approximations to $\sigoneone(R)$
and $\kzoneone(R)$ obtained using the \iv\ scale, using the optimal
radii derived from the PDFs for the \iv\ scale are
\begin{align}
  \sigoneone(R) & = 70\,M_\odot\,\mathrm{pc}^{-2}\,\exp\left(-(R-R_0)/ 2.5\kpc\right)\nonumber\\
  & \qquad \qquad \qquad \mathrm{(\iv\ distance\ scale)}\,,\\
  \frac{\kzoneone(R)}{2\pi G} & = 68\,M_\odot\,\mathrm{pc}^{-2}\,\exp\left(-(R-R_0)/ 2.7\kpc\right)\nonumber\\
  & \qquad \qquad \qquad \mathrm{(\iv\ distance\ scale)}\,,
\end{align}
which is very close and within the statistical uncertainties of the
result obtained using the \an\ scale
(\equationname~[\ref{eq:bestfitsig}]). Therefore, the impact of
systematic distance uncertainties on the measurement of
$\sigoneone(R)$ and $\kzoneone(R)$ is insignificant, at the level of
$\approx2\%$.

To determine the impact of the rather constrained fiducial model for
the gravitational potential, we repeat the measurement of
$\sigoneone(R)$ for different choices for the most important fixed
parameters of the potential. Foremost among these is the normalization
of the potential, characterized by the local circular velocity in our
parameterization. We have fixed this to $V_c(R_0) = 230\kms$ in the
fiducial model. In the top panel of \figurename~\ref{fig:systematics1}
we compare the results when using $V_c(R_0) = 250\kms$ to those
obtained for the fiducial model, again at the radius determined using
the fiducial model. We see that the impact of changing the potential
normalization is close to zero for almost all \map s and in all cases
the shift is much smaller than the random uncertainties. The dashed
line in this figure shows the expected difference if $\sigoneone
\propto V_c^2$, the expected scaling if the measured surface-density
were wholly dependent on the normalization of the potential (or
equivalently, of the rotation curve), that is, if we were not really
measuring \sigoneone\ or \kzoneone.  $\kzoneone(R)$ behaves the same
when changing $V_c(R_0)$.  Similarly, in the second panel we compare
the results obtained using $V_c(R_0) = 210\kms$ to those from the
fiducial model; these two sets of results are again nearly
indistinguishable and the difference is much smaller than the
$\sigoneone \propto V_c^2$ expectation (dashed line). Note in this
case that the measured value for $V_c(R_0) = 230\kms$ for the bins
that measure \sigoneone\ at $R \approx5\kpc$ is at the edge of the
prior range for $V_c(R_0) = 210\kms$, which artificially lowers the
inferred \sigoneone\ for these \map s (this is clear from an
inspection of the PDFs). The best-fitting exponential approximations
to $\sigoneone(R)$ and $\kzoneone(R)$ measured from the $V_c(R_0) =
250\kms$ results are
\begin{align}
  \sigoneone(R) & = 72\,M_\odot\,\mathrm{pc}^{-2}\,\exp\left(-(R-R_0)/ 2.5\kpc\right)\nonumber\\
  & \qquad \qquad \qquad (V_c(R_0) = 250\kms)\,,\\
  \frac{\kzoneone(R)}{2\pi G} & = 69\,M_\odot\,\mathrm{pc}^{-2}\,\exp\left(-(R-R_0)/ 2.7\kpc\right)\nonumber\\ & \qquad \qquad \qquad (V_c(R_0) = 250\kms)\,,
\end{align}
 and those from the $V_c(R_0) =210\kms$ results are
\begin{align}
  \sigoneone(R) & = 65\,M_\odot\,\mathrm{pc}^{-2}\,\exp\left(-(R-R_0)/ 2.5\kpc\right)\nonumber\\& \qquad \qquad \qquad (V_c(R_0) = 210\kms)\,,\\
  \frac{\kzoneone(R)}{2\pi G} & = 63\,M_\odot\,\mathrm{pc}^{-2}\,\exp\left(-(R-R_0)/ 2.8\kpc\right)\nonumber\\
  & \qquad \qquad \qquad (V_c(R_0) = 210\kms)\,,
\end{align}
Therefore, the scale length of $\sigoneone(R)$ hardly changes and the
normalization is only affected by a few percent, which is similar to
the statistical uncertainties in the fit.

We have also performed an extreme test where we set $V_c(R_0) =
280\kms$, far beyond any reasonable setting for this
parameter. Setting $V_c(R_0)$ to $280\kms$ shifts the prior on
$\sigoneone(R)$ in \figurename~\ref{fig:prior-surface-range} upward
such that many of the best-fit \sigoneone\ in
\figurename~\ref{fig:surf} lie on the edge or slightly outside the
prior range. We find that the measured $\sigoneone$ for $V_c(R_0)$ are
all on the lower edge of the prior, such that they clearly indicate
that \sigoneone\ is in reality lower than what is allowed by the $V_c
= 280\kms$ prior. Even in this extreme case (which should not be
trusted, since all measured \sigoneone\ are on the edge of the prior),
the scale length of the best-fitting exponential is $2.5\kpc$ and the
normalization only changes to $76\,M_\odot\,\pc^{-2}$. This change in
normalization is only $10\,\%$, while $V_c^2$ has changed by
$50\,\%$. This and the results for $V_c = 210\kms$ and $V_c = 250\kms$
confirm that we are primarily measuring \sigoneone\ (or \kzoneone)
such that the assumed normalization of the potential is unimportant as
long as it permits the inclusion of the actual \sigoneone.

In the fiducial model, the scale height of the stellar disk is fixed
to $z_h = 400\pc$, which is the mass-weighted scale height inferred
from the \map\ decomposition of \bo\ and agrees with dynamical
measurements of the disk scale height based on $\Sigma(R_0,Z)$
\citep{Siebert03a,Zhang13a}. However, the mass-weighted scale height
of the disk is still quite uncertain. Because we are primarily
measuring \sigoneone\ using the vertical kinematics, we do not expect
our results to depend on the disk scale height, which for any
reasonable value changes the vertical distribution of disk matter
below approximately $1\kpc$, but not above it (as at most a few disk
$M_\odot\,\pc^{-2}$ are above $1\kpc$). We have repeated the analysis
above using a scale height of $200\pc$, which is at the low end of
plausible values of this parameter. The middle panel of
\figurename~\ref{fig:systematics1} demonstrates that the inferred
$\sigoneone$ for all \map s barely change when using a different
stellar-disk scale height, and all changes are within the random
uncertainties of the measurement. The inferred $\sigoneone(R)$ and
$\kzoneone(R)$ are almost exactly the same as those measured using the
fiducial potential model:
\begin{align}
  \sigoneone(R) & = 70\,M_\odot\,\mathrm{pc}^{-2}\,\exp\left(-(R-R_0)/ 2.6\kpc\right)\nonumber\\& \qquad \qquad \qquad \mathrm{(scale\ height}\ = 200\pc)\,,
\end{align}
and
\begin{align}
  \frac{\kzoneone(R)}{2\pi G} & = 67\,M_\odot\,\mathrm{pc}^{-2}\,\exp\left(-(R-R_0)/ 2.7\kpc\right)\nonumber\\ & \qquad \qquad \qquad \mathrm{(scale\ height}\ = 200\pc)\,.
\end{align}

The final important parameter of the gravitational-potential model is
the slope of the rotation curve, set to $\dd \ln V_c(R_0) / \dd \ln R
= 0$ in the fiducial model. From the vertical Poisson equation it is
clear that the slope of the rotation curve changes the surface density
measured from vertical kinematics by adding approximately $|Z|\,(\dd
V^2_c / \dd R)/2\pi G R$ \citep{Kuijken89a,BovyTremaine}. Locally this
amounts to a systematic uncertainty of 
\[\approx 7\,M_\odot\pc^{-2}
(|Z| / 1.1\kpc) ([\dd \ln V_c / \dd \ln R]/0.1)\]\[\qquad \quad (V_c / 230\kms)^2(R_0 /
8\kpc)^{-2}\]
 given the current uncertainty in the slope of the
rotation curve (see below). However, we expect the vertical force
$K_Z$ to be unaffected by this, because the vertical Jeans equation
shows that $K_Z$ is constrained by the vertical dynamics only and that
it does not depend on the slope of the rotation curve.

The bottom two panels of \figurename~\ref{fig:systematics1} show the
change in the measured \sigoneone\ and \kzoneone\ (gray diamond
symbols) when changing the assumed slope of the rotation curve. This
changes the slope of the model rotation curve at all $R$ and the
dashed lines show the expectation from the simple calculation given in
the previous paragraph assuming that the change is such that $\dd \ln
V_c / \dd \ln R = +/- 0.1$ at all $R$. It is clear that the
measurements of \sigoneone\ are strongly and systematically affected
by the change in the slope of the rotation curve, especially near
$R_0$. The inferred $\sigoneone(R)$ profile changes by about twice the
random uncertainties:
\begin{align}
  \sigoneone(R) & = 65\,M_\odot\,\mathrm{pc}^{-2}\,\exp\left(-(R-R_0)/ 2.3\kpc\right)\nonumber\\& \qquad \qquad \qquad (\dd \ln V_c(R_0) / \dd \ln R= -0.1)\,,\\
  \sigoneone(R) & = 72\,M_\odot\,\mathrm{pc}^{-2}\,\exp\left(-(R-R_0)/ 2.7\kpc\right)\nonumber\\ & \qquad \qquad \qquad (\dd \ln V_c(R_0) / \dd \ln R= 0.1)\,.
\end{align}
However, the measured $\kzoneone$ is much less affected by the change
in the slope of the rotation curve and the change in \kzoneone\ is
well within the random uncertainties for all \map s. The inferred
$\kzoneone(R)$ profile changes only by about $1\sigma$:
\begin{align}
  \frac{\kzoneone(R)}{2\pi G} & = 69\,M_\odot\,\mathrm{pc}^{-2}\,\exp\left(-(R-R_0)/ 2.6\kpc\right)\nonumber\\& \qquad \qquad \qquad (\dd \ln V_c(R_0) / \dd \ln R = -0.1)\,,\\
  \frac{\kzoneone(R)}{2\pi G} & = 63\,M_\odot\,\mathrm{pc}^{-2}\,\exp\left(-(R-R_0)/ 2.8\kpc\right)\nonumber\\& \qquad \qquad \qquad (\dd \ln V_c(R_0) / \dd \ln R = 0.1)\,.
\end{align}
Therefore, the inferred \kzoneone\ are more robust to changes in the
assumed rotation curve. When using our measurements in analyses where
the rotation curve is varied or assumed to be very different from the
flat rotation curve in our fiducial model, we recommend using the
\kzoneone\ measurements instead of the \sigoneone\ measurements.

We have also investigated the effect of changing the value of
$R_0$. One would expect that our measurements, which are based on a
volume centered on the Sun, are at constant distances from the assumed
$R_0$. Therefore, we should find that a measurement of $\sigoneone$ or
$\kzoneone$ at $R$ for $R_0 = 8\kpc$ is at $R + \Delta R$, where
$\Delta R = R_0-8\kpc$, for a different value of $R_0$. We have
performed the full $\sigoneone$ measurement for the two \map s in
\figurename~\ref{fig:pdfs} using $R_0 = 8.5\kpc$ and have found that
the inferred $\sigoneone$ and $\kzoneone$ is indeed shifted by $R_0 -
8\kpc = 0.5\kpc$ for these \map s, such that their measured
\sigoneone\ and \kzoneone\ are the same as for the fiducial $R_0$, but
at Galactocentric radii that are $0.5\kpc$ larger. We are therefore
confident that this holds for the measurements from all of the \map
s. \tablename~\ref{table:surf} contains a column that lists the radius
$R_0-R$ at which \sigoneone\ and \kzoneone\ is measured; this value
should be kept constant when changing the assumed value of $R_0$.

The effect of changing $R_0$ is only to shift the measured
$\sigoneone(R)$ and $\kzoneone(R)$ profiles without changing their
(logarithmic) slopes. Below, we are primarily interested in
determining the mass scale length of the stellar disk and this
measurement is mainly dependent on the logarithmic slope of the
surface density profile. As this slope does not depend on the assumed
value of $R_0$, we do not vary $R_0$ in the analysis below.

\section{Measurement of the disk scale length and constraints on the mass profile in the inner Milky Way}\label{sec:potentialfit}

The measurement of the vertical force profile between $4.5\kpc$ and
$9\kpc$ presented in the previous Section provides a strong new
constraint on mass models for the inner Milky Way. We explore these
constraints in this Section. We find that the stellar disk scale
length is largely constrained by the measurements of $\kzoneone(R)$,
but we also consider additional data on the rotation curve and the
local vertical mass distribution to separate the disk and halo
contributions to the total mass. These additional data are described
in \sectionname~\ref{sec:adddata}. The results of mass-model fits to
these data are presented in \sectionname~\ref{sec:fitresults}.

\subsection{Additional data}\label{sec:adddata}

\begin{figure}[t!]
  \includegraphics[width=0.475\textwidth,clip=]{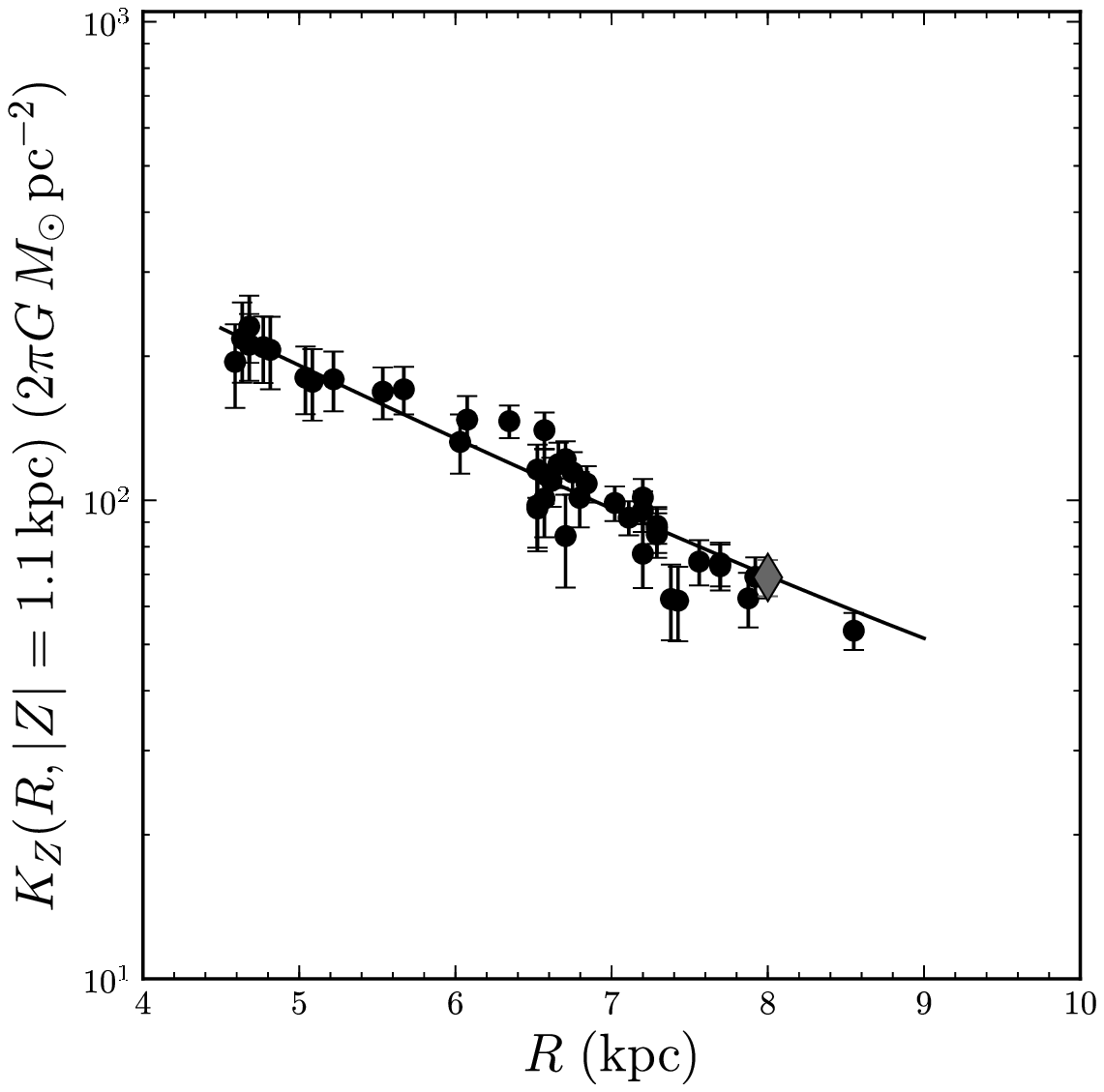}\\
  \includegraphics[width=0.475\textwidth,clip=]{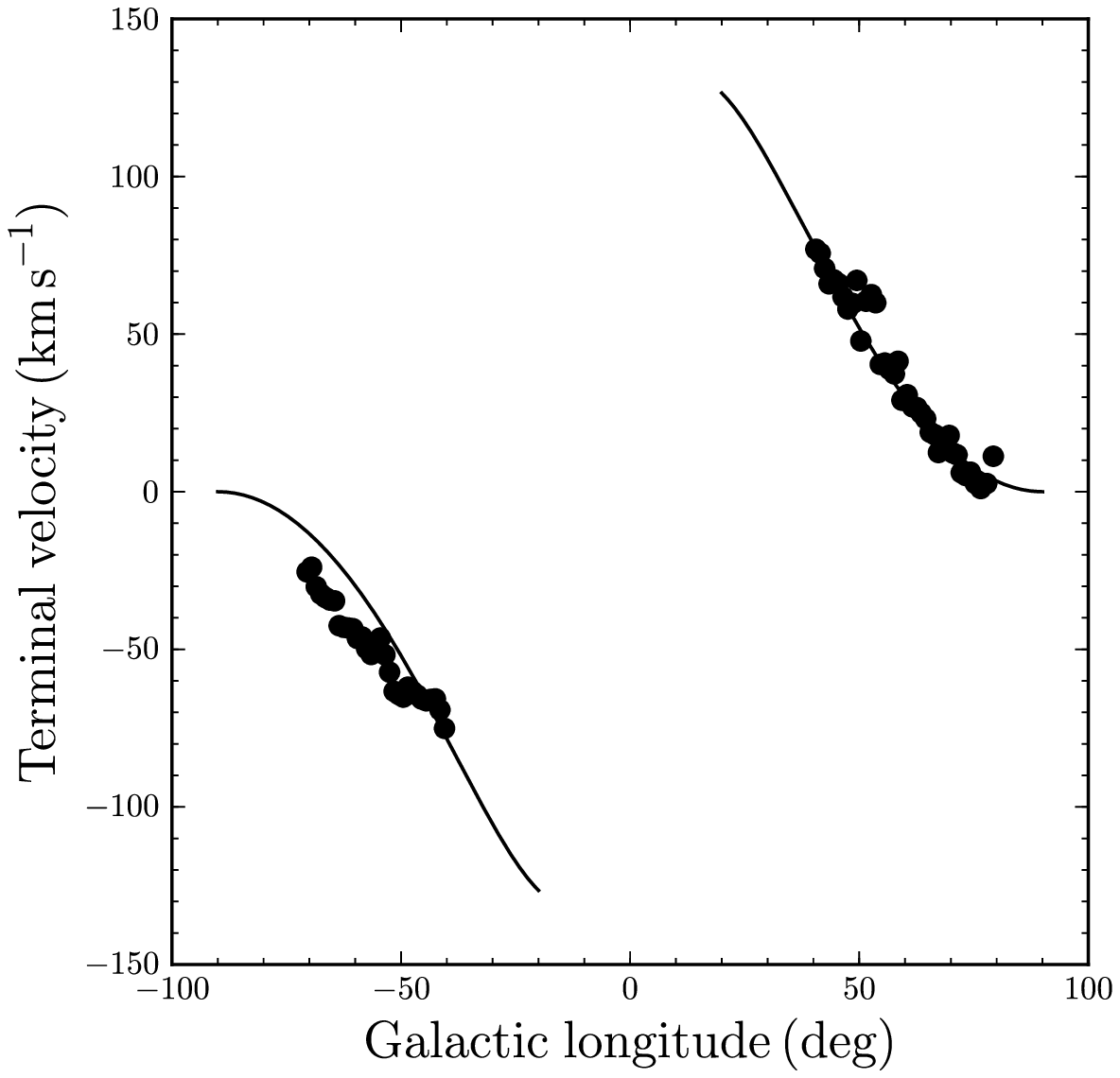}
  \caption{The upper panel shows $\kzoneone(R)$ of the best-fit mass
    model for the inner Milky Way when fitting the $\kzoneone(R)$
    measurements from this paper, the terminal velocity data, the
    measurement of the slope of the rotation curve from
    \citet{BovyVc}, and the measurement of the local contribution from
    the stellar disk from \citet{Zhang13a}. The black dots are the
    \kzoneone\ measurements from this paper (\figurename~\ref{fig:kz}
    and \tablename~\ref{table:surf}) and the measurement from
    \citet{Zhang13a} is indicated by a gray diamond. The lower panel
    compares the terminal velocity curve for the best-fitting mass
    model with the terminal velocity data of \citeauthor{Clemens85a}
    (\citeyear{Clemens85a}; longitudes $40^\circ$ to $80^\circ$) and
    \citeauthor{McClure07a} (\citeyear{McClure07a}; longitudes
    $-80^\circ$ to $-40^\circ$).}\label{fig:bestfit}
\end{figure}

As additional constraints on the mass distribution in the inner Milky
Way, we use data on the rotation curve and the vertical mass
distribution near the Sun. The rotation-curve data that we use are in
the form of terminal velocities measured through HI and CO
emission. In the fourth quadrant we use the HI data from
\citet{McClure07a} and in the first quadrant we use the CO data from
\citet{Clemens85a}, both binned into bins of $\Delta l =
1^\circ$. These terminal velocity measurements are shown in the bottom
panel of \figurename~\ref{fig:bestfit}. Because the terminal
velocities are characterized by wiggles on the order of $10\kms$ that
are likely due to the effects of non-axisymmetry, we model these data
very conservatively by assuming that they have an uncertainty of
$7\kms$ correlated over $\Delta \sin l = 0.125 \approx 1\kpc$. This
makes sure that wiggles typical of non-axisymmetry do not affect our
results. Because the terminal velocities depend on the Sun's peculiar
velocity in the plane of the Milky Way through terms proportional to
$\sin l$ and $\cos l$, we further marginalize over such terms in the
fits below, so that any effects of non-axisymmetry and the Sun's
peculiar velocity do not affect the analysis.

We further use constraints on the local slope of the rotation
curve. Such constraints exists, \eg, from the measurement of the Oort
constants \citep[\eg,][]{Feast97a}, from the kinematics of masers in
the disk of the Milky Way \citep{Reid09a,Bovy09b}, or from the
kinematics of stars throughout the disk \citep{BovyVc}. Since the
measurement of $\dd \ln V_c (R_0)/ \dd \ln R$ of \citet{BovyVc}
encompasses the results from these different analyses, we use it to
represent the current uncertainty in the logarithmic slope of the
rotation curve. We represent the \citet{BovyVc} measurement with the
simple analytic form
\begin{align}\label{eq:dlnvcdlnrprior}
  p\left(\frac{\dd \ln V_c(R_0)}{\dd \ln R} \right)& = 0\qquad
  \mathrm{if}\ \frac{\dd \ln V_c(R_0)}{\dd \ln R} > 0.04\,,\\ 
  p\left(\frac{\dd  \ln V_c(R_0)}{\dd \ln R} \right)& = W\,e^{-W}\,,\ \mathrm{otherwise},\nonumber\\
& \quad
  \mathrm{where}\ W = \left(1 - \frac{1}{0.04}\,\frac{\dd \ln V_c(R_0)}{\dd
    \ln R}\right)\,,\nonumber
\end{align}
which approximately matches the PDF for $\dd \ln V_c (R_0)/ \dd \ln R$
of \citet{BovyVc}. We stress that we do \emph{not} use any
measurements of the circular velocity itself. This way we can assess
what direct measurements of the mass distribution combined with
measurements of the \emph{shape} of the rotation curve (but not its
normalization) constrain the circular velocity to be.

Finally, we also include measurements of the vertical mass
distribution near the Sun, $\Sigma(R_0,Z)$. A number of such
measurements exist in the literature and we use the results from
\citet{Zhang13a}, which are the best measurements of $\Sigma(R_0,Z)$
to date and they are consistent with all other measurements. We only
use the measurement of $\kzoneone(R_0) = 67\pm6\,M_\odot\pc^{-2}$ and
the measurement of the stellar disk surface density $\Sigma_{*}(R_0) =
42\pm5\,M_\odot\pc^{-2}$. The measurement of $\kzoneone(R_0)$ is
consistent with our much tighter measurement in
\sectionname~\ref{sec:surfresults} and therefore it does not contain
much additional information, but the measurement of the stellar disk
surface density is useful for separating the contribution from the
disk and the dark halo to the mass budget.

\subsection{Results}\label{sec:fitresults}

\begin{figure}[t!]
  \includegraphics[width=0.475\textwidth,clip=]{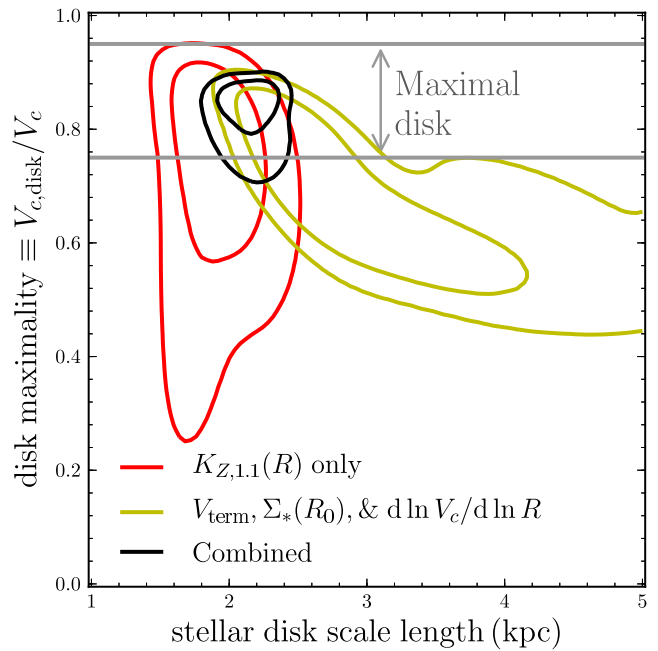}
  \caption{Contours of the joint PDF for the stellar disk scale length
    and the contribution of the disk to the circular velocity at 2.2
    scale lengths (the parameter describing whether the disk is
    maximal following \citealt{Sackett97a}'s definition; this
    definition is indicated in the figure as ``Maximal disk''). One
    and two sigma contours of the PDFs based on 3 combinations of the
    dynamical data are shown: (a) the $\kzoneone(R)$ measurements from
    this paper, (b) the terminal velocity data $V_{\mathrm{term}}$,
    constraints on $\dd \ln V_c(R_0) / \dd \ln R$, and the
    measurements from \citet{Zhang13a} (denoted as $\Sigma_{*}(R_0)$),
    and (c) the combination of (a) and (b). This Figure shows that the
    $\kzoneone(R)$ measurements of this paper are the most informative
    data for the dynamical measurement of the disk scale length. The
    combination of the new measurements in this paper and the existing
    dynamical constraints indicate that the Milky Way's disk is
    maximal.}\label{fig:pdf_rd_vcdvc}
\end{figure}

\begin{figure}[t!]
  \includegraphics[width=0.475\textwidth,clip=]{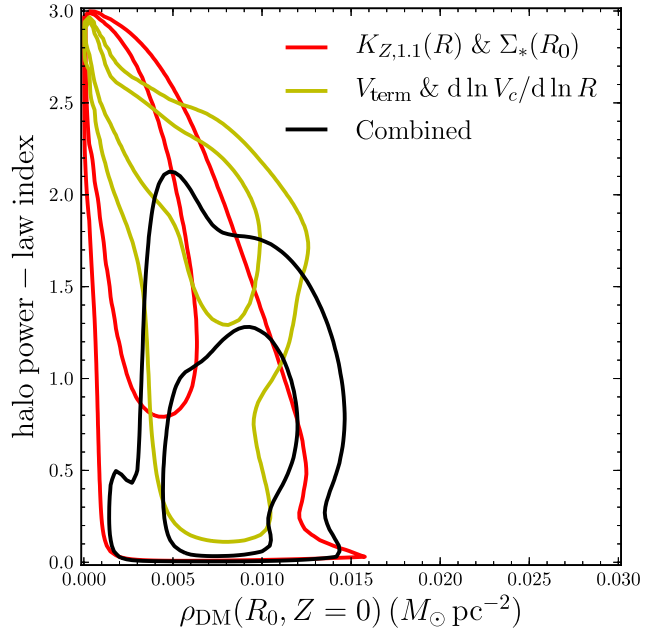}
  \caption{Contours of the joint PDF for the local dark-matter halo
    density and its local logarithmic radial slope. The halo is
    modeled with a power-law density profile $\rho_{\mathrm{DM}}(r) =
    \rho_{\mathrm{DM}}(R_0,Z=0)\,(R_0/r)^\alpha$. Neither the
    measurements of the vertical mass profile (local and
    $\kzoneone(R)$) nor the measurements of the shape of the circular
    velocity curve (terminal velocities $V_{\mathrm{term}}$ and local
    slope $\dd \ln V_c(R_0) / \dd \ln R$) constrain the radial profile
    of the halo very much. However, the combination of these two
    dynamical probes give a first constraint on the local halo
    profile: $\alpha < 1.53$ at 95\,\% confidence. This combination is
    different from multiplying the red and yellow PDFs, as both of
    these PDFs are marginalized over the other parameters of the
    Galactic potential. In particular, the steep halo-density peaks in
    the red and yellow PDFs correspond to very different and
    mutually-inconsistent slopes of the rotation curve, which is why
    they are disfavored in the combined
    PDF.}\label{fig:pdf_rhodm_plhalo}
\end{figure}

\begin{figure}[t!]
  \includegraphics[width=0.475\textwidth,clip=]{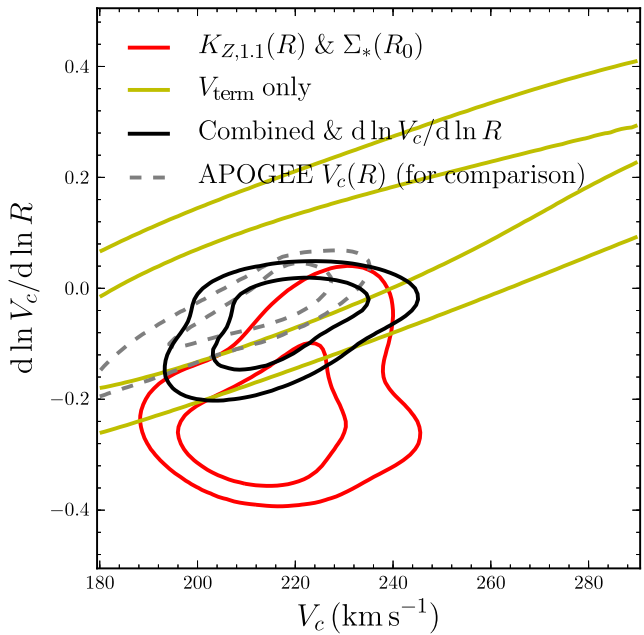}
  \caption{Joint PDF for the circular velocity at $R_0$ and the local
    logarithmic slope of the circular velocity curve. The terminal
    velocity curve alone constrains $V_c$ to be around $220\kms$ for a
    flat rotation curve and a larger (smaller) $V_c$ for a rising
    (falling) rotation curve). The measurements of the vertical mass
    distribution (red curves) give the opposite constraint: for $V_c$
    to be larger than $220\kms$, the rotation curve needs to be
    falling near $R_0$. The combination of rotation-curve shape
    constraints and surface-density measurements requires $V_c =
    218\pm10\kms$ and a gently falling rotation curve $\dd \ln V_c /
    \dd \ln R = -0.06 \pm 0.05$. This is consistent with the recent
    direct measurements of these quantities from stellar kinematics in
    the plane by APOGEE \citep{BovyVc}, which are shown for
    comparison.}\label{fig:pdf_vc_dlnvcdlnr}
\end{figure}

We use the same mass model as described in
\sectionname~\ref{sec:potential}, except that we replace the bulge
model with an exponentially cut off power-law with a power-law
exponent of $-1.8$, a cut-off radius of $1.9\kpc$, and a mass of
$6\times10^{9}\,M_\odot$, because this is a more realistic model for
the mass distribution of the bulge (see
\citealt{binneytremaine,McMillan11a}). We fit this to various
combinations of (a) the measurements of $\kzoneone(R)$ of
\sectionname~\ref{sec:surfresults}, (b) the terminal velocity data,
(c) the constraint on the local slope of the rotation curve from
\equationname~\ref{eq:dlnvcdlnrprior}, and (d) the measurements of
$\kzoneone(R_0)$ and $\Sigma_*(R_0)$ from \citet{Zhang13a}. We now
vary all five of the basic parameters of the mass model, that is, the
stellar disk mass scale length and scale height, the circular
velocity, the relative halo-to-disk contribution to $V_c^2(R_0)$, and
the local $\dd \ln V_c / \dd \ln R$. None of the considered data
really constrains the stellar scale height, but its value is
uncorrelated with the value of the other parameters; we let it vary
between $100\pc$ and $500\pc$. PDFs showing the primary results from
these fits are shown in
\figurenames~\ref{fig:pdf_rd_vcdvc}-\ref{fig:pdf_vc_dlnvcdlnr}. A
comparison between the best-fit model using all of the dynamical data
and the $\kzoneone(R)$ measurements of this paper and the terminal
velocities is shown in \figurename~\ref{fig:bestfit}.

\figurename~\ref{fig:pdf_rd_vcdvc} shows the joint PDF for the stellar
disk scale length and the contribution of the stellar disk to the
circular velocity at 2.2 scale lengths ($\approx$ the peak of the disk
rotation curve). The latter parameter determines whether the Milky
Way's disk is maximal using the definition of \citet{Sackett97a},
according to which a disk is maximal when its contribution to $V_c$ at
2.2 scale lengths is $85\pm10\,\%$. We show the constraints from
different combinations of the dynamical constraints. The red curves
show the contours of the PDF based on the $\kzoneone(R)$ measurements
from this paper only, while the black curves give the constraints
based on all of the dynamical data. It is clear that the measurements
of $\kzoneone(R)$ from this paper alone measure the scale length,
largely independent of the contribution to the mass from the other
Galactic components. \figurename~\ref{fig:pdf_rd_vcdvc} also shows
that the additional dynamical data from \sectionname~\ref{sec:adddata}
does not constrain the stellar disk scale length. The joint PDF for
the scale length and disk-maximality parameter based on the rotation
curve and $\Sigma(R_0,Z)$ does show the familiar relation that for the
disk to be maximal, the disk scale length needs to be small.

\subsubsection{Constraints on the Galactic disk mass distribution}

Therefore, we conclude that the $\kzoneone(R)$ measurements from this
paper are the single most important existing dynamical constraint on
the stellar disk scale length of the Milky Way. The result from the
combined fit to all dynamical data gives
\begin{equation}
  \mathrm{stellar\ disk\ scale\ length} = 2.15\pm0.14\kpc\,.
\end{equation}
The combination of the $\kzoneone(R)$ measurements and the additional
dynamical data shows that the disk is maximal since
\begin{equation}
  \frac{V_{c,*}}{V_c}\Bigg|_{2.2\,R_d} = 0.83\pm0.04\,.
\end{equation}

These measurements of the stellar disk scale length and its
contribution to the rotation curve allow us to derive the mass of the
disk. We can measure the surface density of the stellar disk because
the additional data on the rotation curve described in
\sectionname~\ref{sec:adddata} combined with the $\kzoneone(R)$
measurements from this paper allow us to disentangle the stellar and
dark-halo contributions to $\sigoneone(R)$. We find that the surface
density of the stellar disk at $R_0$ is $\Sigma_{*}(R_0) =
38\pm4\,M_\odot\pc^{-2}$, for a total surface density to $1.1\kpc$ of
$\sigoneone(R_0) = 68\pm4\,M_\odot\pc^{-2}$, $13\,M_\odot\pc^{-2}$ of
which is assumed to be in the thin ISM layer. As a consequence of ours
being a full 3D dynamical model, this measurement of $\sigoneone(R_0)$
is a real measurement of $\sigoneone(R_0)$ as opposed to a measurement
of $\kzoneone(R_0)$ converted to $\sigoneone(R_0)$. The fact that it
agrees so well with the local normalization of our measured
$\sigoneone(R)$ profile in \equationname~(\ref{eq:bestfitsig}) is due
to the fact that the local slope of the circular velocity curve is
very close to flat in our best-fit model (see below). Our measurement
of the local surface density to $1.1\kpc$ is consistent with all
previous measurements (which are really $\kzoneone(R_0)$ measurements)
but with a smaller uncertainty
\citep{Kuijken89b,Siebert03a,Holmberg04a,Bienayme06a,Garbari12a,Zhang13a}. We
note in particular that the measurement of \citet{Garbari12a} of
$\sigoneone(R_0) = 105\pm24\,M_\odot\,\pc^{-2}$ is consistent with our
measurement; our much smaller errorbar leads to a much tighter
measurement of the local dark matter density (see below).

As mentioned above, none of the dynamical data in
\sectionname~\ref{sec:adddata} constrain the mass scale height of the
stellar disk: we obtain a flat PDF for $z_h$ over the prior range
$100\pc < z_h < 500\pc$. To illustrate how $z_h$ can be constrained
using measurements of the (surface density) at heights different from
$|Z| = 1.1\kpc$, we fit the dynamical data from
\sectionname~\ref{sec:adddata} together with the constraint on the
total midplane density at $R_0$ from \citet{Holmberg00a}:
$\rho_{\mathrm{total}}(R_0,|Z| = 0) =
0.102\pm0.010\,M_\odot\pc^{-3}$. While all of the other dynamical
parameters are unchanged, in this case we do constrain the mass scale
height: $z_h = 370\pm60\pc$, in good agreement with the
measurements from star counts ($z_h \approx 400\pc$, see above and \bo
d) and constraints from the vertical profile of $K_Z(R_0,Z)$ ($z_h <
430\pc$ at $84\,\%$ confidence; \citealt{Zhang13a}).

Using our measurements of the disk mass scale length and the local
disk normalization, we can derive the total mass of the disk, to the
extent that a characterization of the disk with a single scale length
makes sense. We find that the total stellar disk mass is $M_{*} =
4.6\pm0.3\times10^{10}\,M_\odot$. Under our assumption that the local
ISM column density is $13\,M_\odot\pc^{-2}$ and that the ISM layer has
a scale length twice that of the stellar disk, the ISM contributes
$\approx0.7\times10^{10}\,M_\odot$ for a total (stars+gas) disk mass
of $M_{\mathrm{disk}} = 5.3\pm0.4\times10^{10}\,M_\odot$. Note that in
our model the ISM layer's scale length is tied to that of the stellar
disk and that the local normalization is fixed. Therefore, the mass of
the ISM layer is perfectly correlated with that of the stellar disk,
and the uncertainty in the total disk mass would be
$0.3\times10^{10}\,M_\odot$. For this reason, we have added an
uncertainty of $0.3\times10^{10}\,M_\odot$, or almost $50\,\%$ of the
ISM mass, in quadrature to the formal uncertainty in the disk
mass. The bulge contributes another $\approx6\times10^{9}\,M_\odot$
\citep{binneytremaine}, while the stellar halo mass is negligible,
such that the total baryonic mass of the Milky Way is
\begin{equation}
  M_{\mathrm{baryonic}} = 5.9\pm0.5\times10^{10}\,M_\odot\,,
\end{equation}
assuming an uncertainty of $0.3\times10^{10}\,M_\odot$ on the bulge
mass. To derive these masses, we have assumed that $R_0 =
8\kpc$. Changing $R_0$ to a different value does not change the
measurements made in this subsection or in subsequent subsections,
except for the total masses, which all increase by
$1.5\times10^{10}\,M_\odot$ when increasing $R_0$ to $8.5\kpc$; this
change is approximately linear in $(R_0 - 8\kpc)$. The change is so
large because of the short scale length of the disk: a small change in
$R_0$ leads to a big change in $R_0/R_d$. All other measurements are
independent of $R_0$, in particular the disk scale length, disk
maximality $V_{c,\mathrm{disk}} / V_{c,\mathrm{total}}$ at $R =
2.2\,R_d$, and the local surface densities.

Our finding that the Milky Way's disk has a short scale length raises
the question of whether such a massive disk is stable to axisymmetric
perturbations, \ie, whether Toomre's $Q > 1$ \citep{Toomre64a}. We do
not discuss this here in detail as this needs to be looked at
carefully, accounting for the mix of different components of the disk
and their radial dispersion profiles (which are poorly constrained
currently), and for the finite thickness of the disk.

\subsubsection{Constraints on the dark matter halo}

\begin{figure}[t!]
  \includegraphics[width=0.475\textwidth,clip=]{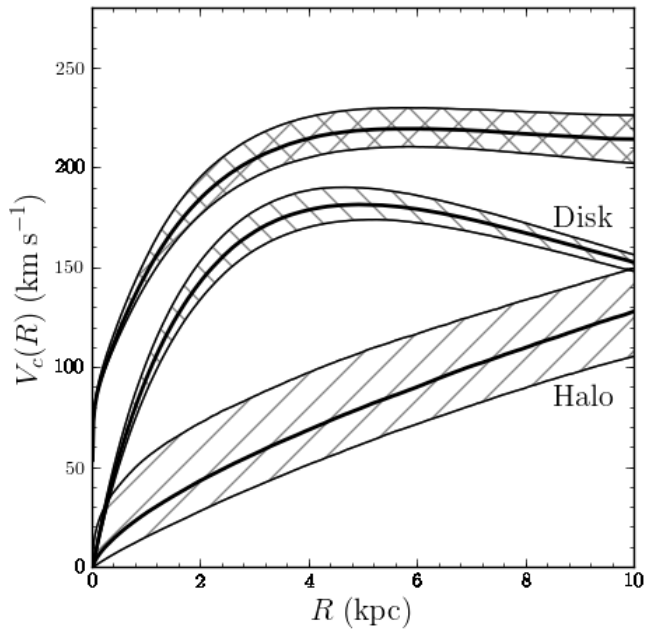}
  \caption{The Milky Way's rotation curve at $R < 10\kpc$ and its
    decomposition into stellar-disk and dark-halo contributions when
    using all of the dynamical data (terminal velocities,
    $\kzoneone(R)$, $\dd \ln V_c(R_0) / \dd \ln R$, and
    $\Sigma_{*}(R_0)$). The bulge is included in the total rotation
    curve, but we stress that the bulge is largely unconstrained by
    the dynamical data used here and all of its parameters are
    therefore held fixed. The thick lines are the median rotation
    curves and the hatched regions indicate 68\,\% confidence
    regions. Both the disk and halo rotation curves are highly
    constrained by the data.}\label{fig:rotcurves}
\end{figure}

A plausible way toward measuring the local dark halo density profile
is to combine the rotation curve, which measures the total mass as a
function of radius but is relatively insensitive to the flattening and
therefore cannot separate the disk's contribution from the halo's,
with independent measurements of the disk contribution as a function
of radius, as provided in this paper. The constraints on the halo
parameters---the local normalization $\rho_{\mathrm{DM}}(R_0)$ and
power-law index $\alpha$ in $\rho_{\mathrm{DM}} \propto 1 /
r^\alpha$---from our data marginalized over all other mass-model
parameters (including $V_c(R_0)$) are shown in
\figurename~\ref{fig:pdf_rhodm_plhalo}. We show the constraints from
measurements of the vertical dynamics ($\kzoneone(R)$ and
$\Sigma_*(R_0)$) and those from the rotation curve (terminal
velocities and $\dd \ln V_c / \dd \ln R$) alone; neither of these
constrains the radial profile of the dark halo and the entire prior
range $0 < \alpha < 3$ is allowed at $2\sigma$. However, the
combination of these two dynamical probes allows us to put a first
constraint on the radial profile. The combination gives
\begin{equation}
  \alpha \leq 1.53\quad\ \mathrm{at}\ 4\kpc<R<9\kpc\quad (\mathrm{95\,\%\ confidence})\,.
\end{equation}
This encompasses a cored halo as well as an NFW halo, although very
steep halo density profiles are ruled out. We constrain the local dark
matter density to be $\rho_{\mathrm{DM}}(R_0) =
0.008\pm0.0025\,M_\odot\pc^{-3}$, consistent with more direct
measurements of this quantity \citep[\eg,][]{BovyTremaine,Zhang13a},
which measure the dark matter density using the vertical dependence of
$K_Z(R_0,|Z|)$ ($\approx \Sigma(R_0,Z$) rather than the radial
dependence of \kzoneone\ as we do here.

\subsubsection{Constraints on the rotation curve}

Finally, while we do not measure the local circular velocity in a
direct manner, we do constrain it indirectly in its guise of providing
the normalization of the forces in our model and thus setting the mass
scale. A comparison between the value of $V_c(R_0)$ measured in this
way and more direct measurements can therefore test whether the
vertical dynamics is consistent with the planar
dynamics. \figurename~\ref{fig:pdf_vc_dlnvcdlnr} shows the joint PDF
for the local $V_c$ and the local logarithmic slope of the rotation
curve, $\dd \ln V_c / \dd \ln R$. The constraints from the terminal
velocities alone show the familiar degeneracy, indicating that the
terminal velocities only measure a combination of $V_c$ and the slope
of $V_c(R)$. The constraints from the vertical dynamics (red curves)
have a different degeneracy and strongly disfavor rising rotation
curves. The combination of the terminal velocities and the vertical
dynamics therefore measures the properties of the circular velocity
curve, and we find that $V_c = 218\pm10\kms$ and $\dd \ln V_c(R_0) /
\dd \ln R = -0.06\pm0.05$, consistent with a flat rotation
curve. These measurements are consistent with the recent APOGEE
measurements of \citet{BovyVc}, which are shown for comparison. They
are also consistent with the measurement of the angular rotation
frequency at the Sun by \citet{Feast97a} who found $V_c / R_0 =
27.19\pm0.87\kms\kpc\inv$ and $\dd \ln V_c(R_0) / \dd \ln R =
-0.09\pm0.05$. We emphasize that our measurement of $V_c$ in this
paper does not rely on the Sun's peculiar rotational velocity. For a
further discussion of how a measurement of $V_c = 218\kms$ compares
with the literature we refer the reader to Section 5.3 of
\citet{BovyVc}; suffice it to say that all previous measurements are
consistent with this measurement. A combination of the data considered
in this paper with the APOGEE results gives $V_c = 219\pm4\kms$ and
$\dd \ln V_c(R_0) / \dd \ln R= -0.06 \pm 0.04$. As the measurements of
this paper and the APOGEE measurements are very different in the way
that they probe the dynamics of the disk, the fact that these two
measurements agree on $V_c$ strongly argues that $V_c \approx
220\kms$.

\figurename~\ref{fig:rotcurves} shows a different representation of
all of the results described in this Section. Shown are the total
rotation curve and its decomposition into stellar-disk and halo
contributions. The total and stellar-disk rotation curves are quite
tightly constrained by our dynamical data. This is mostly due to our
precise measurement of the stellar disk scale length, which was made
possible by our measurements of $\kzoneone(R)$ over $4.5\kpc < R <
9\kpc$. The dark halo contributes significantly less to $V_c(R)$ than
the stellar-disk at all $R < 10\kpc$. \figurename~\ref{fig:rotcurves}
decidedly shows that we have for the first time clearly---and through
direct dynamical measurement---separated the disk and halo
contributions to the Milky Way's rotation curve.

\section{Discussion}\label{sec:discuss}

\subsection{First dynamical measurement of the Milky Way's scale length}

We believe that this paper presents the first dynamical measurement of
the Milky Way disk's mass profile. Other measurements of the scale
length are either based on star counts and it is therefore unclear
whether they trace all of the mass in the disk (e.g.,
\citealt{Juric08a}, \bo d), or they are based on previous dynamical
data that leave the scale length essentially unconstrained
(\citealt{Dehnen98a}; \figurename~\ref{fig:pdf_rd_vcdvc}) unless
strong priors are used \citep[\eg,][]{McMillan11a}. It turns out that
our best-fit model for the mass distribution in the inner $10\kpc$ of
the Milky Way is similar to that of model I in \citet{binneytremaine},
which has a maximal disk with a scale length of $2\kpc$.

If star counts do trace the underlying mass distribution, then we can
compare our dynamically-inferred scale length with that measured from
star counts. There have been many measurements over the last few
decades of the radial scale lengths of the thin and thick-disk
components spanning a wide range between $2$ and $5\kpc$. These
measurements have greatly improved over the last few years with the
advent of larger-area surveys with precise multi-band photometry
leading to better photometric distances. For example, \citet{Gould96a}
measured a scale length of $3.0\pm0.4\kpc$ from HST star counts of M
dwarfs, whose distribution is expected to trace that of the underlying
stellar mass, and \citet{Juric08a} found from an analysis of SDSS star
counts that the thin disk scale length is $2.6\kpc$. Both of these are
somewhat larger than the scale length measured in this paper. This
offset may be due to systematic uncertainties in the photometric
distances used by analyses of star counts.  Another, in our view more
likely explanation is that these analyses did not take into account
that the radial scale lengths of different stellar disk components
vary strongly. In particular, the scale lengths of the old, thick
components of the disk are only $\approx2\kpc$ (\citealt{Bensby11a};
\bo d; \citealt{Cheng12a}). This lowers the scale length of the mass
profile compared with that of the thin-disk components, an effect
which is larger at $R < R_0$, where most of our
\sigoneone\ measurements lie.

\begin{figure}[t!]
  \includegraphics[width=0.475\textwidth,clip=]{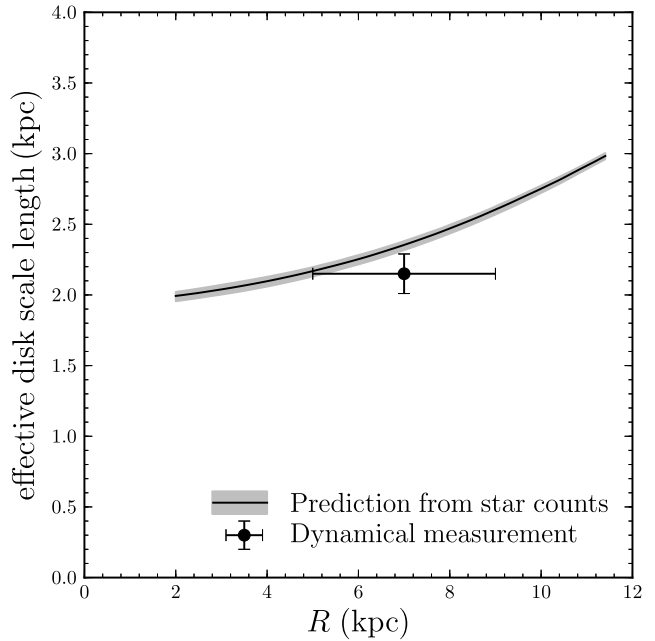}
  \caption{Comparison between the effective disk scale length
    determined from star counts and the dynamically measured disk mass
    scale length from this paper. The prediction from star counts is
    obtained by taking the scale length and stellar surface density
    measurements of MAPs of BO12 and measuring the scale length of the
    effective radial profile that is obtained by summing over all
    MAPs. The errorbar on the radius of the dynamically measured scale
    length indicates the range over which it is measured in this paper
    (cf.~\figurename~\ref{fig:surf}). The scale length determined by
    star counts is in excellent agreement with the
    dynamically-measured scale length.}\label{fig:map_rd}
\end{figure}

A proper comparison between the dynamically-inferred scale length and
that measured from star counts should therefore take into account the
full, complex structure of the disk. \bo\ showed that the disk is made
up of many components with scale lengths ranging from $2\kpc$ for the
oldest populations to $>4.5\kpc$ for the younger populations. The mass
scale length of the stellar disk is determined by the sum of all of
these components, whose combined density profile defines an effective
disk scale length at every radius. This effective disk scale length
based on the re-assembly of the \map\ decomposition of \bo\ is shown
in \figurename~\ref{fig:map_rd} (the \map\ scale lengths of \bo d have
been re-scaled to the \an\ distance scale). The stellar mass profile
derived from summing over all \map s is not a single exponential, but
has a profile whose effective disk scale length increases smoothly
from $\approx2\kpc$ in the inner disk to $\approx3\kpc$ at
$R=12\kpc$. The dynamical measurement from this paper compares well
with that inferred from star counts: the dynamical estimate falls
short of the star-counts measurement by about $1\sigma$. Nevertheless,
this comparison lends credence to the interpretation that the stars in
the Galactic disk are indeed the dominant contributors to the
`dynamically inferred disk mass' derived here. With measurements in
the outer disk at $R > 10\kpc$, the \map\ star-counts model predicts
that one should dynamically measure a mass scale length $> 2.5\kpc$.

We can further compare our dynamical measurement of the stellar-disk
surface density at $R_0$, $\Sigma_*(R_0) = 38\pm4\,M_\odot\pc^{-2}$,
with estimates based on direct star counts. Estimates of the total
amount of ordinary stellar matter are $\Sigma_{\mathrm{visible}}(R_0)
\approx 30\,M_\odot\pc^{-2}$ (\citealt{Flynn06a};\bo b) with an
additional $7\,M_\odot\,\pc^{-2}$ contributed by stellar remnants and
brown dwarfs \citep{Flynn06a}. These estimates have uncertainties of a
few $M_\odot\pc^{-2}$, which includes the uncertainty in the shape of
the initial mass function at low masses (see \bo b). Thus, our
dynamical measurement of the disk mass properties is in excellent
agreement with direct star-counts.

That the dynamical and star-counts measurements of the stellar-disk
scale length and of the local stellar surface density agree so well is
evidence that basically all of the mass in the stellar disk is
accounted for by ordinary stellar matter orbiting under the influence
of Newtonian gravity. Our results leave little room for a dark disk
component in the Milky Way \citep{Read08a}, because the presence of
such a component would increase our estimates of the local surface
density and of the mass scale length. Similarly, in the MOND theory of
modified gravity the local surface-density to $1.1\kpc$ is predicted
to be enhanced with respect to the contribution from baryonic matter
by $\gtrsim60\,\%$ \footnote{A simple way to see this is that the
  vertical force in MOND is enhanced by the same amount as the radial
  force. From \figurename~\ref{fig:rotcurves} it is clear that in
  Newtonian gravity the disk only provides about $60\,\%$ of the
  radial force at $R_0$, such that MOND needs to enhance the radial,
  and consequentially the vertical, force by $\approx60\,\%$.}
\citep{Nipoti07a,Bienayme09a,Famaey12a}, such that we expect for
$\Sigma_{\mathrm{baryonic}}(R_0) = 51\pm4\,M_\odot\pc^{-2}$ that
$\sigoneone^{\mathrm{MOND}}(R_0) \gtrsim 82\pm6\,M_\odot\pc^2$. This
prediction is in $\gtrsim2\sigma$ tension with our measurement of
$\sigoneone(R_0) = 68\pm4\,M_\odot\pc^{-2}$. Furthermore, in MOND the
dynamically-inferred disk scale length (\ie, that inferred from
$K_Z(R)$ measurements) is predicted to be $25\,\%$ larger than that
measured from star counts \citep{Bienayme09a}, so we would expect to
dynamically measure a scale length of $\approx2.9\kpc$. This is
$5\sigma$ removed from our measurement of the mass scale length. We
emphasize that these are preliminary tests of MOND based on the
Bekenstein--Milgrom \citep{Bekenstein84a} formulation of MOND and that
these tests should more fully take into account the uncertain
structure of the baryonic disk. However, it is clear that the
measurements from this paper and further improved measurements of the
vertical forces in the Milky Way are key to improved tests of modified
gravity models such as MOND on galaxy scales.

\subsection{The mass of the Galactic disk}

Our measurement of the mass scale length of the disk now makes our
measurement of the Galactic disk mass the most accurate estimate to
date. Our measurement of $M_{*} = 4.6\pm0.3\times10^{10}\,M_\odot$
compares well with the range found by \citet{Flynn06a} (who also
assume $R_0 = 8\kpc$), who estimated the stellar-disk mass as a
function of an assumed scale length. Our measurement of the scale
length falls at the lower end of their considered range where their
stellar disk mass is $\approx5\times10^{10}\,M_\odot$. Our estimate of
the baryonic mass in the Galaxy of $M_{\mathrm{baryonic}} =
5.9\pm0.5\times10^{10}\,M_\odot$ also agrees with their estimate of
$M_{\mathrm{baryonic}} = 6.1\pm0.5\times10^{10}\,M_\odot$. However, it
is important to note that ours is a purely dynamical measurement,
while Flynn's value relies on assumptions about the stellar mass
function in the disk and about the manner in which mass traces
light. Both measurements are similarly affected by the uncertainty in
$R_0$.

\subsection{Disk maximality and the Milky Way compared to external galaxies}\label{sec:cfexternal}

The measurements of the Milky Way's disk mass scale length and stellar
mass from this paper allow us to make comparisons of the Milky Way
with external galaxies, \eg, through the Tully-Fisher relation,
initially put forth by \citet{Flynn06a}, more precise. A main source
of systematic uncertainty in \citet{Flynn06a} was the uncertainty in
the value of the radial scale length, which a priori could have been
anywhere between $2\kpc$ and $5\kpc$. Our measurement of $R_d =
2.15\pm0.14\kpc$ removes this source of uncertainty. Using
\citet{Flynn06a}'s calculations of the total $I$-band luminosity of
the Milky Way we find that $L_I^{\mathrm{disk}} \approx
3.5\times10^{10}\,L_\odot$ ($M_I = -22.2$). Using a bulge luminosity
of $10^{10}\,L_\odot$ \citep{Kent91a}, we find that
\begin{equation}
  L_I \approx 4.5\times10^{10}\,L_\odot, \ \mathrm{or}\qquad M_I \approx -22.5\,.
\end{equation}
Combined with our measurement that $V_c = 218\pm10\kms$, this allows
us to place the Milky Way onto the Tully-Fisher relation defined by
nearby disk galaxies: it falls well within the $1\sigma$ scatter
\citep[\eg,][]{Dale99a} as this relation predicts $M_I = -22.8$ for
$V_c = 220\kms$ with a scatter of $0.4\,\mathrm{mag}$. Therefore, from
the point of view of the Tully-Fisher relation, the Milky Way is a
typical galaxy.

The decomposition of the Milky Way's rotation curve into the
contributions from the stellar disk and the dark-matter halo in
\sectionname~\ref{sec:fitresults} has shown that the Milky Way's disk
is maximal by the definition of \citet{Sackett97a}. This appears to be
in conflict with arguments based on the lack of correlation between
Tully-Fisher velocity residuals and disk size for external galaxies
\citep{Courteau99a}, which have been used to argue against disks being
maximal. However, whether this really shows that \emph{all} disks are
sub-maximal is far from clear, as dynamical modeling of gas kinematics
in various galaxies and constraints from spiral structure have shown
that some disks, especially those with $V_c > 200\kms$, are maximal
\citep{Atha87a,Weiner01a,Kranz03a}, while being consistent with the
considerations of \citet{Courteau99a}. Measurements of the vertical
velocity dispersions of external galaxies have been interpreted to
indicate that disks are substantially sub-maximal (\eg,
\citealt{Bottema97a} and more recently \citealt{Kregel05a} and
\citealt{Bershady11a}). Such measurements rely on the same dynamical
principles as those employed in this paper's measurement, but these
are much more difficult to apply in external galaxies without strong
assumptions. It is essentially impossible to measure both the scale
height and the vertical velocity dispersion for any individual
external galaxy; further the velocity dispersions obtained from
integrated light do not trace the older, dynamically-relaxed stellar
populations very well. As such, these measurements are afflicted with
systematic uncertainties, which have not been sufficiently
investigated. The fact that the Milky Way would appear substantially
sub-maximal in the analysis of \citet{Bershady11a} while the detailed
dynamical modeling in this paper indicates otherwise may be a sign of
these systematic uncertainties\footnote{A specific example is the
  following: \citet{Bershady11a} use the scale length of $\sigma_Z^2$
  as a proxy for that of $K_Z$ and use the latter as a proxy for the
  disk scale length. Neither of these are good proxies in the Milky
  Way: in \bo c we measured the scale length of $\sigma_Z^2$ to be
  $3.5\kpc$, which is much longer than that of $K_Z$ found in
  \sectionname~\ref{sec:surfresults}; in this paper we find that the
  scale length of $K_Z$ is significantly longer than that of
  $\Sigma$. As \citet{Bershady11a} use the scale length of
  $\sigma_Z^2$ to estimate galaxy disks' central surface density
  $\Sigma_0$, these effects lead one to underestimate $\Sigma_0$ and
  the maximality parameter of external disks.}.

\begin{figure}[t!]
  \includegraphics[width=0.475\textwidth,clip=]{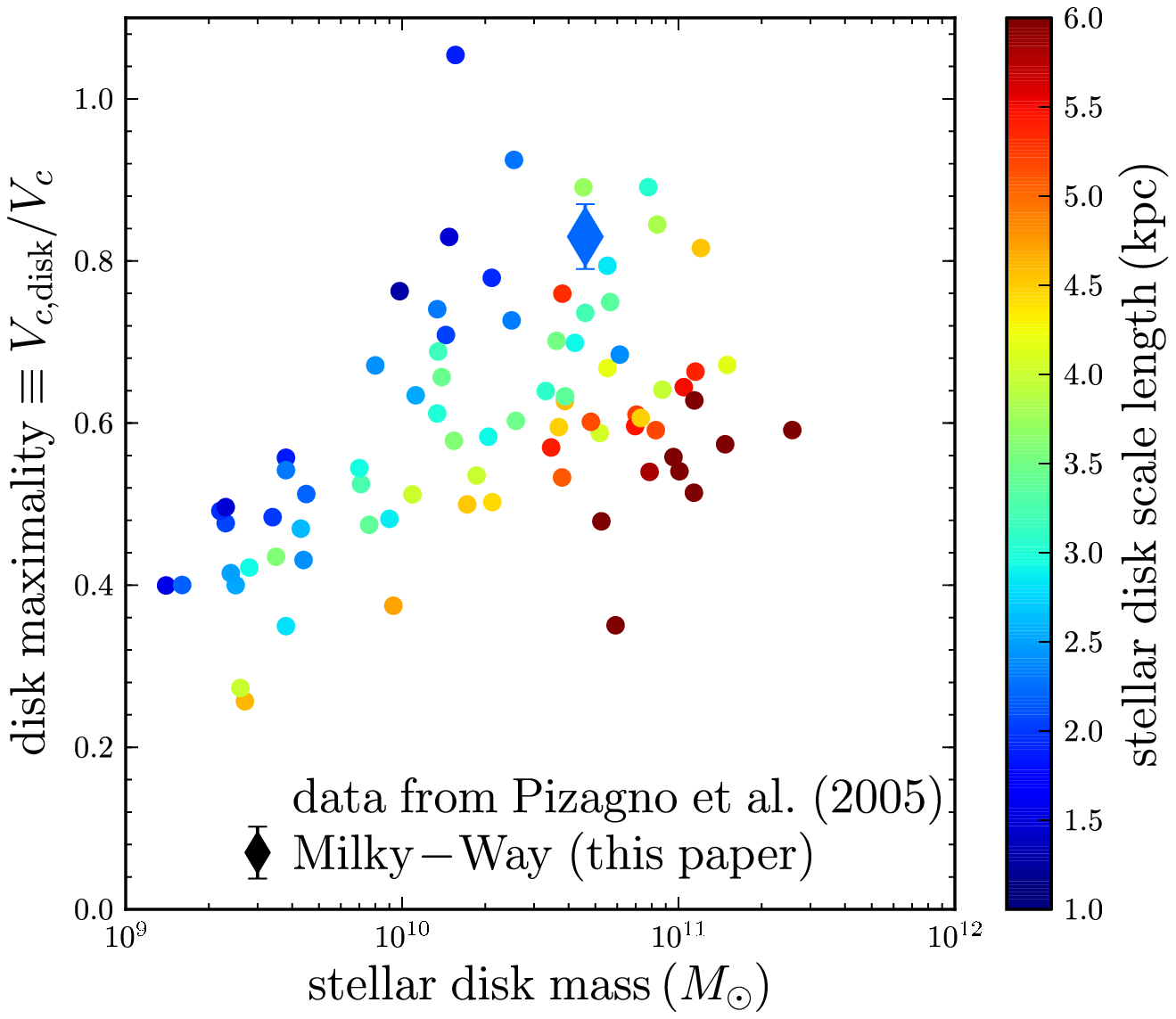}
  \includegraphics[width=0.475\textwidth,clip=]{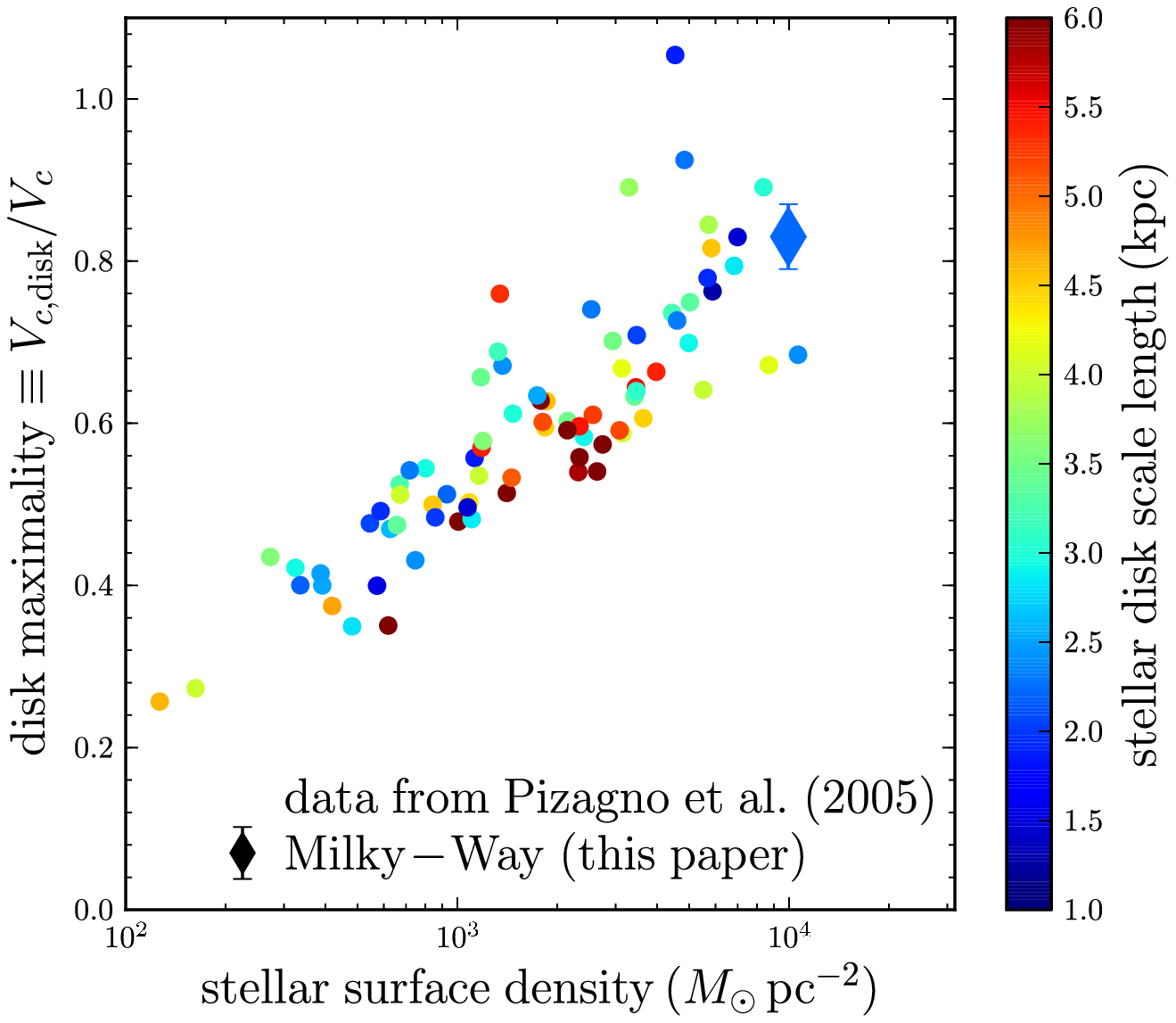}
  \caption{The Milky Way's disk properties compared to 81 external
    galaxies from \citet{Pizagno05a}. The top panel shows the relation
    between stellar mass and the disk's contribution to the rotation
    velocity at 2.2 scale lengths (the extent to which the disk is
    maximal, see \figurename~\ref{fig:pdf_rd_vcdvc}); points are
    color-coded by the radial scale length. The bottom panel shows the
    relation between the surface density ($\equiv M_*\,R_d^{-2}$) and
    the disk maximality. The Milky Way falls along the general trends
    defined by external galaxies, except that its scale length appears
    short compared to that of similar external
    galaxies.}\label{fig:pizagno}
\end{figure}

In \figurename~\ref{fig:pizagno} we compare our measurements of the
Milky Way disk's properties to those derived from a sample of 81
disk-dominated galaxies from \citet{Pizagno05a}. We follow
\citet{Gnedin07a} in using the measurements of the disk scale lengths,
stellar masses, and circular velocities at 2.2 disk scale lengths to
derive the disk's contribution to the rotation curve at 2.2 disk scale
lengths. In detail, we scale the stellar masses down by 20\% to
correct them for the presence of any bulge component and we lower the
observed scale lengths by 10\% as these scale lengths are measured
from $g$ and $r$-band images, which are typically larger than the
$K$-band scale length, which better traces the stellar mass
\citep{deJong96a}. The top panel shows that the Milky Way falls nicely
within the stellar-mass--disk-maximality relation derived from the
\citet{Pizagno05a} data. However, the Milky Way's scale length is
quite different from that of external galaxies with similar stellar
masses: for its stellar mass, the Milky Way would be expected to have
scale length of $\approx4.5\kpc$, albeit it with a large scatter of
$\approx2\kpc$; this is clear from the color-coding of the points in
the top panel of \figurename~\ref{fig:pizagno}. Following
\citet{Gnedin07a}, we also consider the relation between the disk's
surface density ($\equiv M_* \,R_d^{-2}$) and the fraction of the
circular velocity contributed by the disk. This is shown in the bottom
panel of \figurename~\ref{fig:pizagno}. The short scale length that we
measure for the Milky Way has the effect of putting the Milky Way at
the upper edge of the observed surface densities, but the Milky Way
still falls along the general trend determined by external galaxies.

Therefore, if the Milky Way is atypical in any way, it is because of
its short scale length. However, the mass-weighted scale length is
only poorly measured for external galaxies, because variations in the
mass-to-light ratio with radius may hamper the photometrically
measured disk profiles.

\subsection{\map s as dynamically phase-mixed populations}

The analysis in this paper has used the properties of \map s as
measured in \bo\ to justify modeling the \map s as phase-mixed,
steady-state stellar populations that lend themselves to simple
dynamical models. This has worked well and all \map s lead to a
consistent measurement of the vertical mass distribution near the
disk. The single-qDF-per-\map\ model provides a good fit to the
spatial and kinematic properties of \map s, as explicitly shown in the
detailed comparisons between the data and the best-fit dynamical
models in \appendixname~\ref{sec:datamodel}. Thus, the \map s are
indeed well-described by the action-based qDF, as first proposed by
\citet{Ting13a}, although with the caveat that we have not modeled the
radial and azimuthal velocities.

The fact that all \map s can be described by the qDF adds further to
the evidence that \map s are simple dynamical building blocks of the
(local) disk, as first proposed by \bo. In particular the \map s with
scale heights and velocity dispersions intermediate between those of
the canonical thin and thick disk are real dynamical populations. If
these intermediate \map s would have arisen because of abundance
errors, as was already strongly argued against based on their observed
spatial properties and kinematics in \bo, we would not expect to be
able to model them using a consistent dynamical model. For example, we
would expect that the vertical profile measurements would be dominated
by the stars in the thin component, while the dispersion measurements
would be dominated by the stars in the thicker, kinematically-warmer
component, which would strongly overestimate the surface density (by a
factor of 3 or more). The fact that this does not happen strongly
argues that the intermediate \map s are real, phase-mixed stellar
populations in a dynamical steady state.

\section{Conclusions}\label{sec:conclusion}

In this paper we have used six-dimensional dynamical fitting employing
three-action-based distribution functions, in order to model
abundance-selected stellar populations from the SEGUE survey, fully
accounting for the selection function and deriving dynamical
constraints while marginalizing over properties of the DF. We have
used this in particular to obtain a measurement of \sigoneone\ (or
\kzoneone) at a single Galactocentric radius for each \map, thus
dynamically measuring for the first time the radial profile of the
surface density near the disk. These measurements are given in
\tablename~\ref{table:surf} and they present stringent new constraints
on the mass distribution in the inner Milky Way.

We have used these new measurements of $\kzoneone(R)$ in addition to
(weak) existing measurements of the terminal velocity curve between
$4\kpc$ and $R_0$, the contribution of the disk to the local surface
density, and the local slope of the rotation curve to constrain the
gravitational potential between $4\kpc$ and $10\kpc$. We find that our
new measurements of $\kzoneone(R)$ provide the only dynamical
constraint able to measure the dynamical (mass-weighted) disk scale
length; in combination with the other dynamical constraints we can
measure the properties of the Milky Way's disk to great precision and
find
\begin{align*}
  \mathrm{stellar\ disk\ scale\ length} & = 2.15\pm0.14\kpc\,,\\
  \Sigma_{*}(R_0) & = 38\pm4\,M_\odot\pc^{-2}\,,\\
  \Sigma_{\mathrm{disk}}(R_0) & = 51\pm4\,M_\odot\pc^{-2}\,,
\end{align*}
where $\Sigma_{*}(R_0)$ includes the contributions from ordinary
stellar matter, stellar remnants, and brown dwarfs and
$\Sigma_{\mathrm{disk}}(R_0) =
\Sigma_{\mathrm{ISM}}(R_0)+\Sigma_{*}(R_0)$.  We further find that
\begin{align*}
  M_{*} & = 4.6\pm0.3\,(\mathrm{ran.})\,\pm1.5\,(\mathrm{syst.})\,\times10^{10}\,M_\odot\,,\\
  M_{\mathrm{disk}} & = 5.3\pm0.4\,(\mathrm{ran.})\,\pm1.5\,(\mathrm{syst.})\,\times10^{10}\,M_\odot\,,\\
  M_{\mathrm{baryonic}} & = 5.9\pm0.5\,(\mathrm{ran.})\,\pm1.5\,(\mathrm{syst.})\,\times10^{10}\,M_\odot\,.
\end{align*}
The systematic uncertainty is due to the uncertainty in $R_0$:
increasing $R_0$ from our fiducial value of $8\kpc$ to $8.5\kpc$
increases the estimated masses by $1.5\times10^{10}\,M_\odot$. The
disk mass $M_{\mathrm{disk}}$ includes the mass of the stellar disk,
$M_{*}$, as well as the mass of the ISM layer; $M_{\mathrm{baryonic}}$
is the total baryonic mass of the Milky Way, including the stellar and
ISM disks and the bulge.

These direct dynamical measurements of the stellar disk's properties
are in good agreement with measurements derived from star counts,
leaving little room for dark matter in a disk-like configuration. With
a scale length this short, the Milky Way's disk is maximal by the
definition of \citet{Sackett97a}: we find that $V_{c,*} / V_c =
0.83\pm0.04$ at 2.2 disk scale lengths.

This paper's measurement of the disk mass will also be particularly
valuable when measuring the dark halo's flattening from constraints on
the potential at larger heights. For example, \citet{Koposov10a}
measured the total potential flattening at $\approx8\kpc$ from the
plane from fitting an orbit to the cold GD-1 stream, but found that
the uncertainty in the mass of the disk did not allow for this to be
turned into an interesting constraint on the halo's flattening. Using
our measurement of the disk mass, the GD-1 data indicate that the halo
density flattening is $\approx 0.7^{+0.3}_{-0.15}$, but it is clear
that a more rigorous combination of these measurements and further
progress in the dynamical fitting of tidal streams
\citep[\eg,][]{Sanders13a} are necessary to robustly measure the
halo's flattening.

These measurements of the disk's properties allow us to separate the
contribution from the disk and the halo to the rotation curve, as
shown in \figurename~\ref{fig:rotcurves}. The halo does not contribute
much to the Milky Way's rotation curve at $R < 10\kpc$. In turn this
means that our constraints on the radial profile of the dark halo are
relatively weak. Nevertheless, these are the first dynamical
constraints on the radial distribution of dark matter at $R < 10\kpc$
and we find for a model $\rho_{\mathrm{DM}} (r;\approx R_0) \propto 1
/ r^{\alpha}$ that $\alpha < 1.53$ at 95\,\% confidence (where
$\approx R_0$ indicates that our measurement is based on dynamical
data at $4\kpc\lesssim R \lesssim 9\kpc$). Further measurements of the
vertical mass distribution (\eg, $\Sigma(R)$ at $|Z| \approx2\kpc$ and
$|Z| \approx3\kpc$) would significantly improve this constraint.

The dynamical modeling performed in this paper is complex,
computationally expensive, and currently limited to exploring only a
few parameters of the gravitational potential. It provides a full
generative modeling framework in which selection effects,
observational uncertainties, nuisance parameters, outlier models, and
other data issues can be naturally included using a likelihood-based
approach. We have dealt with the computational complexity by focusing
our efforts on the measurement of a single dynamical constraint
derived from each dynamical sub-sample of stellar tracers (\map s):
the vertical force at $1.1\kpc$. However, it is clear that the fitting
procedure to each \map\ can yield much more information than this.  In
particular, each sub-sample could be used to also measure the vertical
profile of the vertical force within the SEGUE sample's spatial
volume, which should lead to better constraints on the radial profile
of the dark-matter halo. Ultimately, this is what is required for the
optimal analysis of Gaia data. The analysis in this paper has
demonstrated 3D, 3-action dynamical modeling of disk populations using
observations of individual stars in the context of a real data set
with all of the complexities of a non-trivial selection function and
data uncertainties. But it is clear that further development of this
technique, especially in more efficiently exploring the dynamical PDF,
is necessary before the Gaia data arrive.


\acknowledgements It is a pleasure to thank James Binney, Chris Flynn,
Yuan-Sen Ting, Scott Tremaine, and the Oxford dynamics group for
helpful comments and assistance. We would like to thank the
\segue\ team for their efforts in producing the \segue\ data set.
Computations were run on the Institute for Advanced Study's School of
Natural Sciences' Aurora cluster. J.B. was supported by NASA through
Hubble Fellowship grant HST-HF-51285.01 from the Space Telescope
Science Institute, which is operated by the Association of
Universities for Research in Astronomy, Incorporated, under NASA
contract NAS5-26555. J.B. and H.-W.R acknowledge support from SFB 881
(A3) funded by the German Research Foundation DFG.

Funding for the SDSS and SDSS-II has been provided
by the Alfred P. Sloan Foundation, the Participating Institutions, the
National Science Foundation, the U.S. Department of Energy, the
National Aeronautics and Space Administration, the Japanese
Monbukagakusho, the Max Planck Society, and the Higher Education
Funding Council for England. The SDSS Web Site is
http://www.sdss.org/.

\appendix


\begin{figure}[t!]
  \includegraphics[width=0.23\textwidth,clip=]{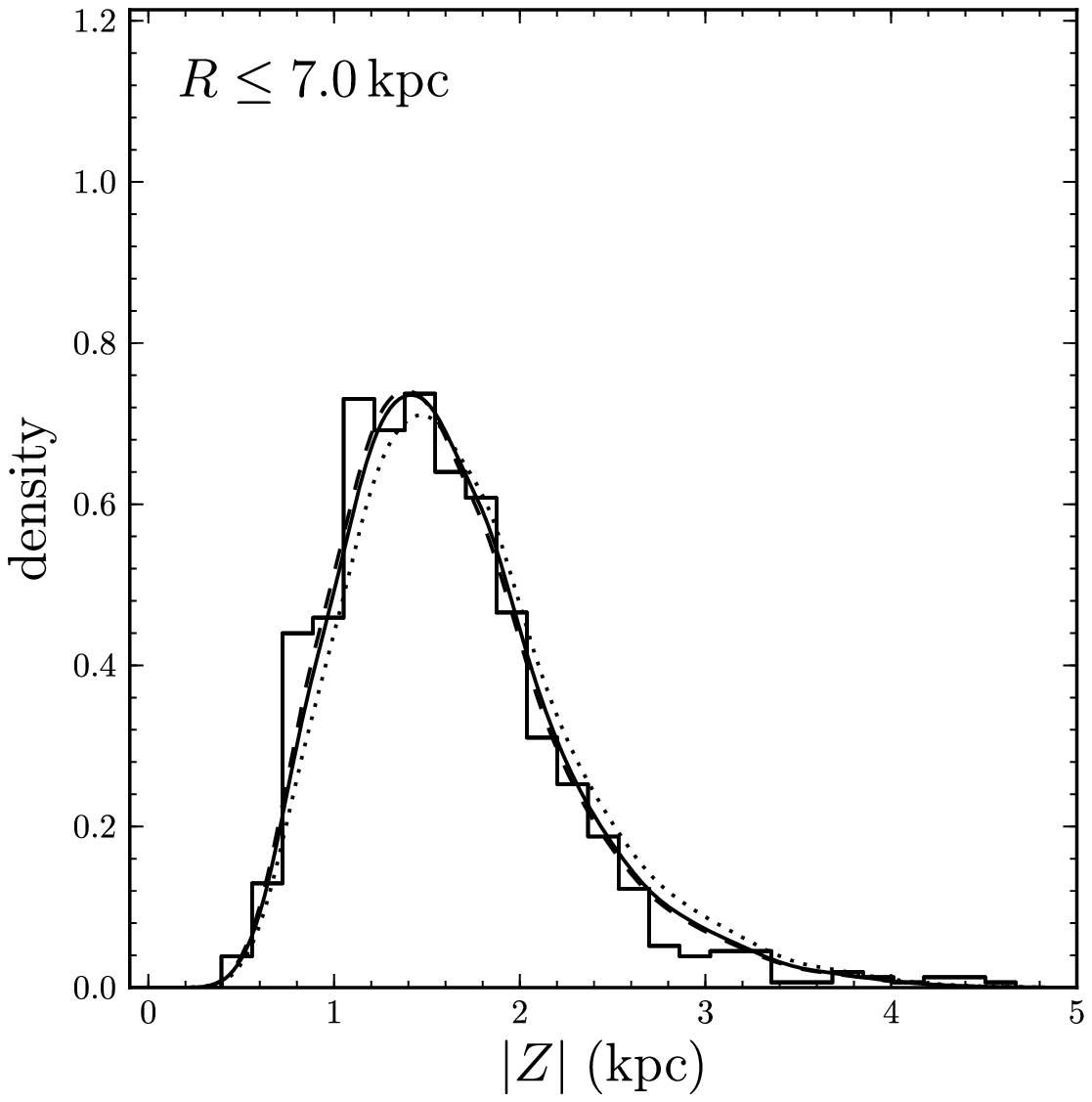}
  \includegraphics[width=0.23\textwidth,clip=]{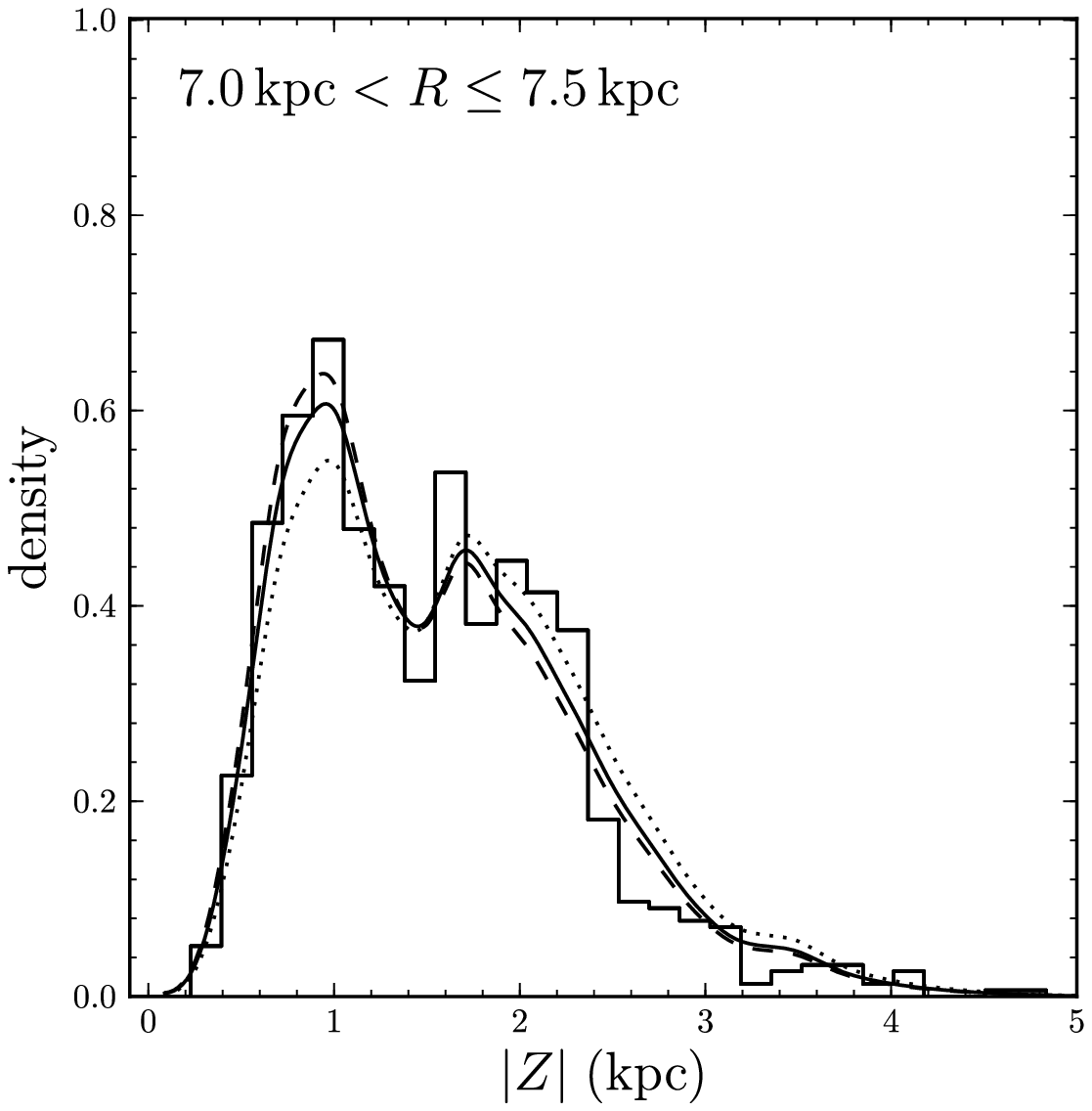}
  \includegraphics[width=0.23\textwidth,clip=]{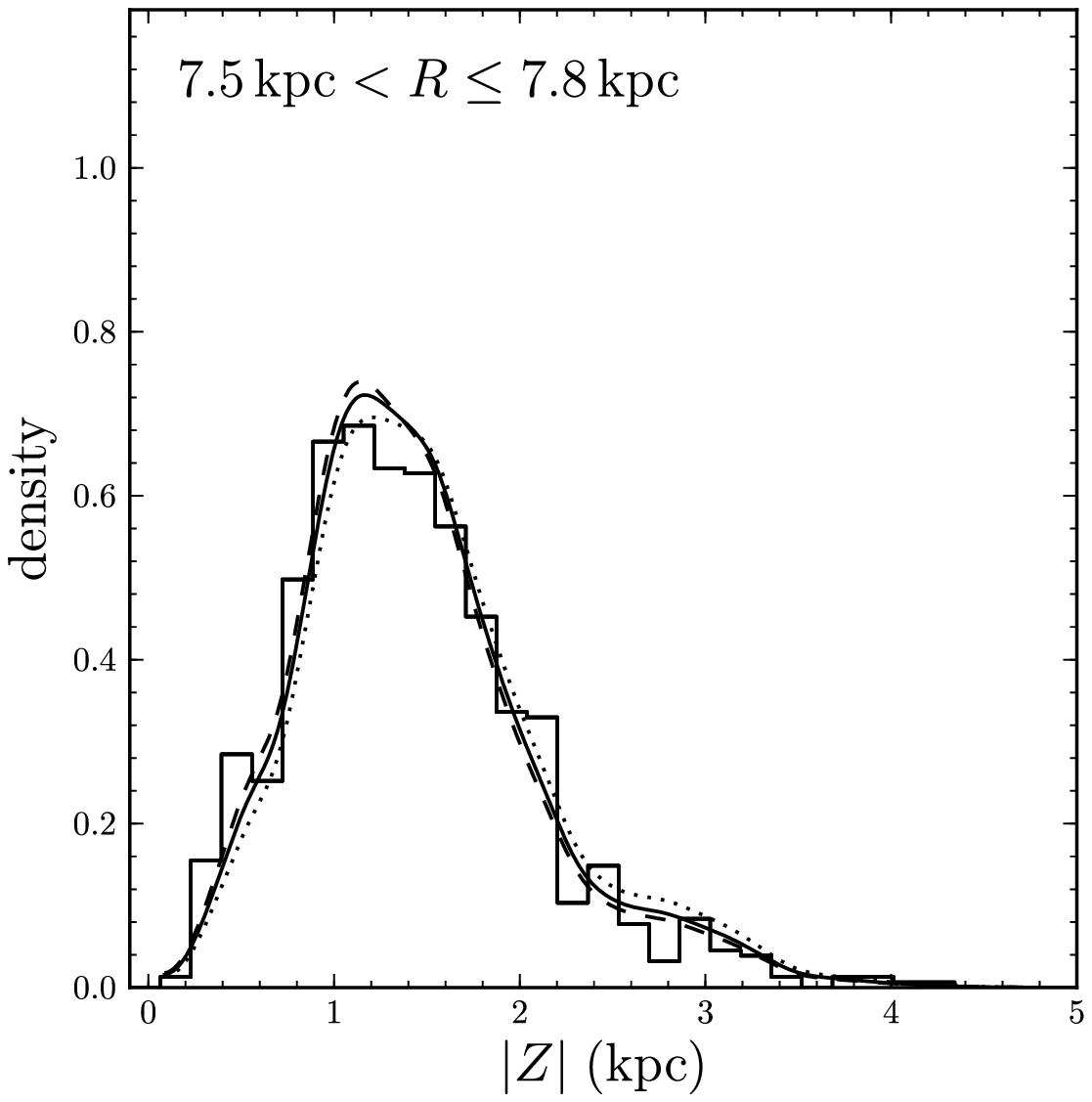}
  \includegraphics[width=0.23\textwidth,clip=]{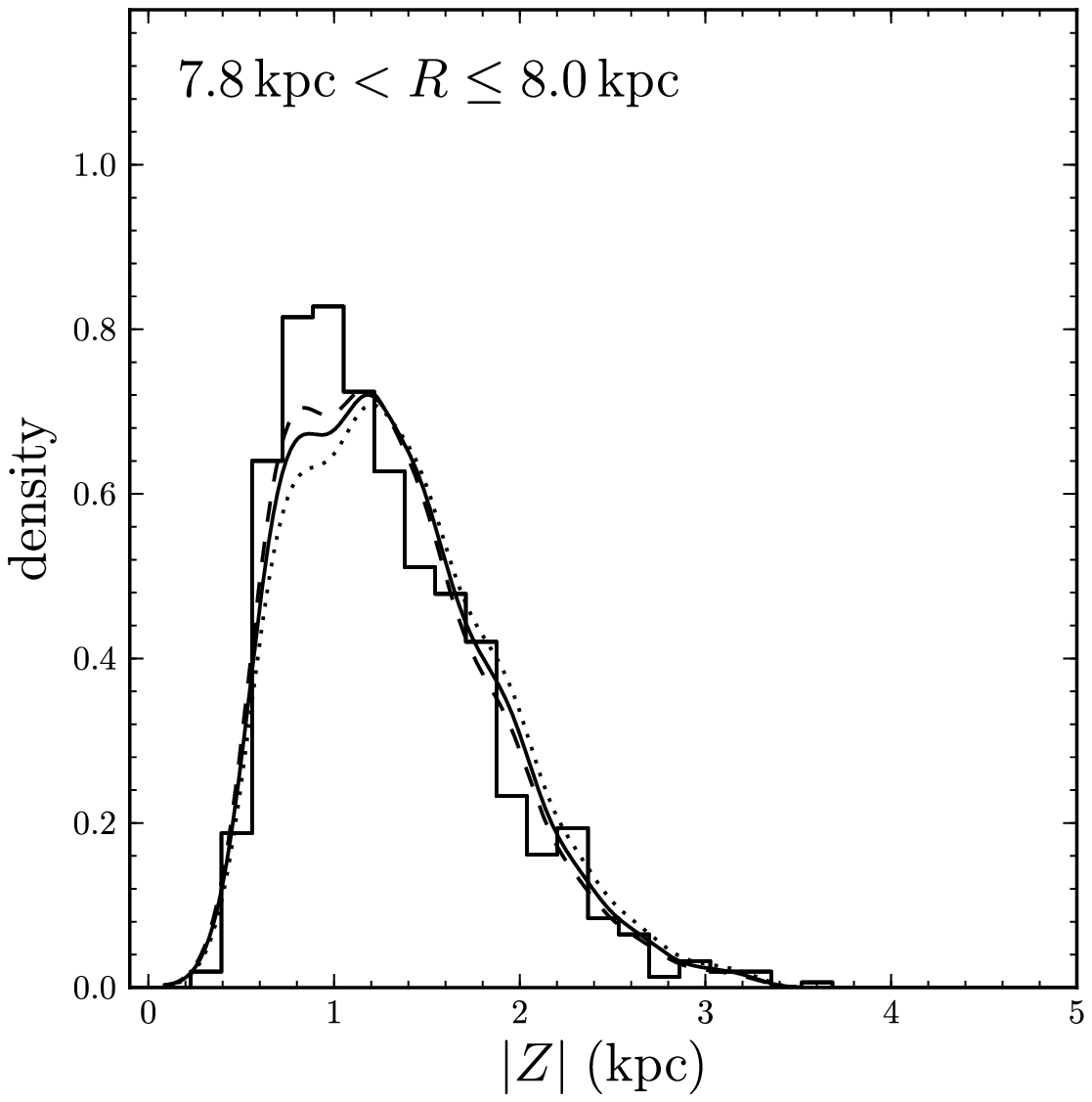}\\
  \includegraphics[width=0.23\textwidth,clip=]{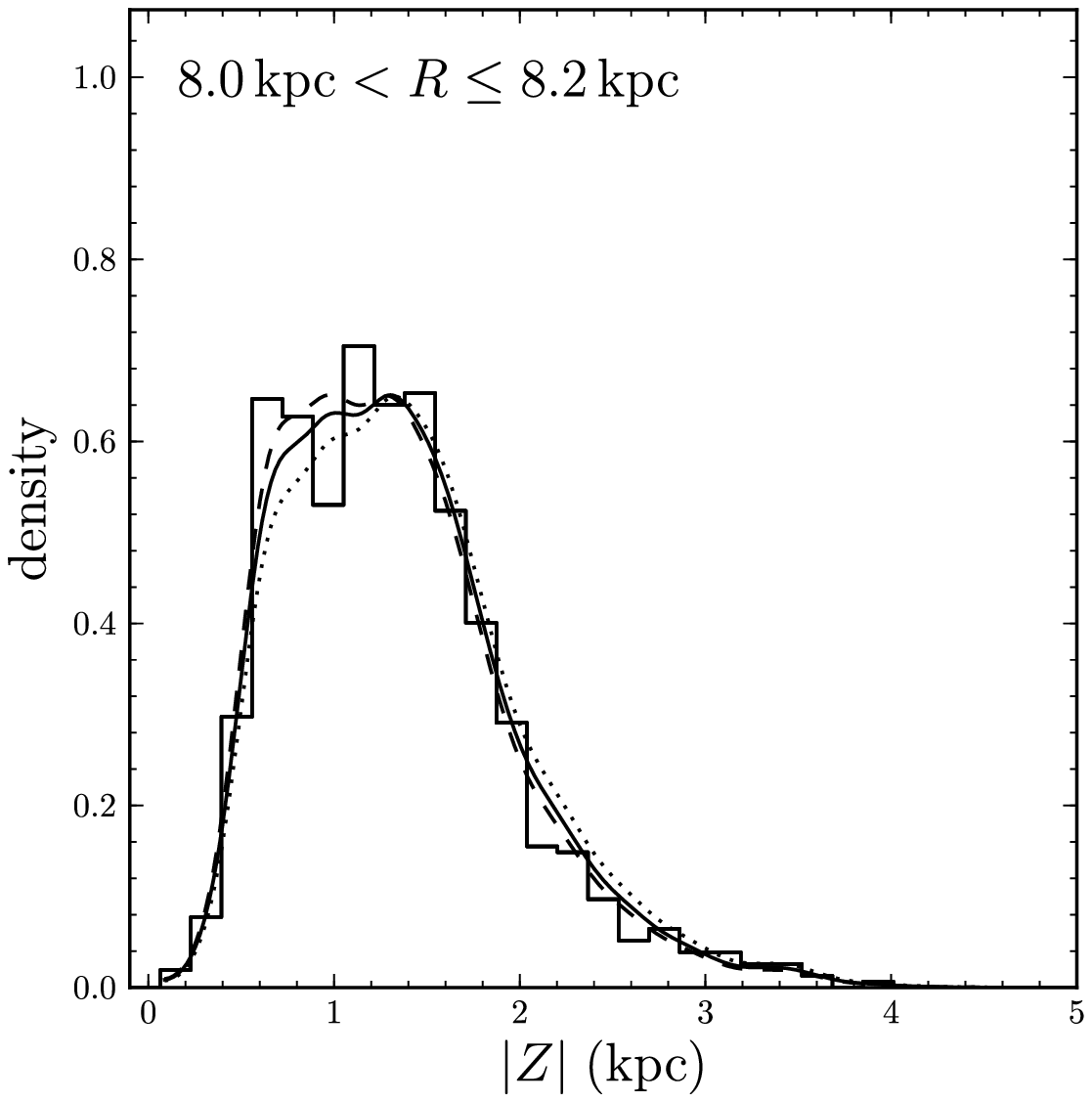}
  \includegraphics[width=0.23\textwidth,clip=]{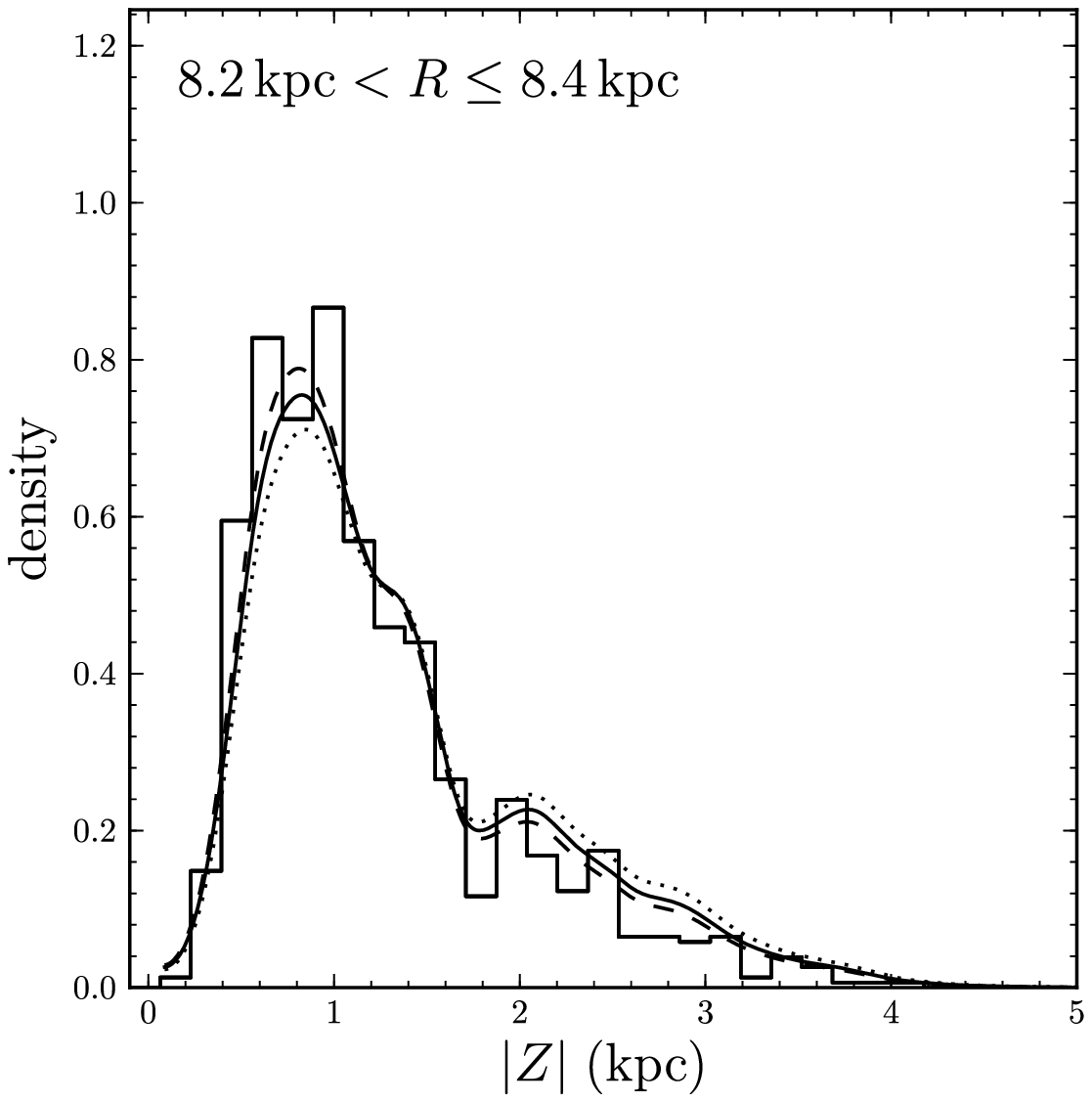}
  \includegraphics[width=0.23\textwidth,clip=]{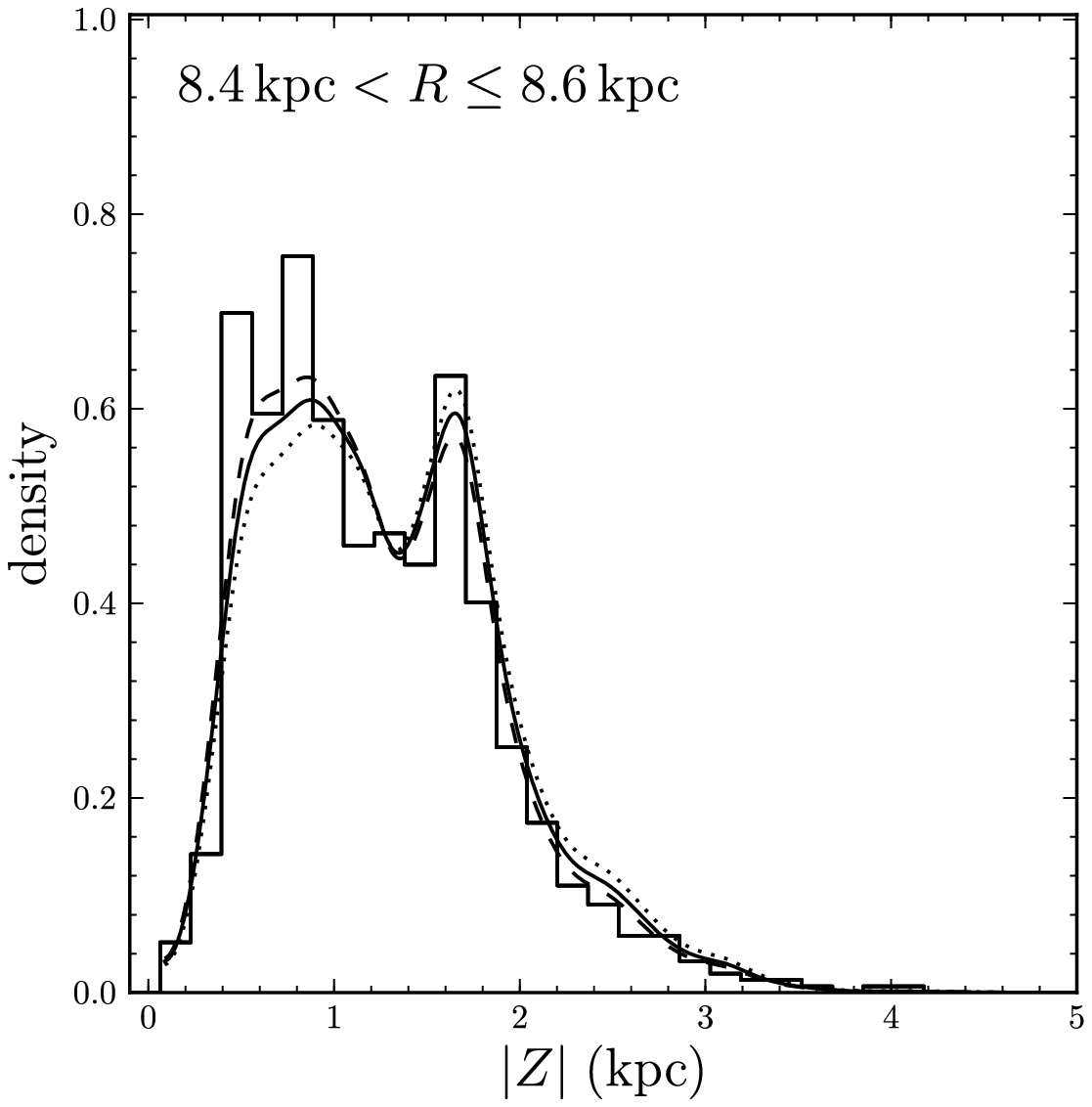}
  \includegraphics[width=0.23\textwidth,clip=]{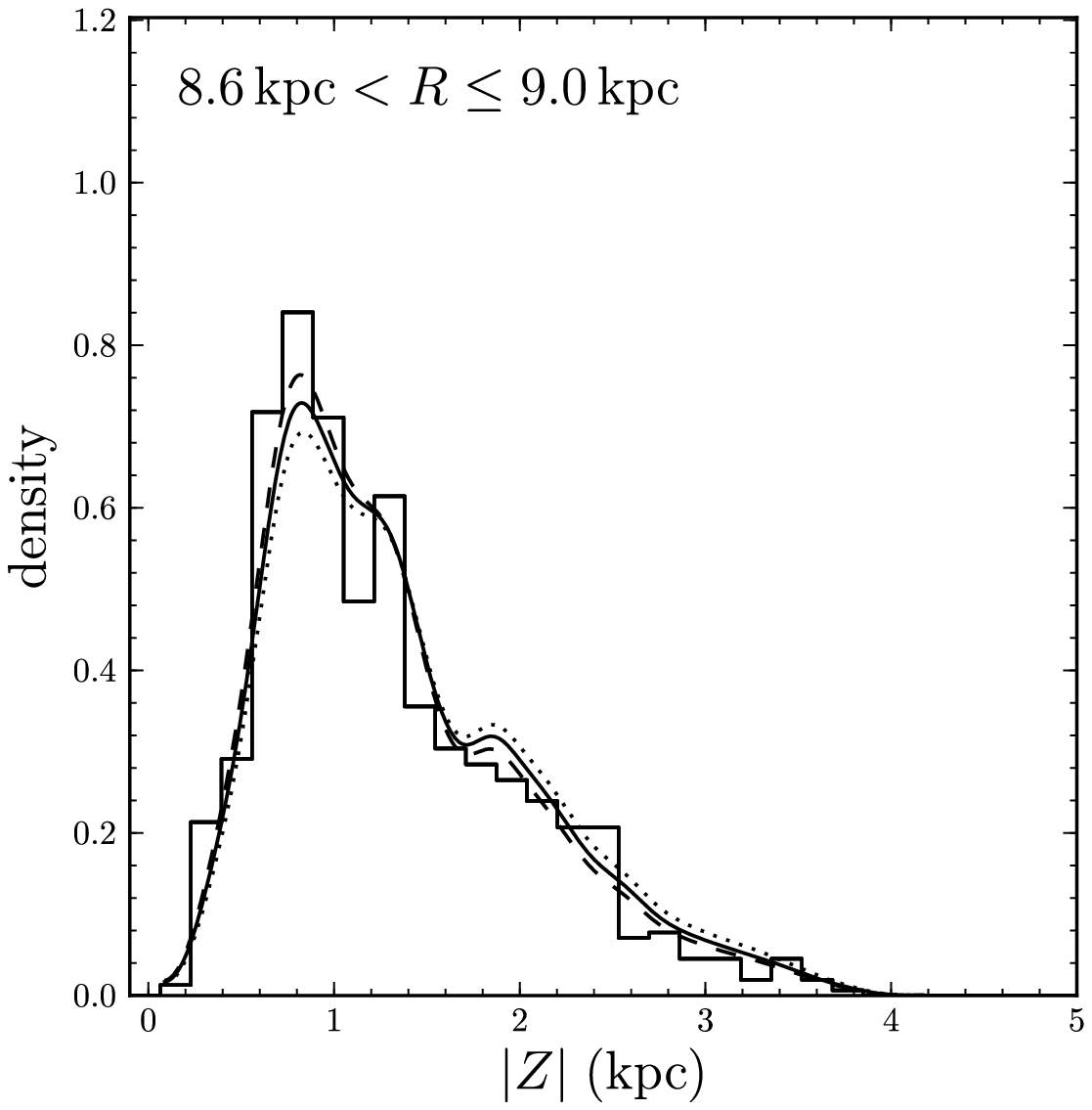}\\
  \includegraphics[width=0.23\textwidth,clip=]{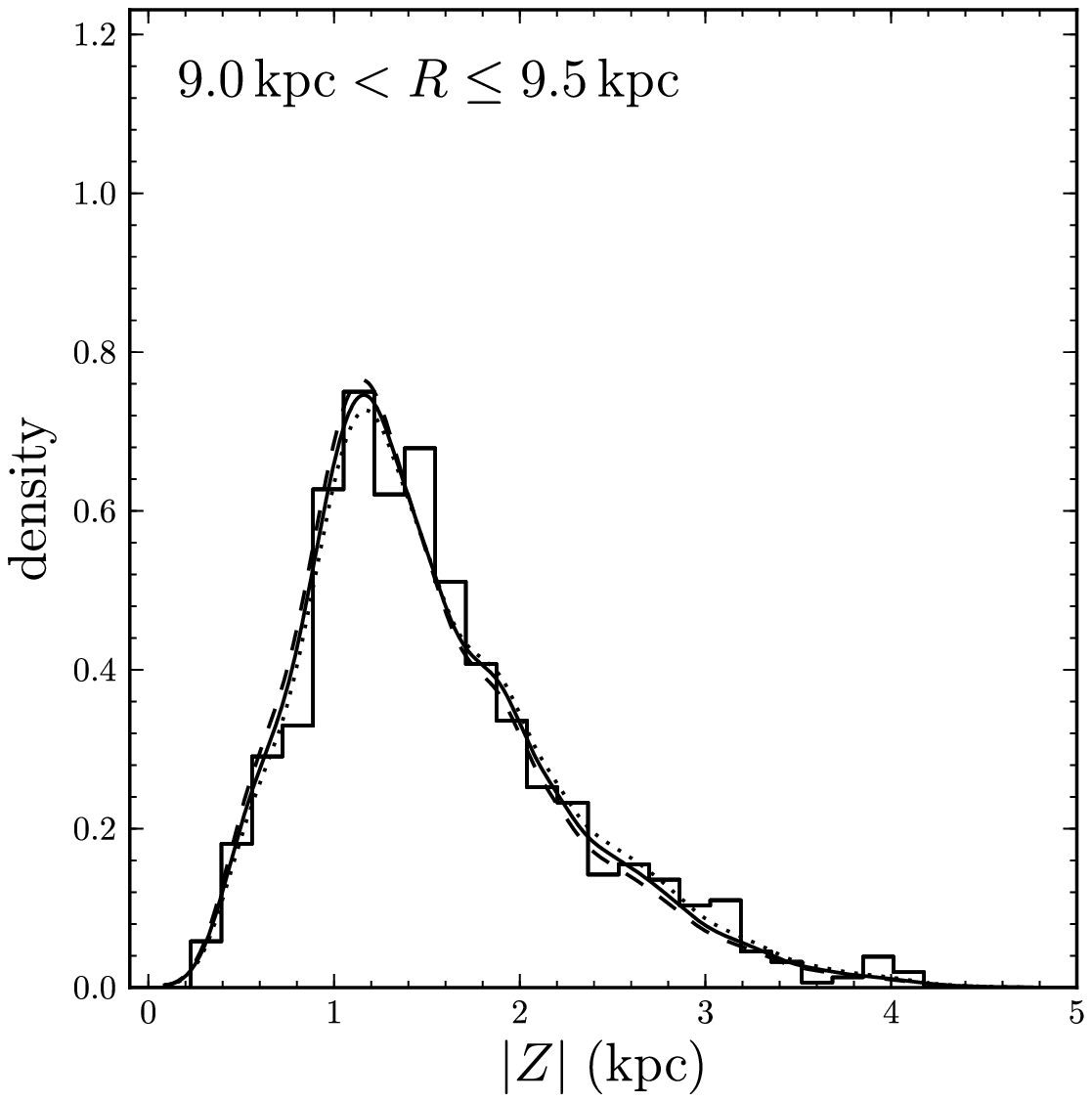}
  \includegraphics[width=0.23\textwidth,clip=]{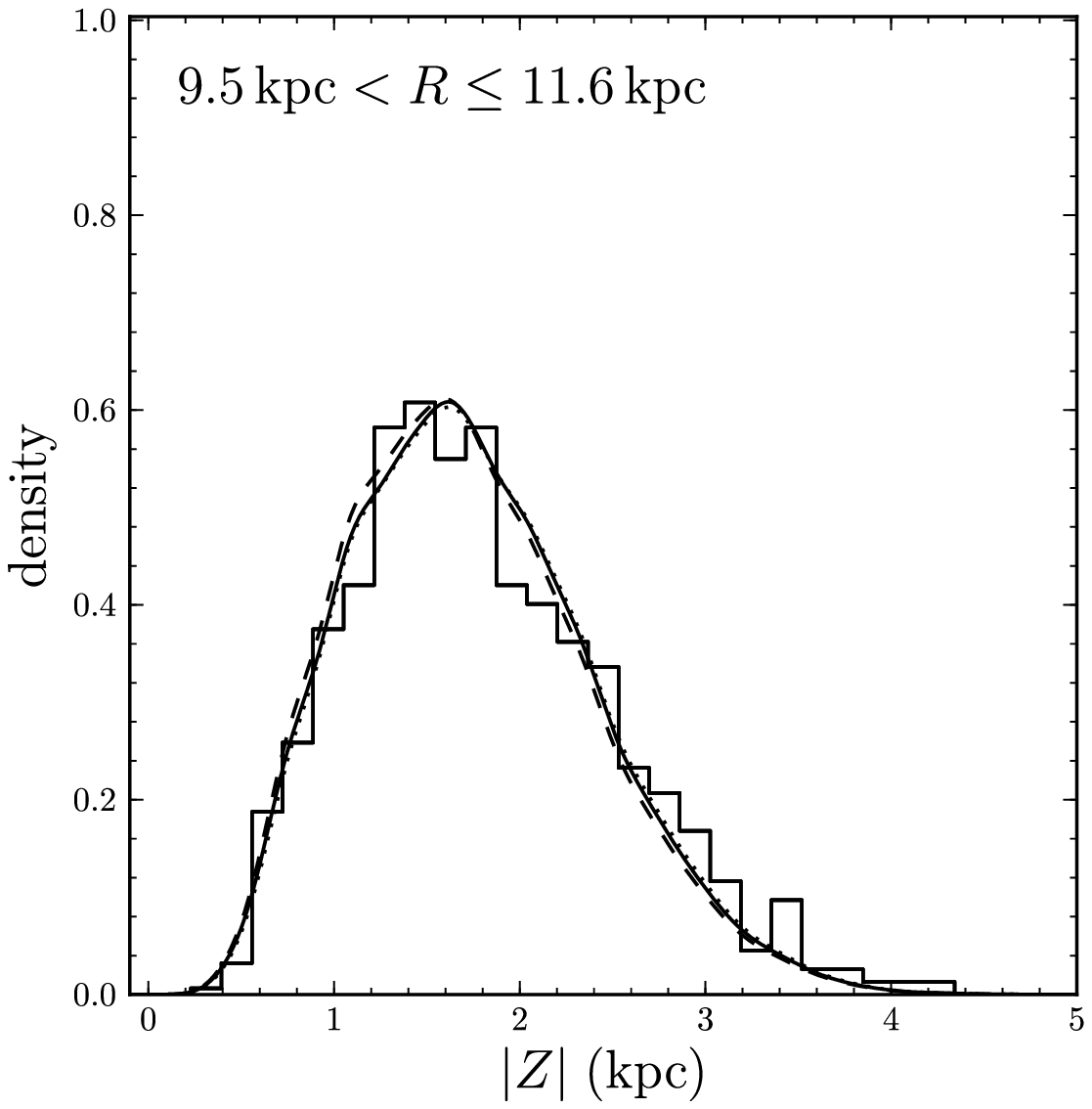}
  \caption{Comparison between the vertical distribution of the data
    and the best-fit dynamical model for \aenhanced\ G-dwarf \map
    s. The \aenhanced\ \map s in this Figure are all of those with
    $\afe > 0.3$; these all have best-fit radial scale lengths from
    \bo d around $2\kpc$. The predictions are calculated for each
    \map\ separately, based on each \map 's best-fit qDF and potential
    parameters, and they are combined to provide better statistics for
    the comparison (because each \map\ only contains a few hundred
    data points). The comparison is shown for 10 different radial bins
    that each contain about 1000 data points. The dashed and dotted
    lines are alternative models that have $\sigoneone(6\kpc) =
    192\,M_\odot\pc^{-2}$ and $\sigoneone(6\kpc) =
    108\,M_\odot\pc^{-2}$, respectively, choosing the best-fitting qDF
    parameters for each \map\ corresponding to those potentials. The
    vertical distribution of the data is excellently matched for all
    radial bins for all three models, indicating that the vertical
    density distribution of the data is so informative that it has to
    be matched by all potential models, regardless of whether the
    velocity dispersion is matched (see
    \figurename~\ref{fig:datamodel_vzdist_aenhanced}). The
    best-fitting model is essentially that for which the best-fitting
    qDF parameters for the tracer density profile also provide the
    best-fit for the velocity
    distribution.}\label{fig:datamodel_zdist_aenhanced}
\end{figure}

\begin{figure}[t!]
  \includegraphics[width=0.23\textwidth,clip=]{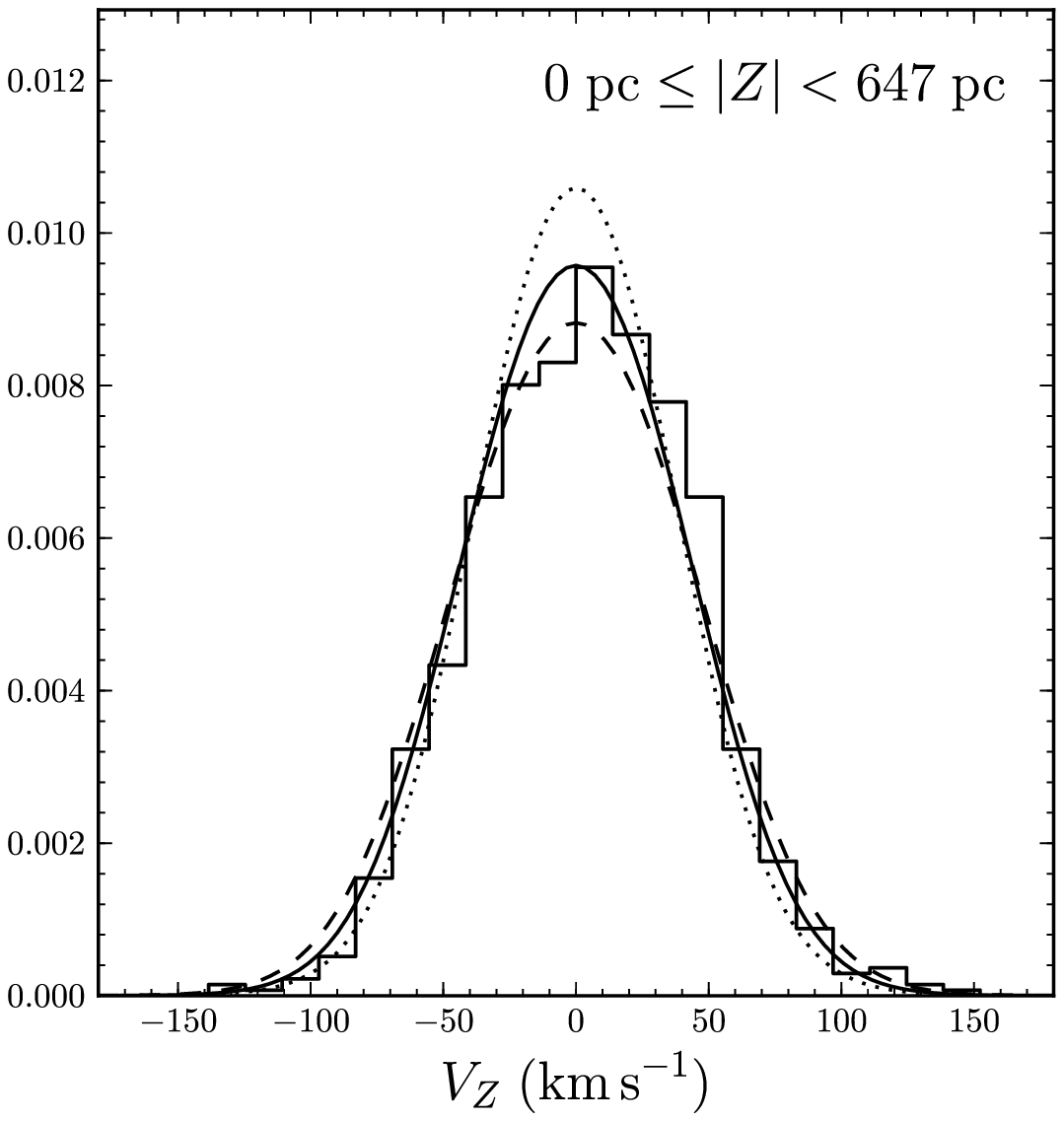}
  \includegraphics[width=0.23\textwidth,clip=]{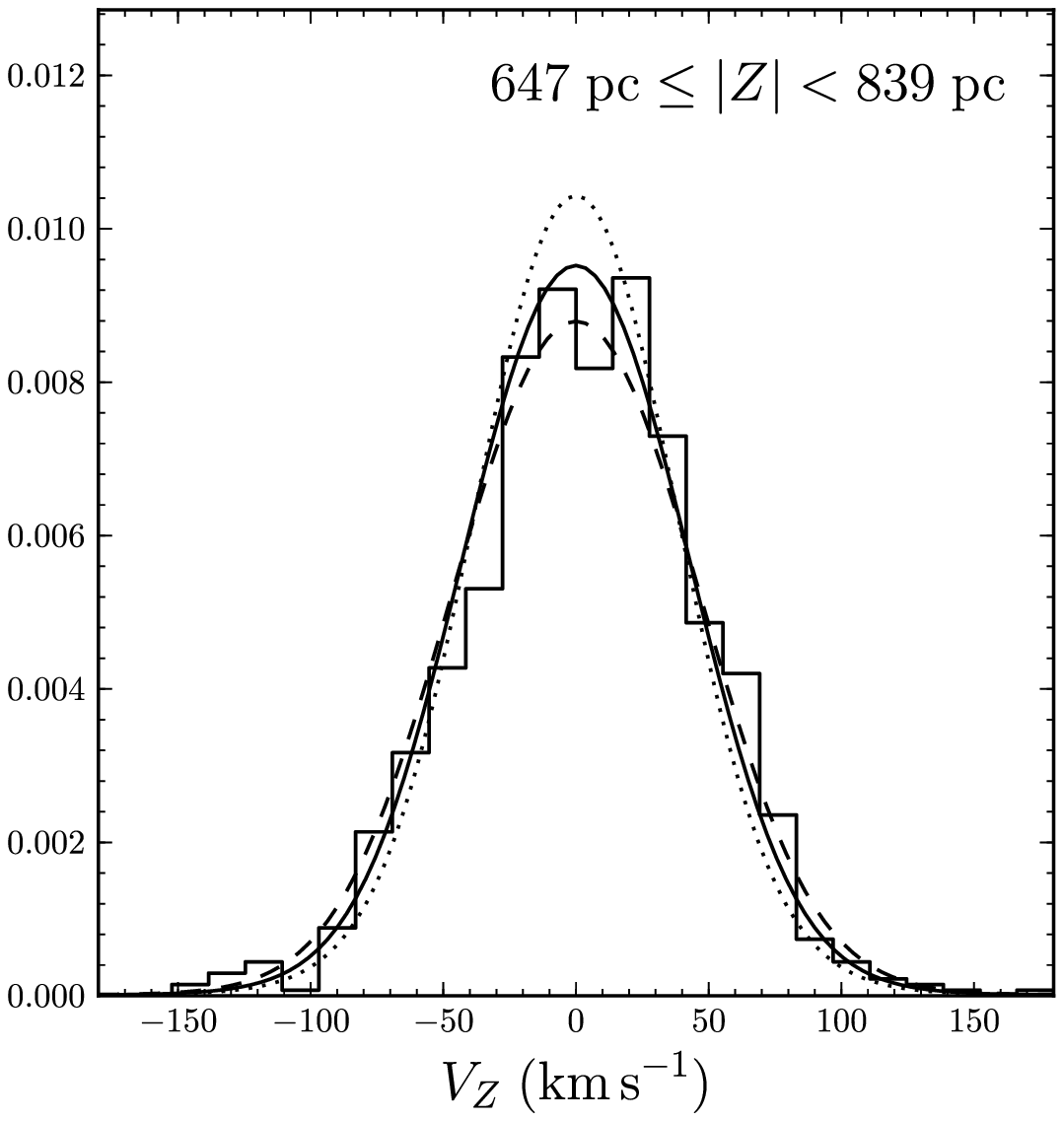}
  \includegraphics[width=0.23\textwidth,clip=]{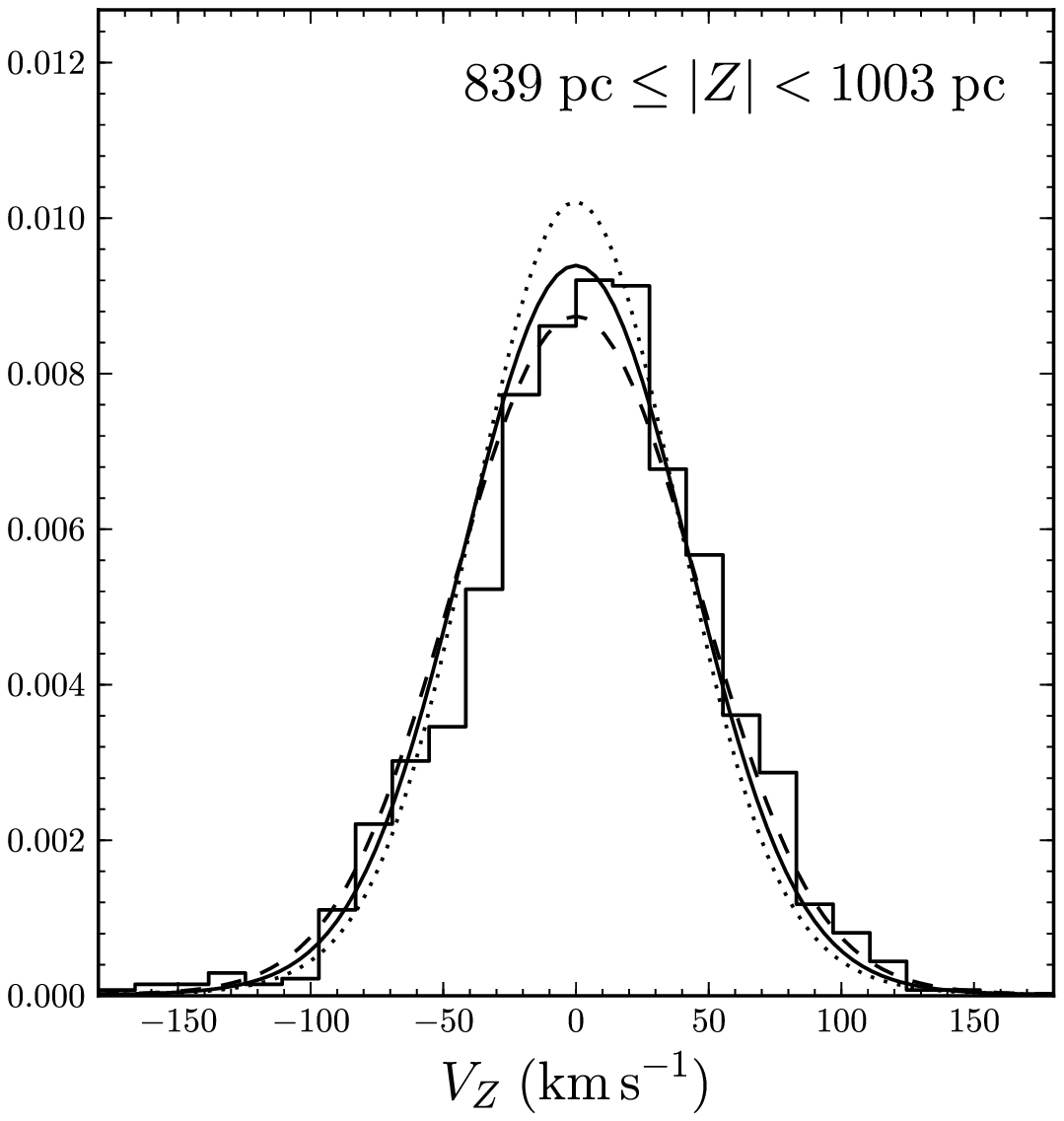}
  \includegraphics[width=0.23\textwidth,clip=]{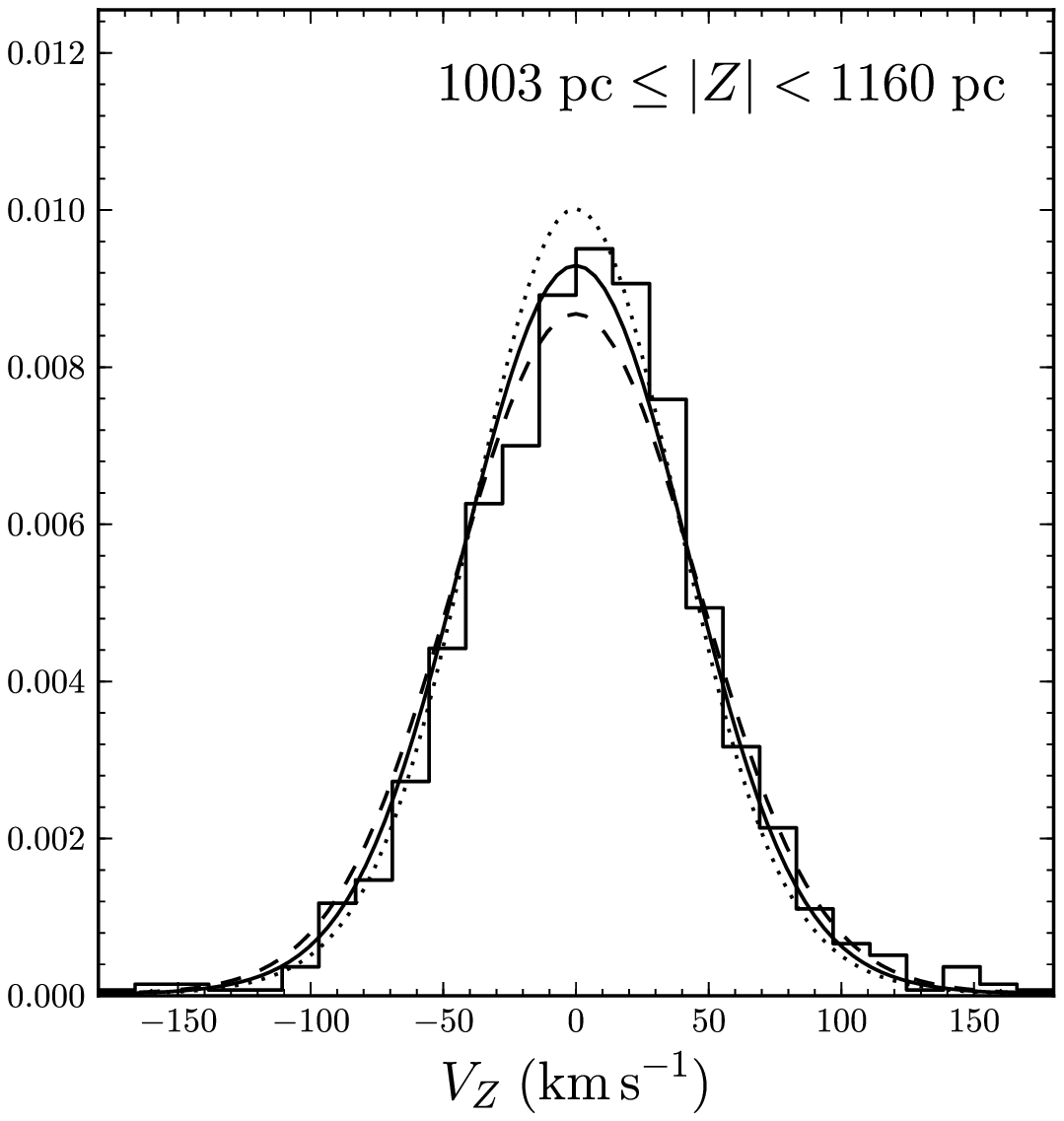}\\
  \includegraphics[width=0.23\textwidth,clip=]{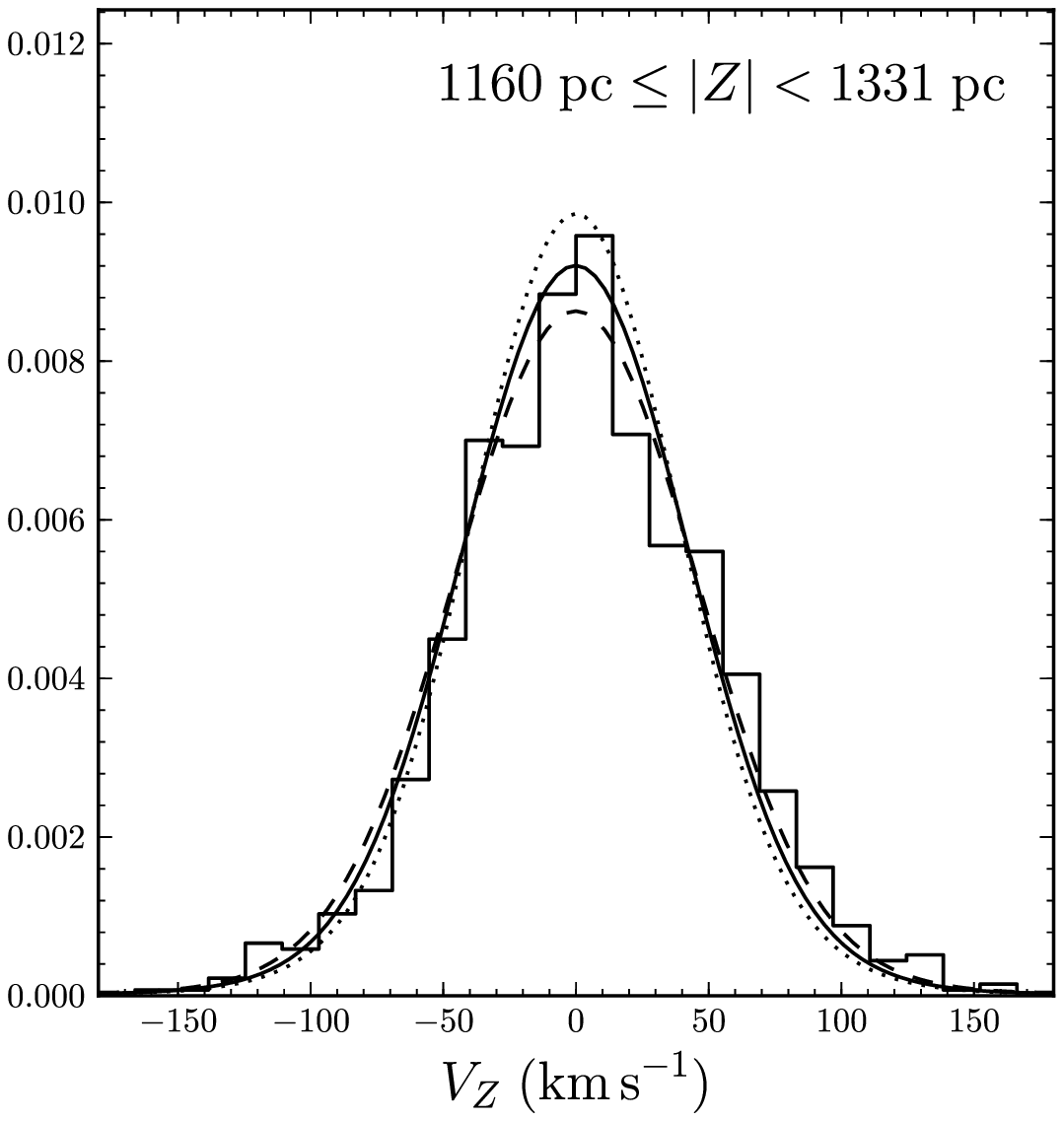}
  \includegraphics[width=0.23\textwidth,clip=]{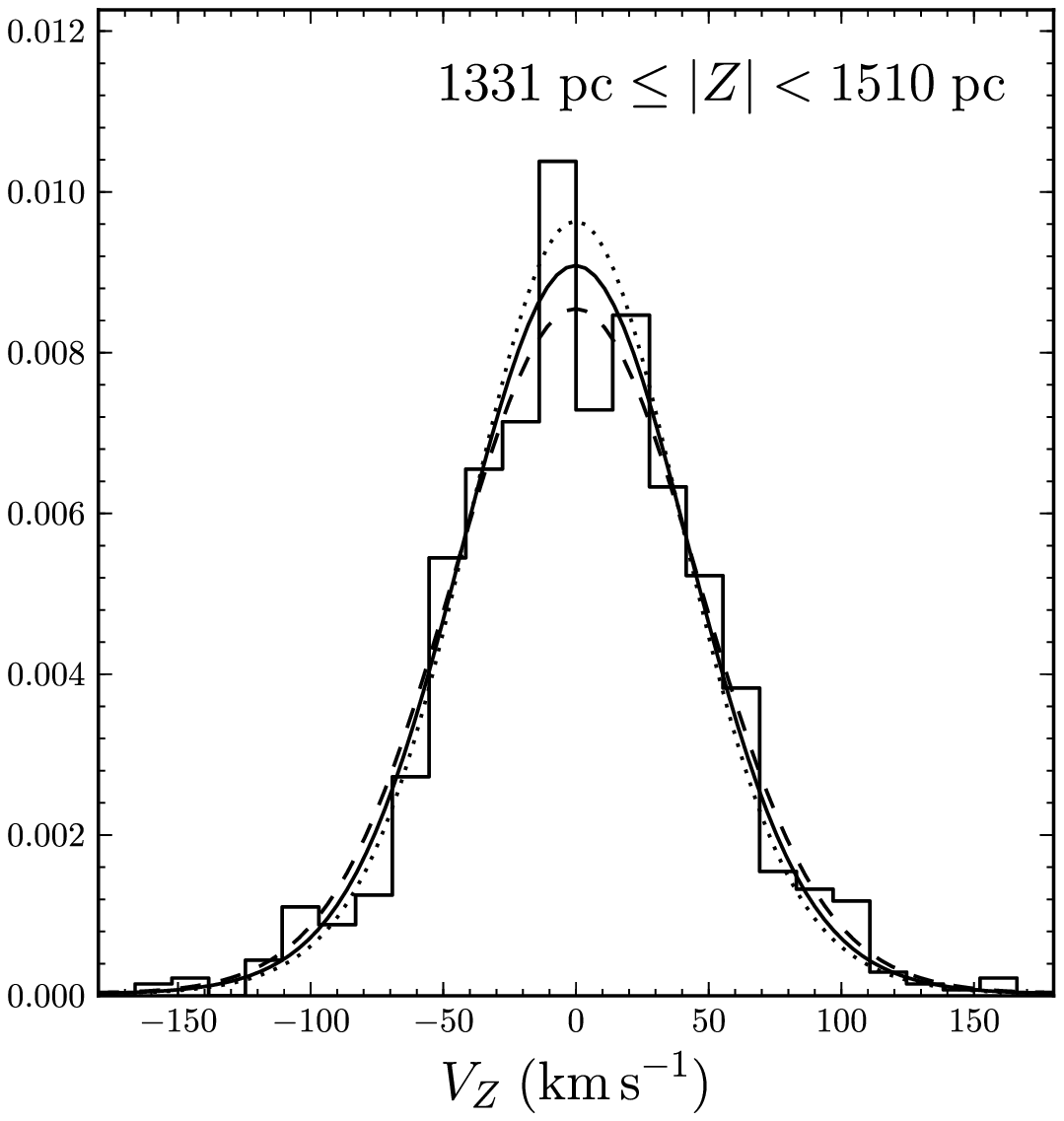}
  \includegraphics[width=0.23\textwidth,clip=]{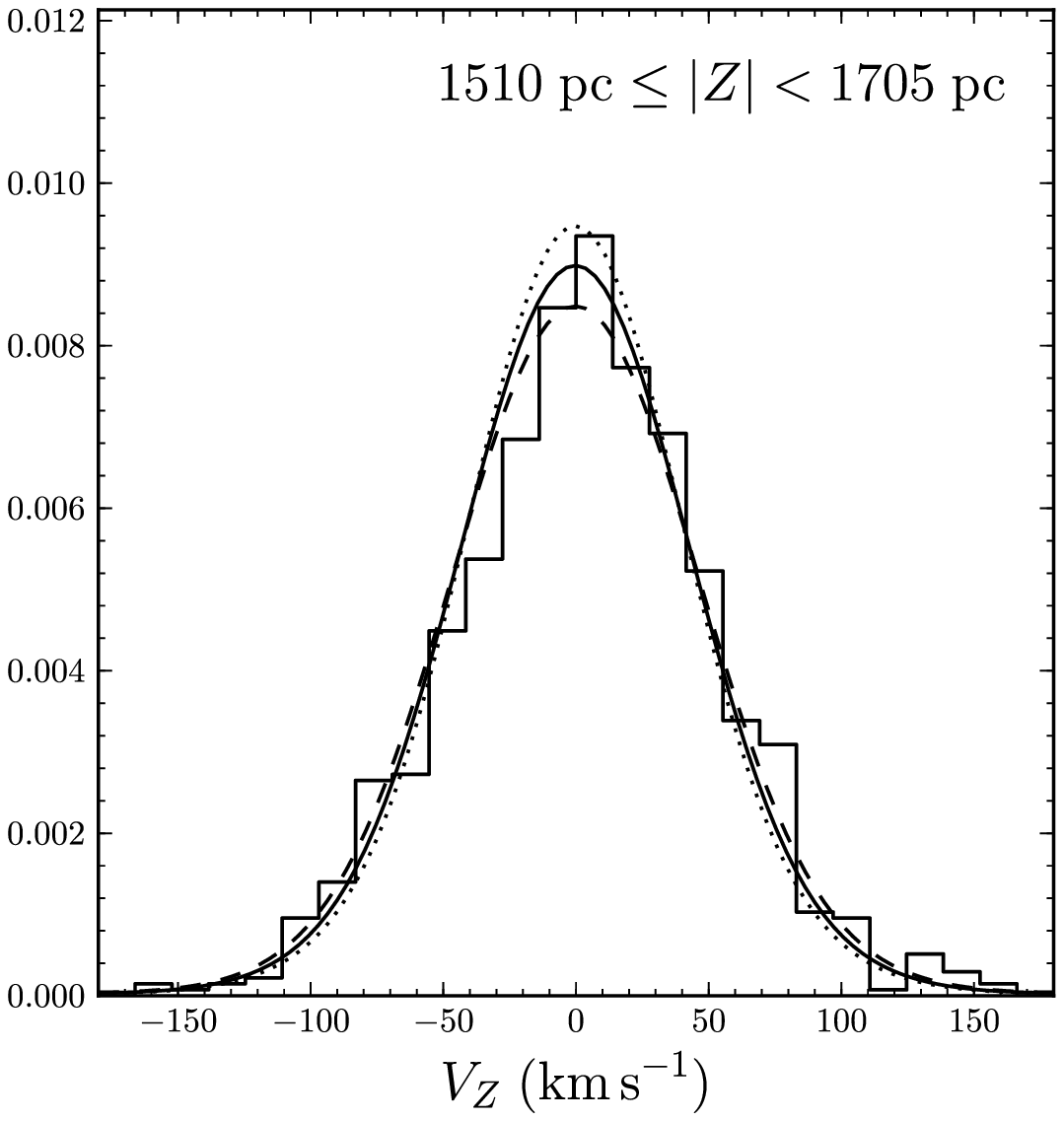}
  \includegraphics[width=0.23\textwidth,clip=]{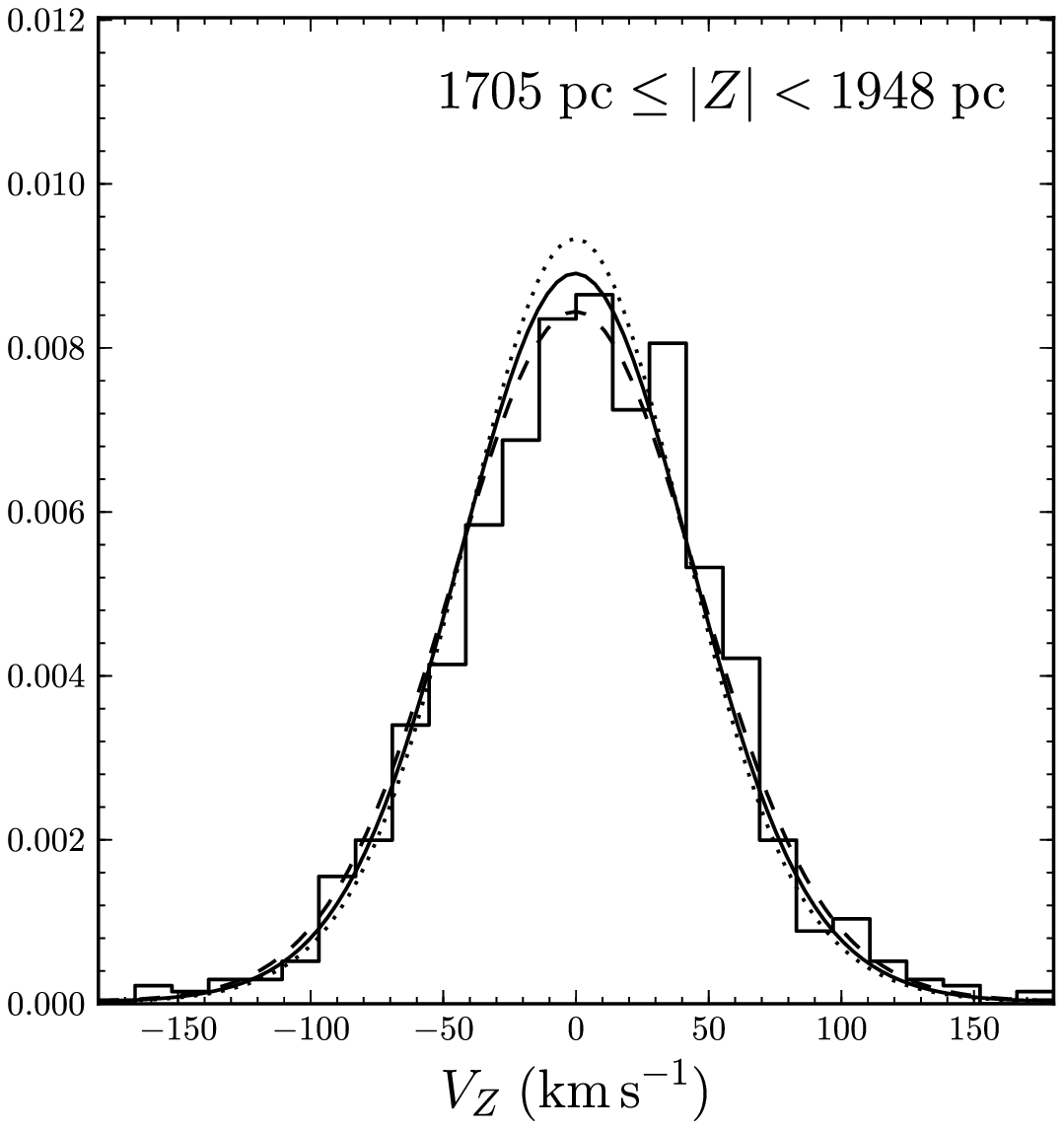}\\
  \includegraphics[width=0.23\textwidth,clip=]{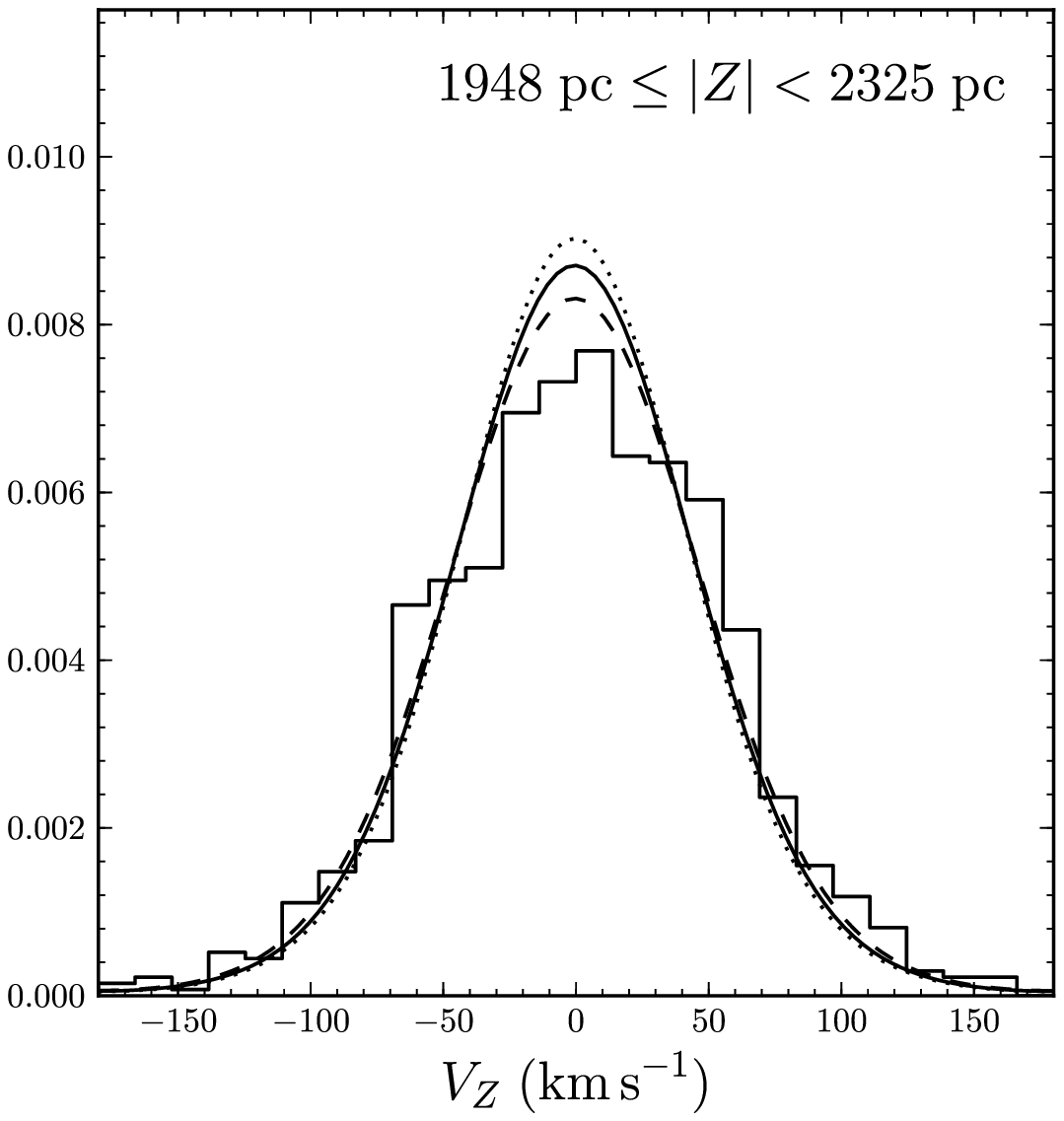}
  \includegraphics[width=0.23\textwidth,clip=]{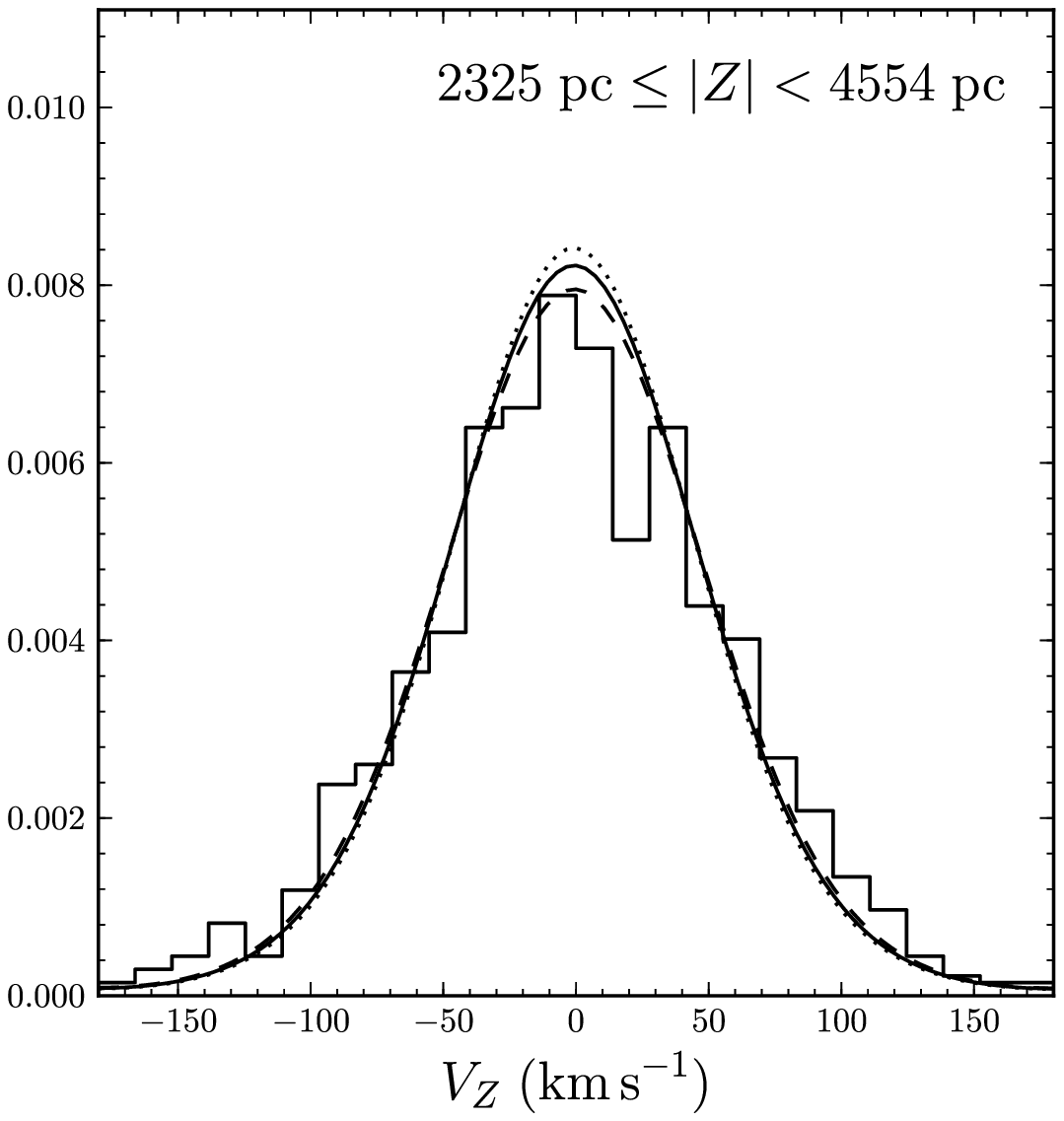}
  \caption{Same as \figurename~\ref{fig:datamodel_zdist_aenhanced},
    but for the distribution of vertical velocities as a function of
    height.}\label{fig:datamodel_vzdist_aenhanced}
\end{figure}

\section{Calculation of the effective survey volume}\label{sec:normint}

In this Appendix we describe how we can efficiently calculate the
8-dimensional normalization integral (the effective survey volume) in
the likelihood in \equationname~(\ref{eq:dflike2}). This calculation
is similar to that in the case of the density fits in \bo d as
described in their Appendix B, and we start from their
\equationname~(B4), re-written to include the velocity integration
\begin{equation}\label{eq:normint4}
\begin{split}
\int & \dd l\,\dd b\,\dd r\,\dd D\, \dd \vecv \,\dd (g-r) \,\dd \feh
\,\lambda(l,b,D,\vecv,r,g-r,\feh|\paramspot,\paramsdf) \\ &=
A_p\,\sum_{\mathrm{plates}\ p}
\sum_{D}\,D^2\,\dens(R,z|l,b,D,\paramspot,\paramsdf)\,\\ & \qquad \sum_{g-r}
\,\sum_{\feh}\,\rho^c(g-r)\,\rho^{\feh}(\feh)\,
S(p,r[g-r,\feh,D],g-r)\,,
\end{split}
\end{equation}
where
\begin{equation}\label{eq:normintdens}
\dens(R,z|l,b,D,\paramspot,\paramsdf) = \int \dd \vecv
\df(\vecj[\vecx,\vecv])|l,b,D,\paramspot,\paramsdf) \,,
\end{equation}
is the spatial density predicted by the DF.

To evaluate the expression in \equationname~(\ref{eq:normint4})
efficiently, we split the calculation in two steps: (1) We calculate
the predicted spatial density in \equationname~(\ref{eq:normintdens})
on a grid in $(R,Z)$ and interpolate it; (2) we use the interpolated
density to evaluate the sums in \equationname~(\ref{eq:normint4}). We
perform the velocity integration in
\equationname~(\ref{eq:normintdens}) by using 20-th order
Gauss-Legendre integration in each direction: from 0 to $3V_c(R_0)/2$
for the tangential velocity and from $-4\,\sigma$ to $4\,\sigma$ for
the radial and vertical velocity. `$\sigma$' is calculated using the
scale dispersion profiles of the DF (see \equationname
s~[\ref{eq:inputprofile}-\ref{eq:inputprofile2}])). We calculate the
density on a grid of $16\times16$ points ranging from 4 to $15\kpc$ in
$R$ and from $0$ to $4\,R_0/5$ in $Z$ and interpolate using 3-th order
two-dimensional spline interpolation.

\section{Detailed data versus model comparisons}\label{sec:datamodel}

In this Appendix we describe the results from detailed comparisons
between the best-fit dynamical models for individual \map s obtained
using the methodology described in \sectionname~\ref{sec:method} and
the SEGUE data. As in \bo d, we do this in a space close to that of
the raw data: star counts uncorrected for selection effects. Even
though our model is a full generative model that could predict the
distribution of any combination of $r,g-r,\feh,\vecx,\vecv$ to be
compared with the data, because we are mainly concerned with measuring
$\sigoneone(R)$, we focus here on (a) vertical number counts and (b)
the distribution of vertical velocities, both as a function of radius,
as these are the main data ingredients for constraining
$\sigoneone(R)$. As described in \sectionname~\ref{sec:method}, we fit
individual \map s that typically only have a few hundred stars, making
visual comparisons between the data and the model highly susceptible
to the influence of Poisson noise. For this reason and because showing
comparisons for 43 different sub-samples would take up the better part
of an ApJ volume, we group \map s into three main divisions with
similar DFs. These are (a) \aenhanced\ \map s, \ie, all \map s with
$\afe > 0.3\dex$; these \map s all have short radial scale lengths
around $2\kpc$, and they are vertically thick. (b) \apoor\ \map s,
\ie, those with $\feh > -0.1\dex$; these are vertically thin and have
longer scale lengths. Finally, we consider \map s intermediate between
(a) and (b), which are all of the \map s with $\afe < 0.3\dex$ and
$\feh < -0.1\dex$. We calculate the predicted number counts for each
\map\ within a sub-division separately and plot the sum of all
predictions, weighted by relative number of data points in different
\map s.

\begin{figure}[t!]
  \includegraphics[width=0.195\textwidth,clip=]{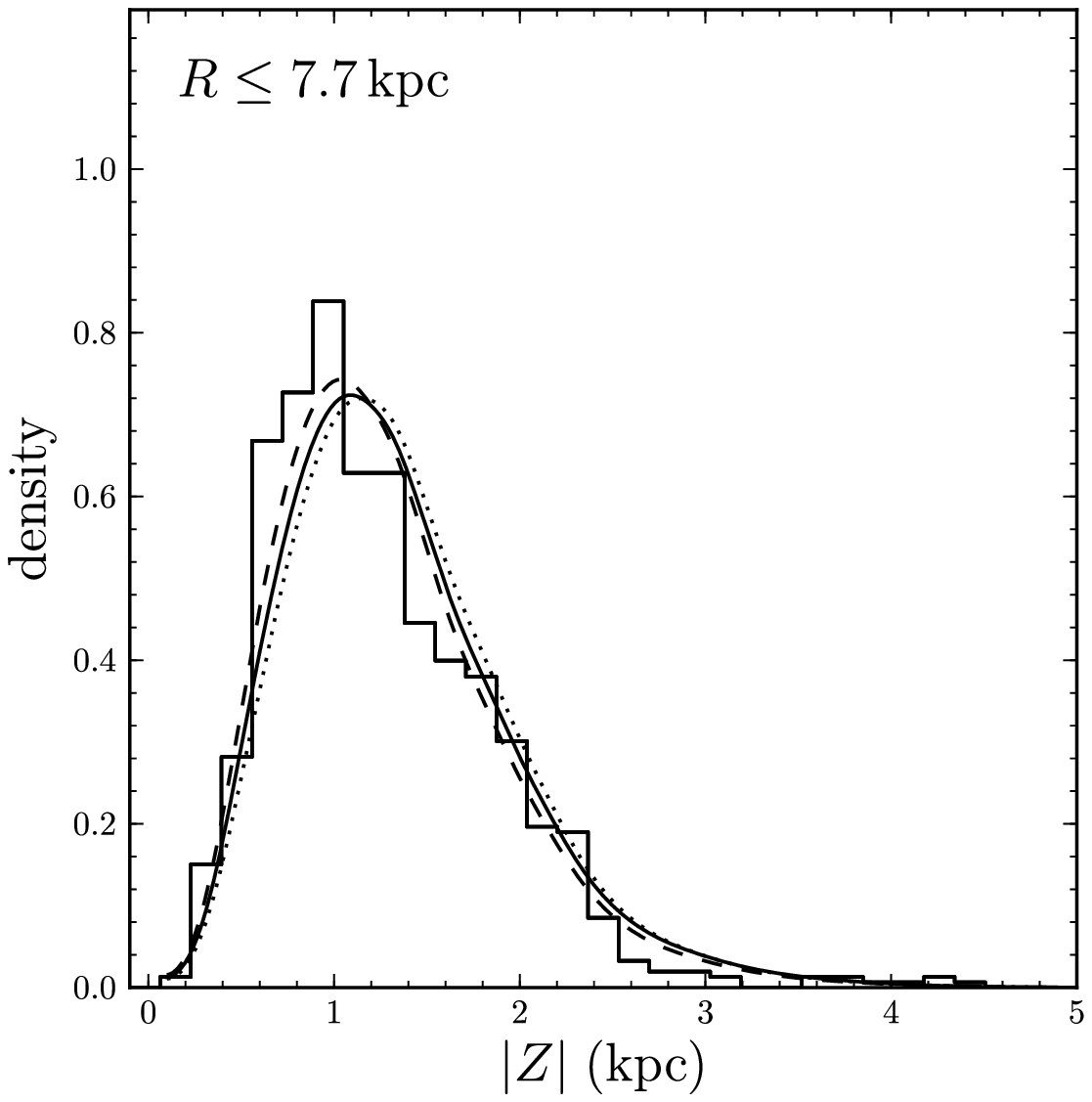}
  \includegraphics[width=0.195\textwidth,clip=]{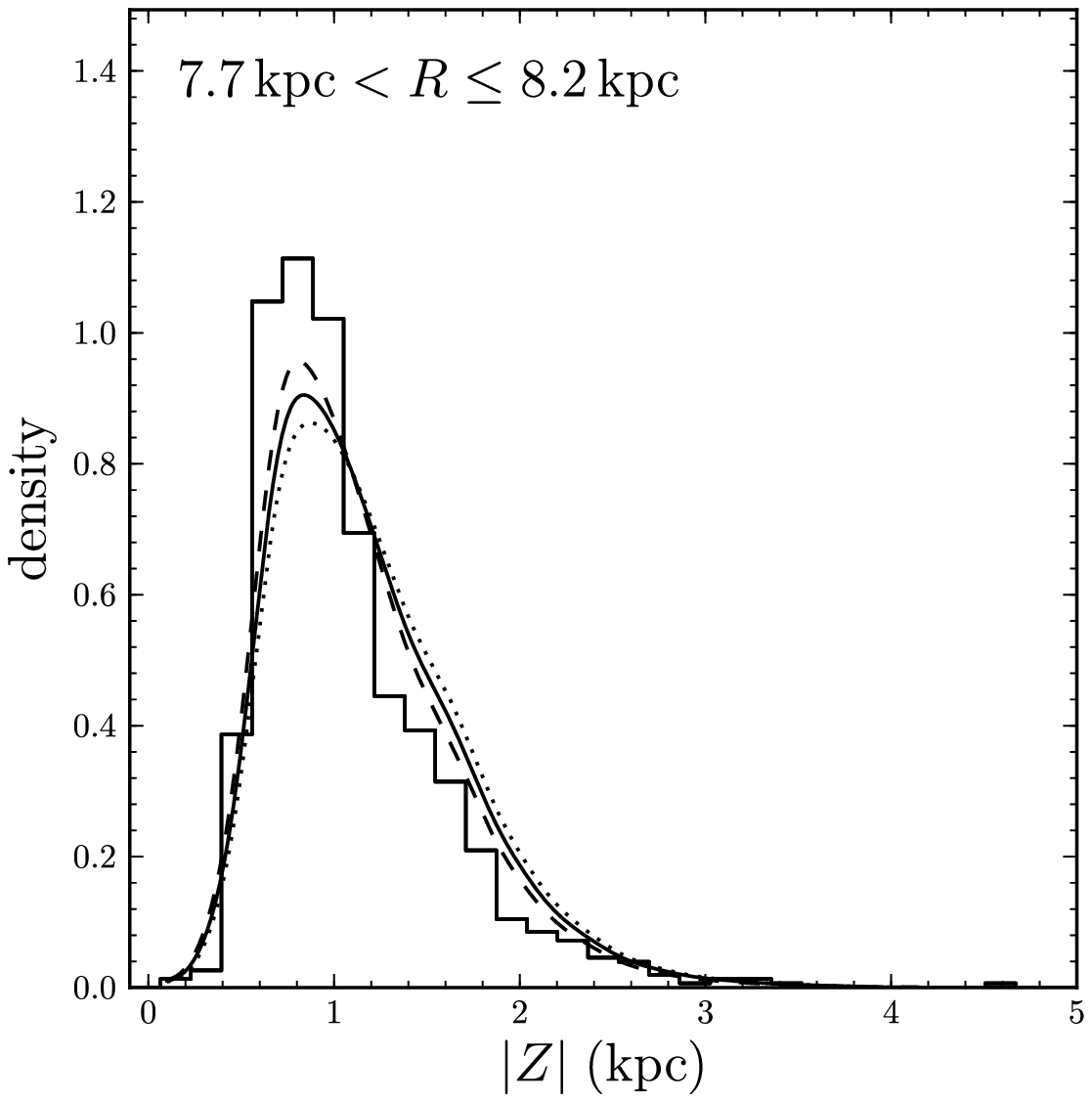}
  \includegraphics[width=0.195\textwidth,clip=]{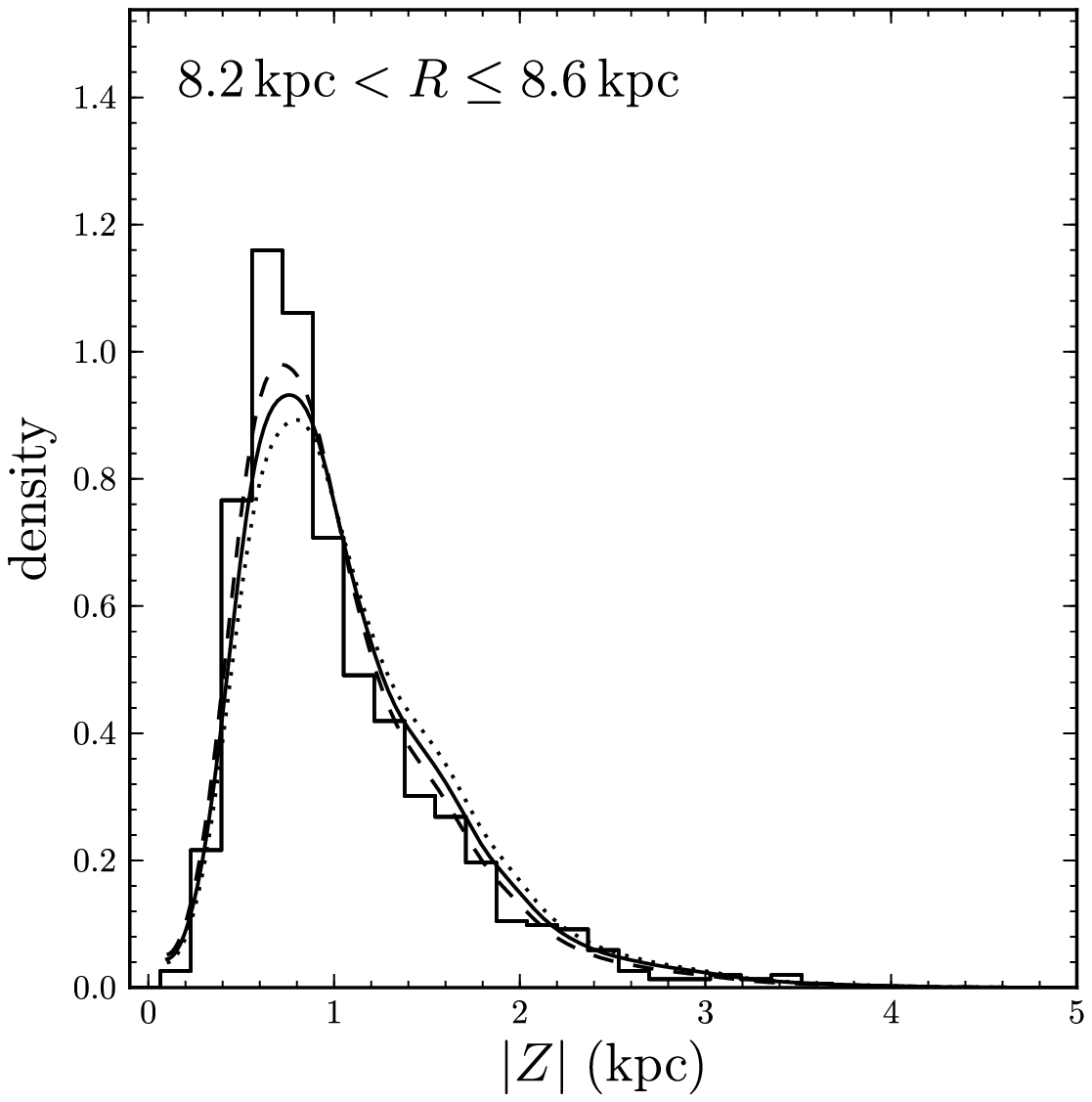}
  \includegraphics[width=0.195\textwidth,clip=]{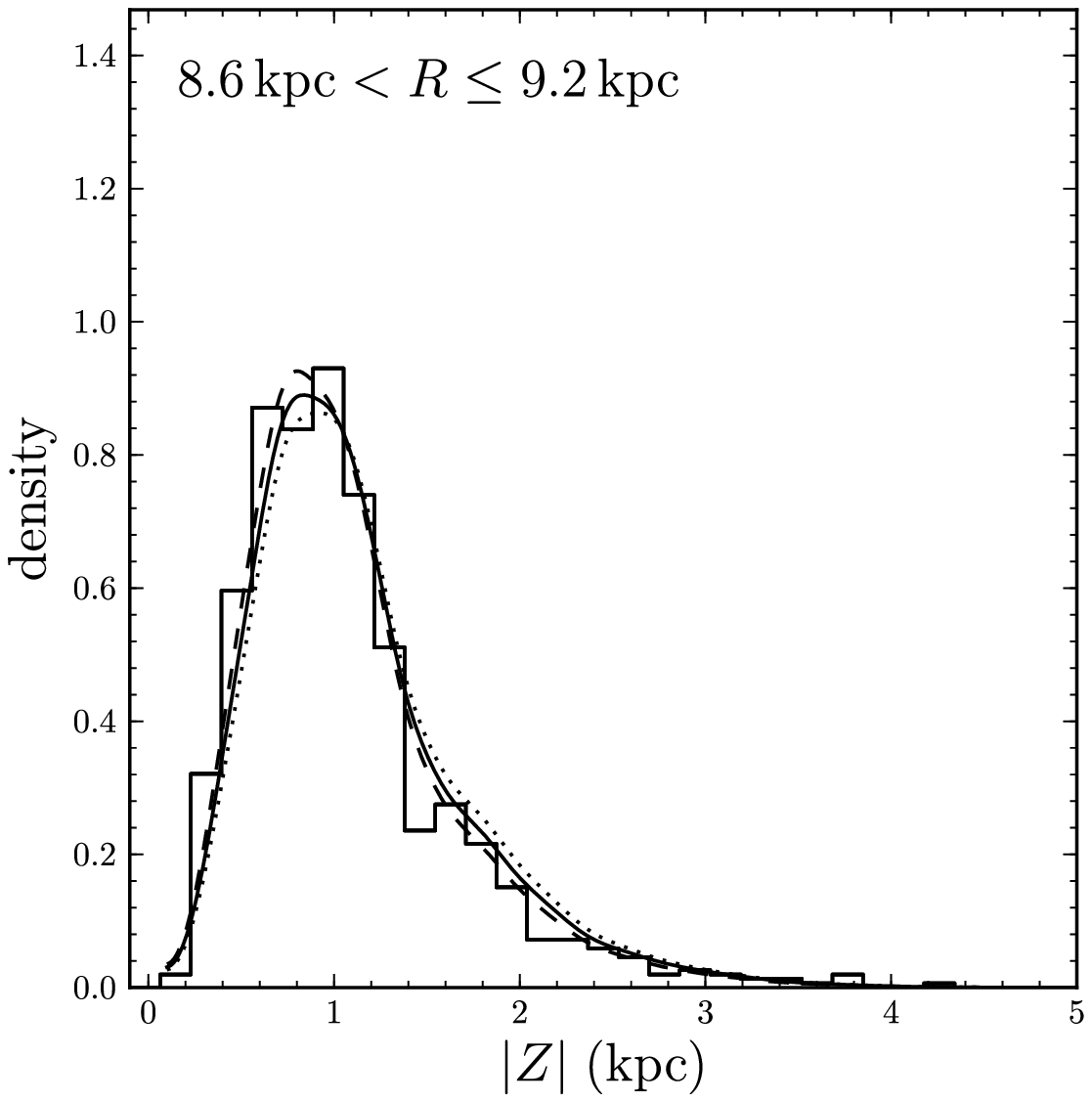}
  \includegraphics[width=0.195\textwidth,clip=]{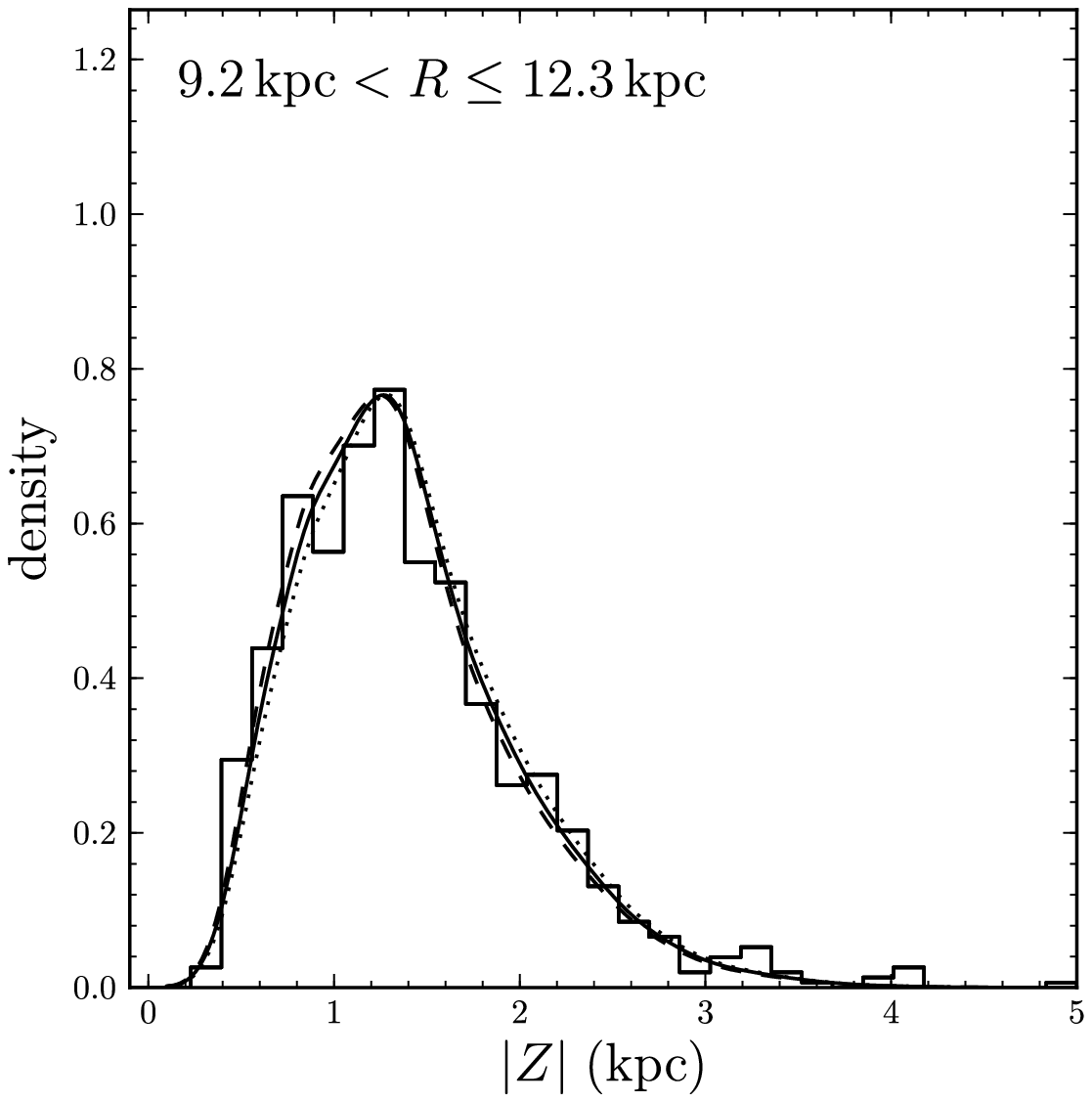}
  \caption{Same as \figurename~\ref{fig:datamodel_zdist_aenhanced},
    but for intermediate \map s ($\afe < -0.3$ and $\feh < -0.1$). The
    alternative dashed and dotted models now correspond to
    $\sigoneone(7\kpc) = 138\,M_\odot\pc^{-2}$ and $\sigoneone(7\kpc)
    = 65\,M_\odot\pc^{-2}$.}\label{fig:datamodel_zdist_aintermediate}
\end{figure}

\begin{figure}[t!]
  \includegraphics[width=0.195\textwidth,clip=]{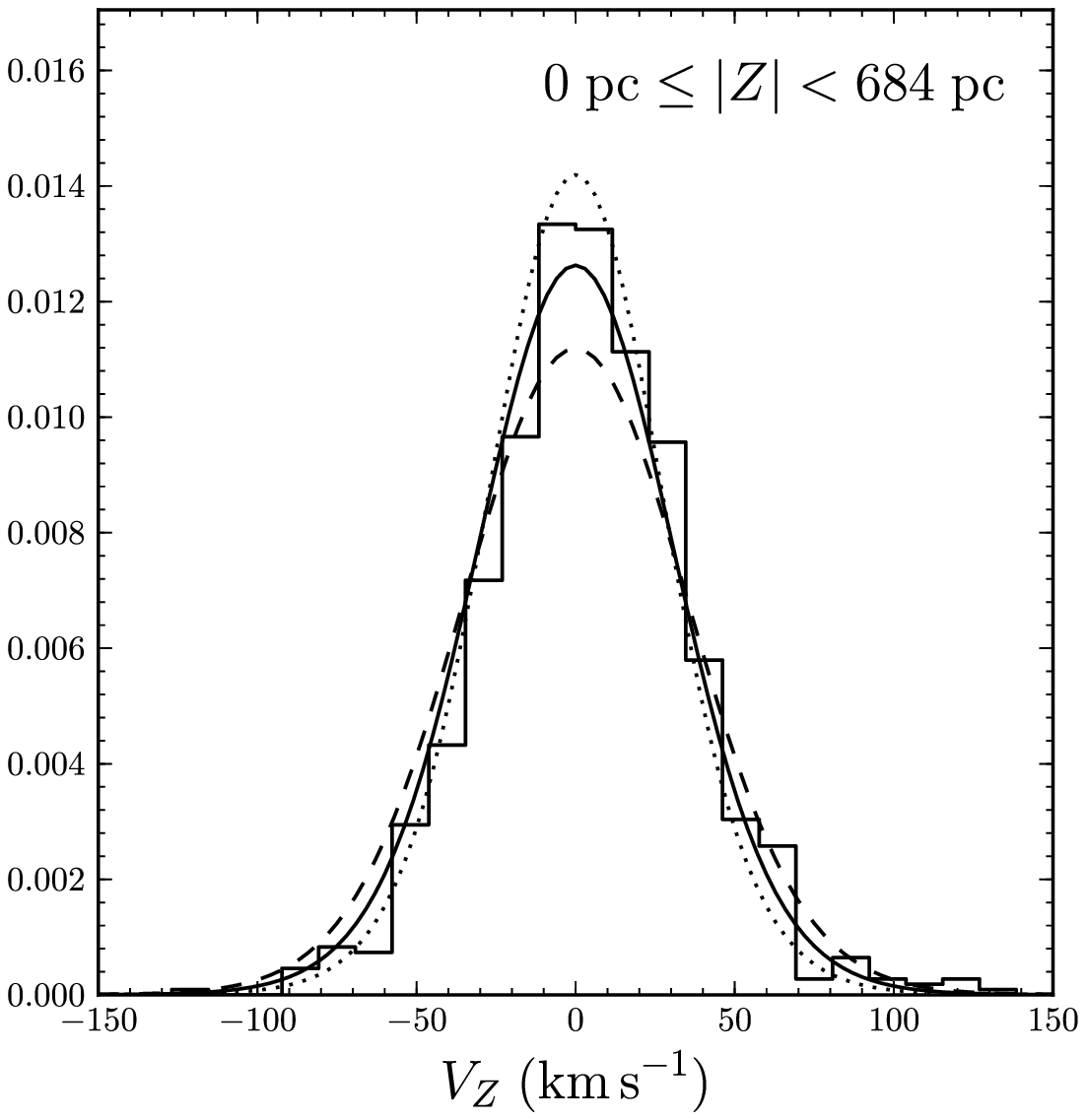}
  \includegraphics[width=0.195\textwidth,clip=]{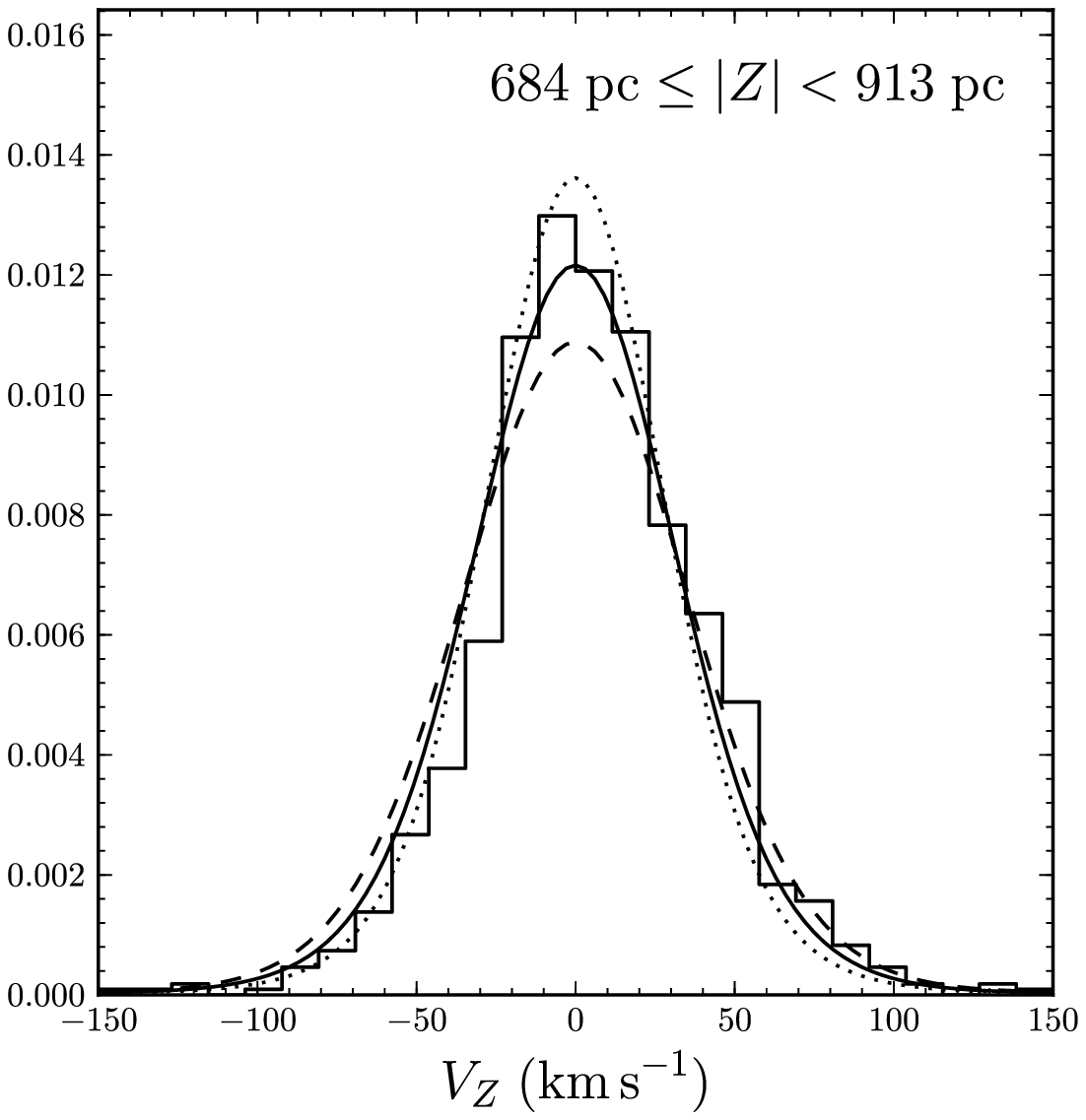}
  \includegraphics[width=0.195\textwidth,clip=]{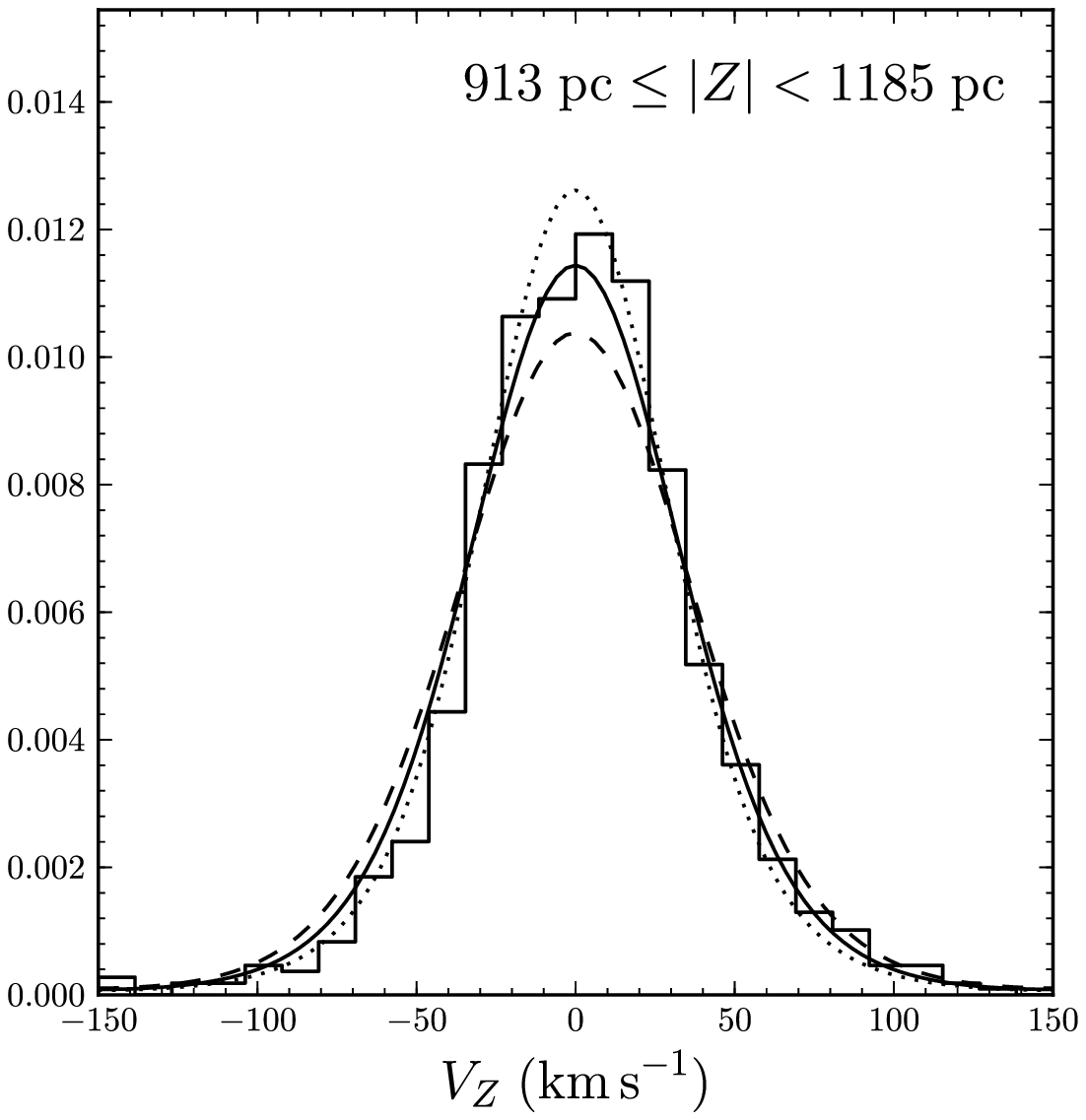}
  \includegraphics[width=0.195\textwidth,clip=]{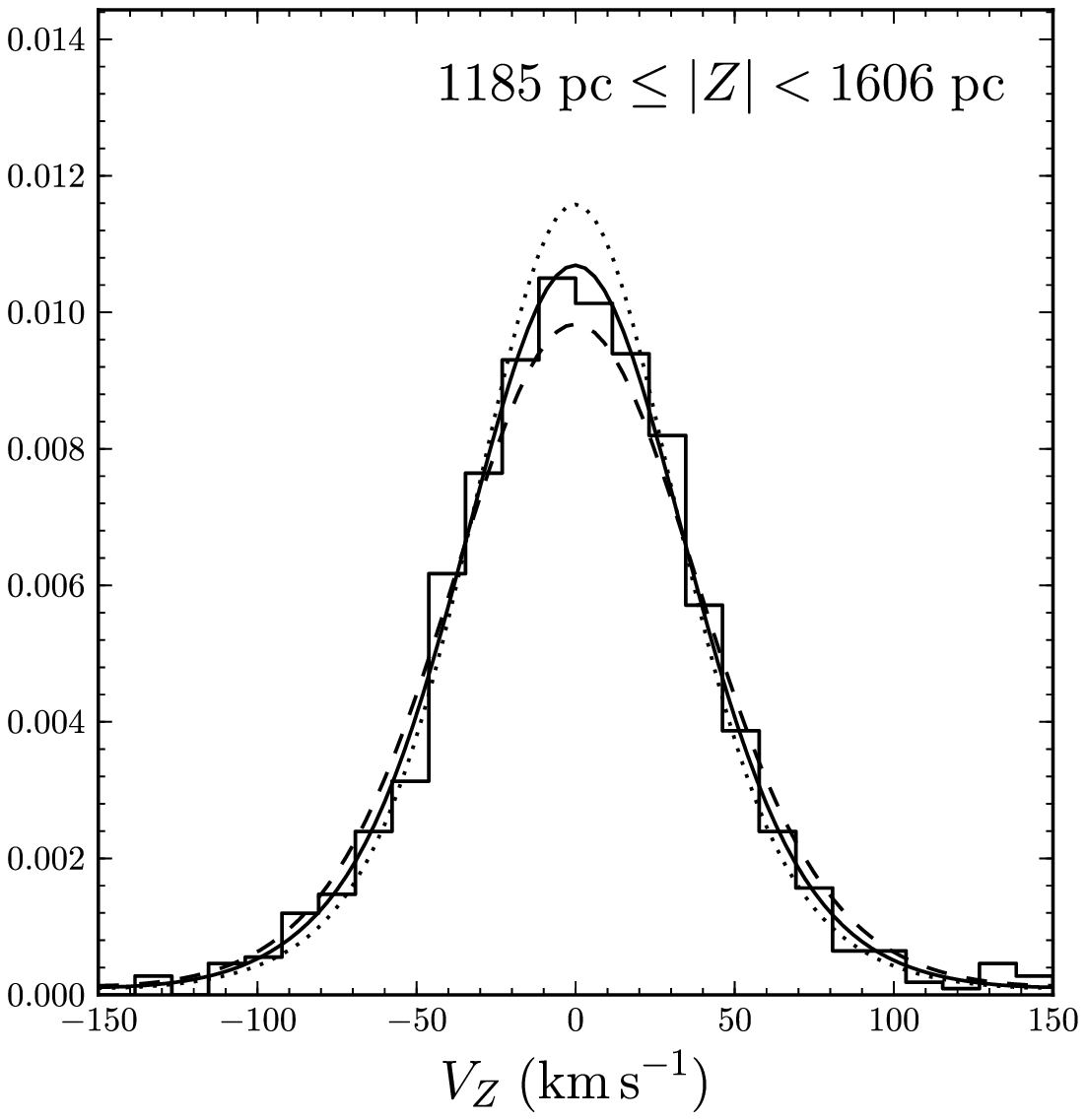}
  \includegraphics[width=0.195\textwidth,clip=]{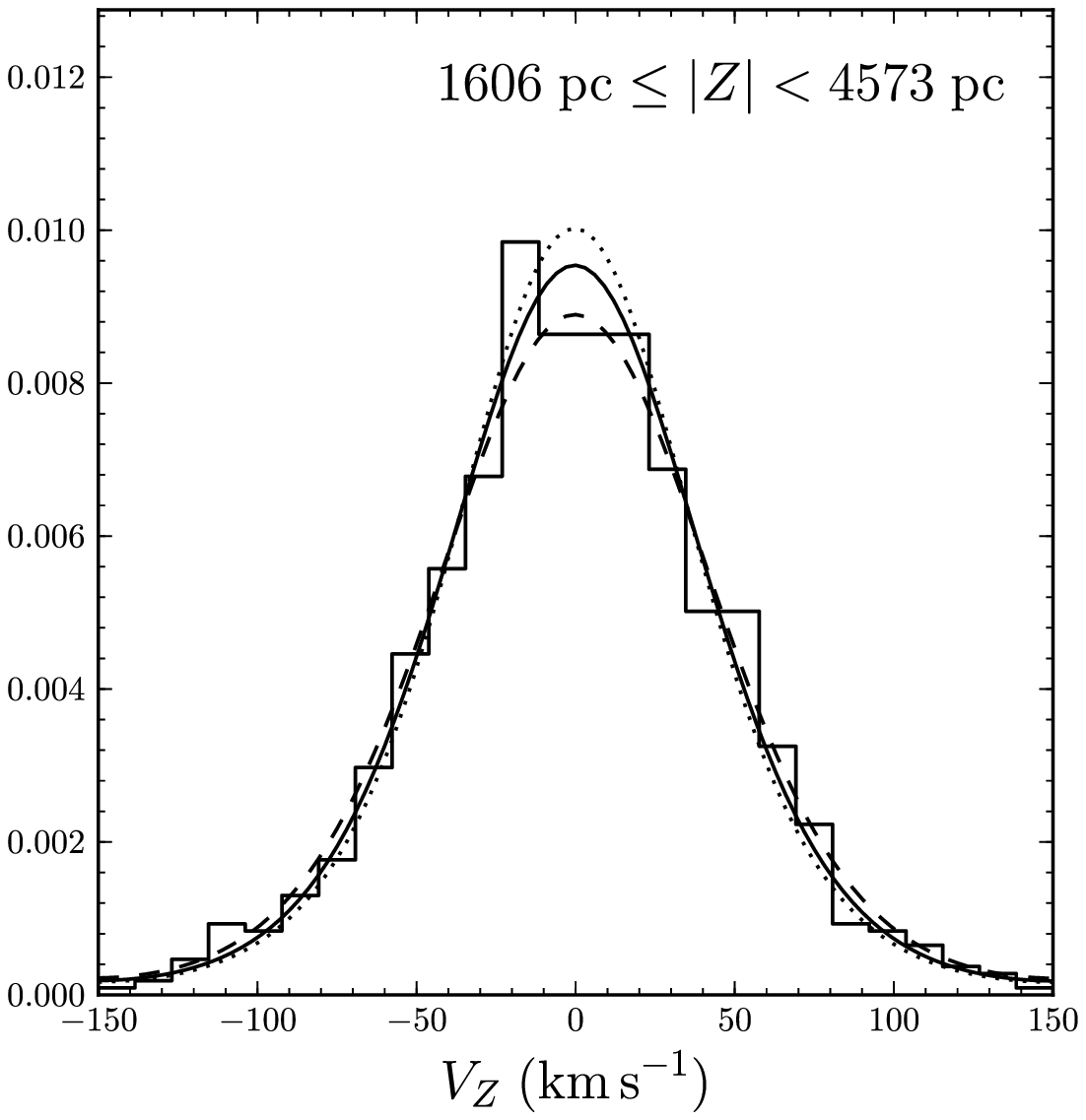}
  \caption{Same as \figurename~\ref{fig:datamodel_vzdist_aenhanced},
    but for intermediate \map s. The alternative models are described
    in the caption of
    \figurename~\ref{fig:datamodel_zdist_aintermediate}.}\label{fig:datamodel_vzdist_aintermediate}
\end{figure}

The sub-division of \aenhanced\ \map s is by far the largest of the
three sub-divisions, allowing us to closely investigate the goodness
of the dynamical fit as a function of
$R$. \figurename~\ref{fig:datamodel_zdist_aenhanced} shows the
comparison between the predicted number counts as a function of
distance from the plane and the data, in 10 different radial bins that
each contain about $1,000$ stars. The number counts for these
\aenhanced\ populations are reproduced excellently by our dynamical
model to almost every single wiggle. As \aenhanced\ \map s primarily
measure $\sigoneone(R)$ at $R \approx6\kpc$ (see
\figurename~\ref{fig:rbest_vs_rmean}), we also include the predictions
from models with very different $\sigoneone(6\kpc)$ from the best-fit
model. All three models shown predict almost the same number-counts
profile.

The distribution of vertical velocities as a function of position for
10 different bins in distance from the midplane is shown in
\figurename~\ref{fig:datamodel_vzdist_aenhanced}. These distributions
show excellent agreement between the data and the best-fit model over
the full range of distances from the plane, from $|Z| \approx500\pc$ to
$|Z| \approx4500\pc$. The alternative models slightly overestimate or
underestimate the width of the distribution, depending on whether
their disk is too heavy or too light. From \figurename
s~\ref{fig:datamodel_zdist_aenhanced} and
\ref{fig:datamodel_vzdist_aenhanced} it is clear that the number-count
measurements are the most informative, such that the dynamical fit
works essentially as follows: (a) for a given potential, the qDF
parameters are adjusted to match the observed number counts without
regard as to whether the vertical velocities are fit well; (b) the
best-fitting potential is the one for which these qDF parameters also
fit the vertical velocity distribution. This is essentially the same
procedure that was used more explicitly by \citet{Kuijken89a}.

\figurenames~\ref{fig:datamodel_zdist_aintermediate} and
\ref{fig:datamodel_vzdist_aintermediate} show similar comparisons for
the intermediate \map s. These demonstrate that the vertical number
counts are matched well, although with a slight overprediction of the
scale height around $R = 8\kpc$. Heavier and lighter disks (with
changes to \sigoneone\ around $7\kpc$ for intermediate \map s) again
predict almost the same vertical
profile. \figurename~\ref{fig:datamodel_vzdist_aintermediate} shows
that the distribution of vertical velocities as a function of height
is excellently fit by the best-fit dynamical model; the predictions of
the heavier and lighter disks are obviously ruled out for these
populations.

\begin{figure}[t!]
  \includegraphics[width=0.23\textwidth,clip=]{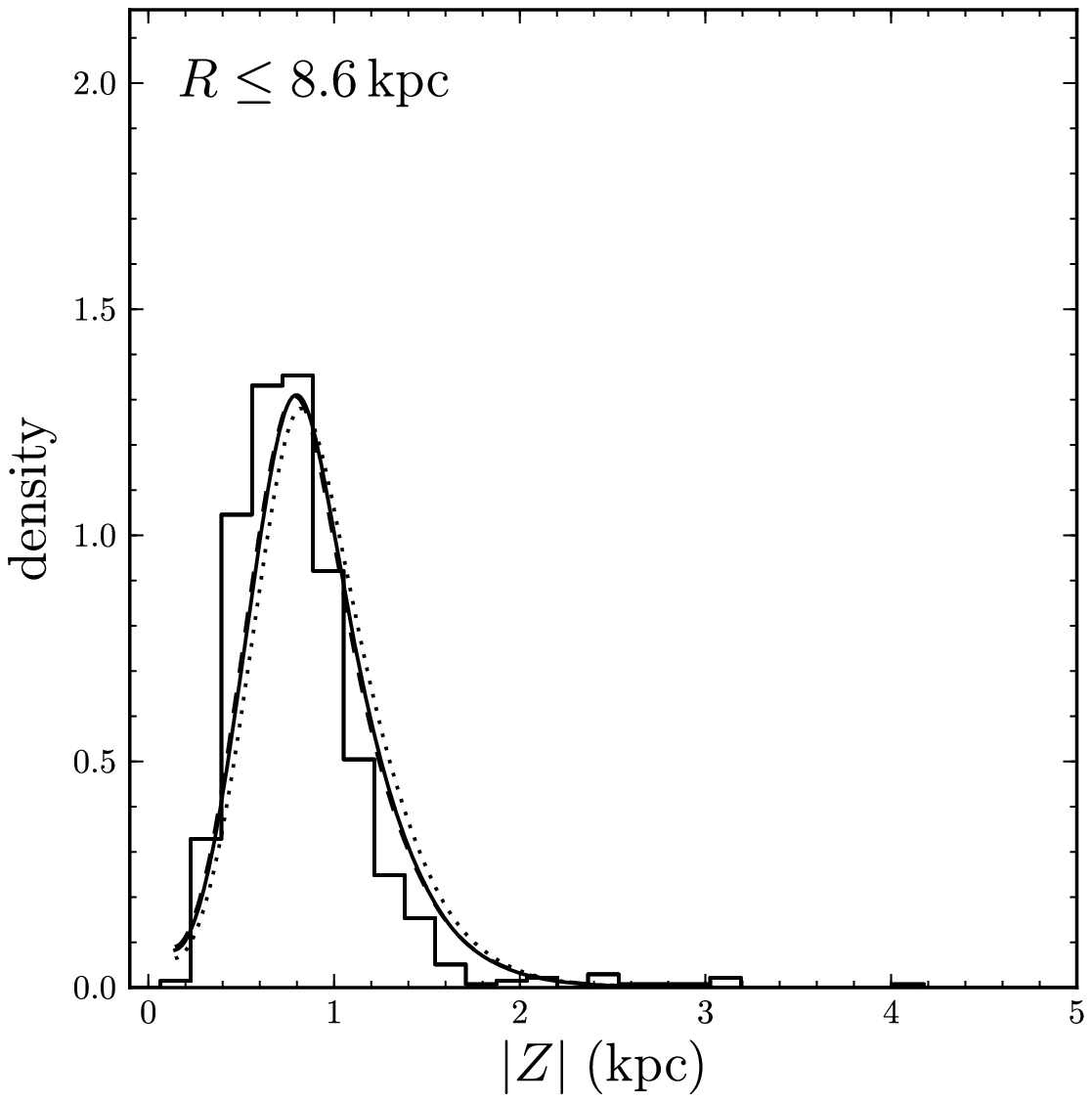}
  \includegraphics[width=0.23\textwidth,clip=]{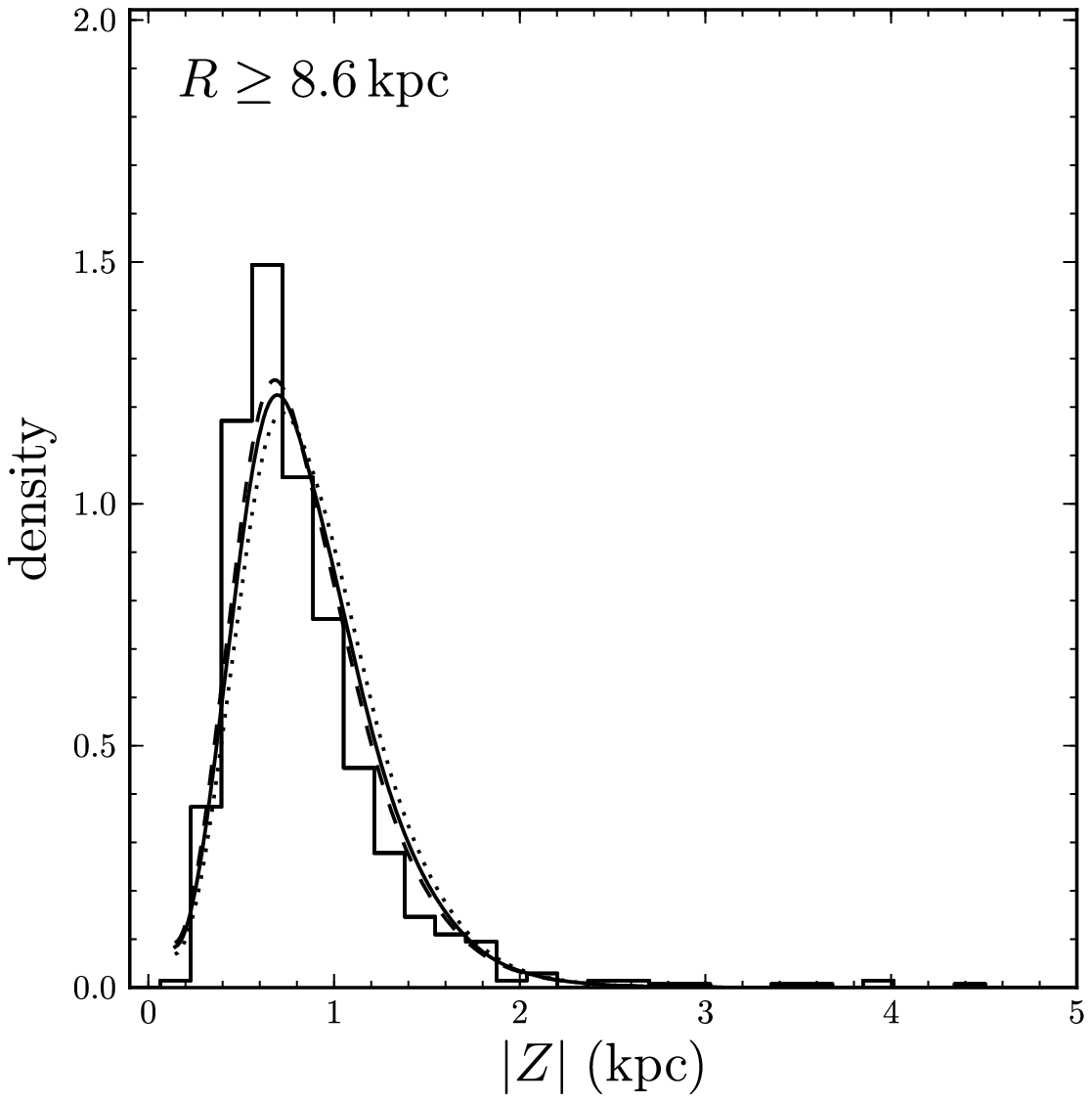}
  \caption{Same as \figurename~\ref{fig:datamodel_zdist_aenhanced},
    but for \apoor\ \map s ($\feh > -0.1$). The
    alternative dashed and dotted models now correspond to
    $\sigoneone(8\kpc) = 87\,M_\odot\pc^{-2}$ and $\sigoneone(8\kpc)
    = 45\,M_\odot\pc^{-2}$.}\label{fig:datamodel_zdist_apoor}
\end{figure}

\begin{figure}[t!]
  \includegraphics[width=0.23\textwidth,clip=]{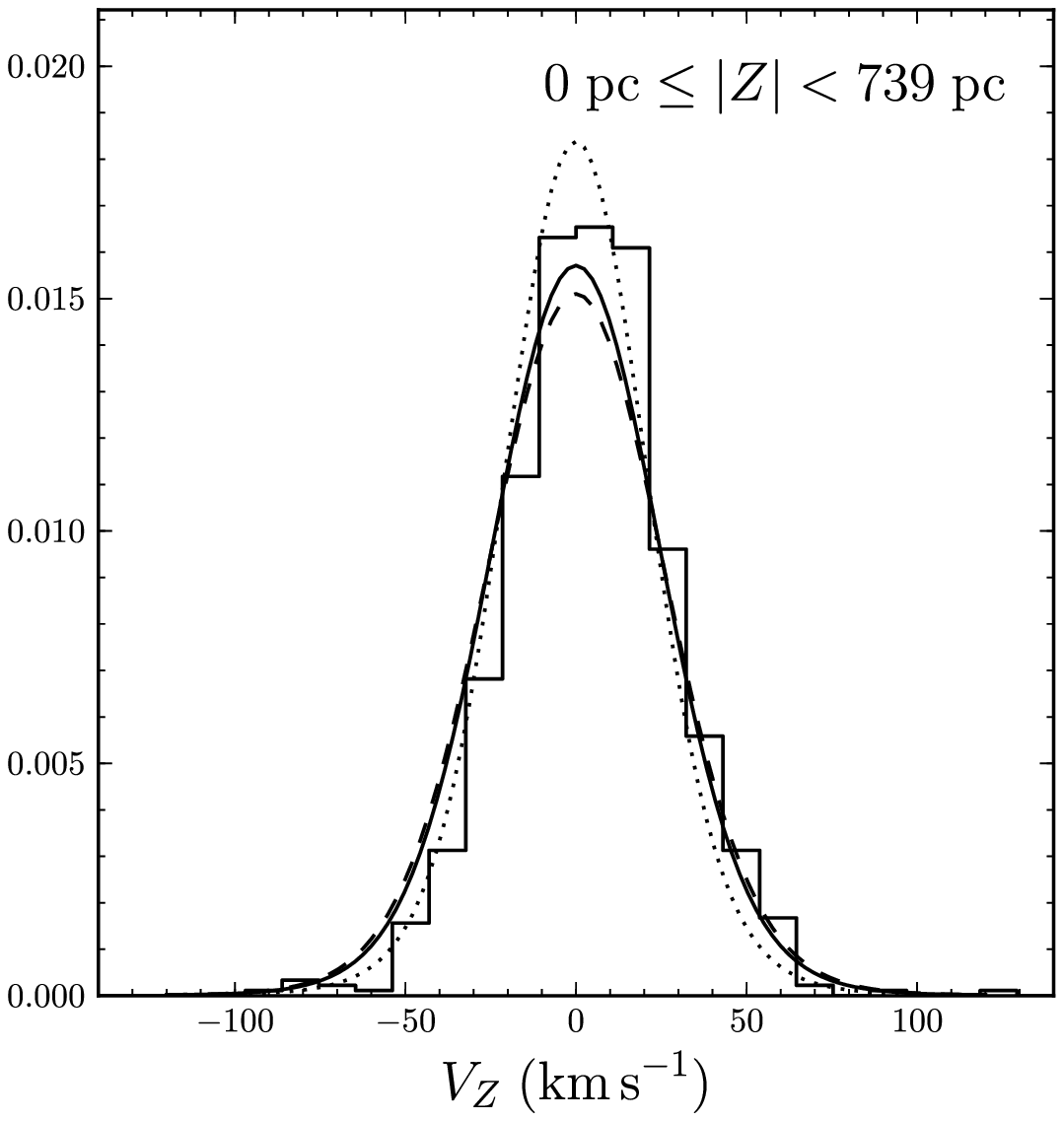}
  \includegraphics[width=0.23\textwidth,clip=]{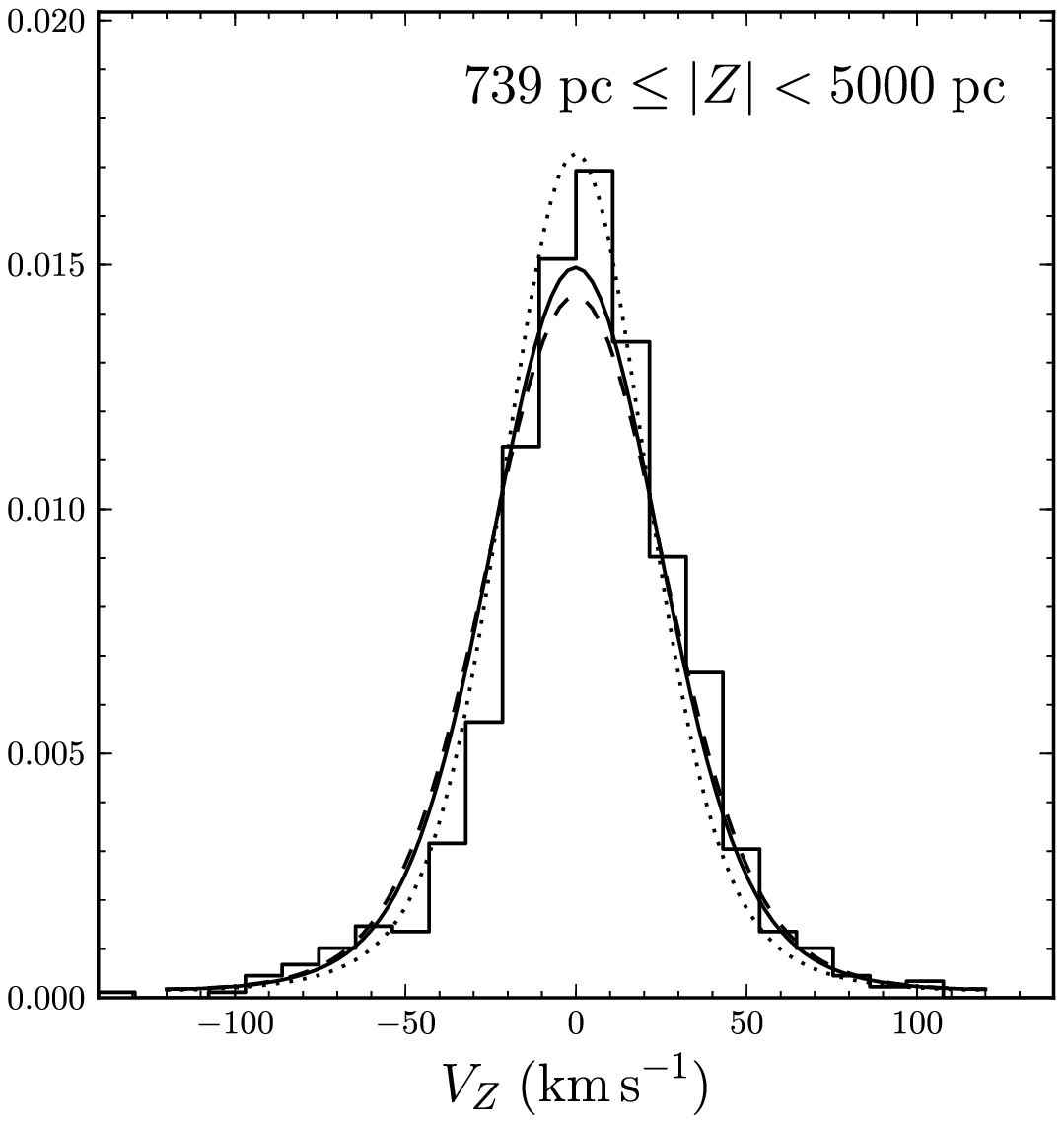}
  \caption{Same as \figurename~\ref{fig:datamodel_vzdist_aenhanced},
    but for \apoor\ \map s. The alternative models are described in
    the caption of
    \figurename~\ref{fig:datamodel_zdist_apoor}.}\label{fig:datamodel_vzdist_apoor}
\end{figure}

Finally, \figurenames~\ref{fig:datamodel_zdist_apoor} and
\ref{fig:datamodel_vzdist_apoor} show data--model comparisons for the
\apoor\ populations, split only into two radial and vertical bins here
because only about $2,000$ stars have $\feh > -0.1\dex$ in our
selection of the SEGUE G-type dwarfs. The comparison is therefore not
as fine-grained as for the \aenhanced\ and intermediate divisions, but
the correspondence between the data and the model are similar as for
the other divisions: the vertical number-count profiles are predicted to
be similar for the best-fit potential and the heavier- and
lighter-disk alternative models; the distribution of the vertical
velocities clearly prefers the best-fit model over the alternatives. 

The fact that both the vertical number counts and the distribution of
vertical velocities is matched in detail by our 3-action based qDF
model proves that a description of these populations as
dynamically-relaxed populations makes sense. However, the detailed
comparison of the vertical number-counts of the best-fit dynamical
model and the data for all three sub-divisions shows a slight
underprediction of the number counts at $|Z| \approx1\kpc$ and $R
\approx 8\kpc$, which is likely due to substructure that is not
captured by our model. Clearly, investigation of the residuals from
the best-fit dynamical models for \map s in the future will be useful
for determining the fine-grained orbital structure of \map s, which
may contain traces of dynamical or accreted substructures.

\end{document}